\documentclass[aps, pre, showpacs, showkeys,amsmath,amssymb,amsfonts,twocolumn,superscriptaddress]{revtex4-1}
\usepackage{amsmath}
\usepackage{multirow}
\usepackage{amsthm}                                  
\usepackage{amscd}   
\usepackage{slashbox} 
\usepackage{graphicx}   
\usepackage{epstopdf}                               
\usepackage{amssymb} 
\usepackage{verbatim}                                 
\usepackage{graphics}                               
\usepackage{subfigure}   
\usepackage{enumitem}                     
\usepackage{bm}                                        
\usepackage{float} 
\usepackage{color} 
\usepackage{soul} 
\usepackage{xcolor} 
\usepackage{textcomp}
\usepackage{stmaryrd}
\usepackage{epstopdf}
\usepackage{mathrsfs}
\usepackage{calligra}
\usepackage{url}
\usepackage[none]{hyphenat}
\usepackage[colorlinks,linkcolor=red,citecolor=blue,urlcolor=blue]{hyperref}

\usepackage{rotating}
\usepackage{array}
\newcolumntype{L}[1]{>{\raggedright\let\newline\\\arraybackslash\hspace{0pt}}m{#1}}
\newcolumntype{C}[1]{>{\centering\let\newline\\\arraybackslash\hspace{0pt}}m{#1}}
\newcolumntype{R}[1]{>{\raggedleft\let\newline\\\arraybackslash\hspace{0pt}}m{#1}}

\usepackage{graphicx}
\usepackage{dcolumn}
\usepackage{bm}
\usepackage[mathlines]{lineno}

\allowdisplaybreaks

\begin{document}

\preprint{APS/123-QED}

\title{Three-dimensional central-moments-based lattice {B}oltzmann method with external forcing: {A} consistent, concise and universal formulation}

\author{Alessandro De Rosis}
\email{derosis.alessandro@icloud.com}
\affiliation{Electric Ant Lab B.V., Science Park 106, 1098 XG Amsterdam, the Netherlands}

\author{Rongzong Huang}
\email{rongzong.huang@tum.de}
\affiliation{School of Mechanical Engineering, Shanghai Jiao Tong University, 200240 Shanghai, China}
\affiliation{Institute of Aerodynamics and Fluid Mechanics, Technical University of Munich, 85748 Garching, Germany}

\author{Christophe Coreixas}
\email{christophe.coreixas@unige.ch}
\affiliation{University of Geneva, Geneva, Switzerland.}

\date{\today}

\begin{abstract}
{The cascaded or central-moments-based lattice Boltzmann method (CM-LBM) is a robust alternative to the more conventional BGK-LBM for the simulation of high-Reynolds number flows. Unfortunately, its original formulation makes its extension to a broader range of physics quite difficult. To tackle this issue, a recent work} [\href{https://journals.aps.org/pre/abstract/10.1103/PhysRevE.95.013310}{A. De Rosis, Phys. Rev. E 95, 013310 (2017)}] {proposed a more generic way to derive concise and efficient three-dimensional CM-LBMs. Knowing the original model also relies on central moments that are derived in an adhoc manner, i.e., by mimicking those of the Maxwell-Boltzmann distribution to ensure their Galilean invariance \emph{a posteriori}, a very recent effort} [\href{https://journals.aps.org/pre/abstract/10.1103/PhysRevE.99.013301}{A. De Rosis and K. H. Luo, Phys. Rev. E 99, 013301 (2019)}] {was proposed to further generalize their derivation. The latter has shown that one could derive Galilean invariant CMs in a \emph{systematic} and \emph{a priori} manner by taking into account high-order Hermite polynomials in the derivation of the discrete equilibrium state.} Combining these two approaches, a compact and mathematically sound formulation of the CM-LBM with external forcing is proposed. More specifically, the proposed formalism fully takes advantage of the D3Q27 discretization by relying on the corresponding set of 27 Hermite polynomials (up to the sixth order) for the derivation of both the discrete equilibrium state and the forcing term. The present methodology is more consistent than previous approaches, as it properly explains how to derive Galilean invariant CMs of the forcing term in an \emph{a priori} manner. Furthermore, while keeping the numerical properties of the original CM-LBM, the present work leads to a compact and simple algorithm, representing a universal methodology based on CMs and external forcing within the lattice Boltzmann framework. To support these statements, mathematical derivations and a comparative study with six other forcing schemes are provided. {The universal nature of the proposed methodology is eventually proved through the simulation of single phase, multiphase (using both pseudo-potential and color-gradient formulations), as well as, magnetohydrodynamic flows.}
\end{abstract}
%
\maketitle


\section{Introduction}
Nowadays, the lattice Boltzmann method (LBM) is a consolidated approach to simulate fluid flows~\cite{benzi1992lattice, SucciBook, succi2015lattice, succi2016chimaera, kruger2016lattice, succi2018lattice}. 
The basic idea is to model the fluid through populations (or distributions) of fictitious particles moving along the links of a fixed Cartesian lattice. 
The space and time evolution of the distributions is predicted by solving the lattice Boltzmann equation that involves a collide-and-stream process, where the collision stage retains the flow physics. 
Then, a proper treatment of the collision process is instrumental to obtain accurate predictions of the fluid dynamics. 

The impressive popularity of the LBM mainly stems from the intrinsic simplicity of the BGK collision operator~\cite{bhatnagar1954model}. 
In short, all the populations are forced to relax (with a common unique rate) towards a discrete equilibrium state derived by applying a Gauss-Hermite quadrature to the continuous Maxwellian distribution~\cite{shan2006kinetic}. Unfortunately, it is well known to be prone to numerical instability in the low-viscosity regime, thus becoming unsuitable for the prediction of turbulent flows. A possible alternative is represented by the multiple-relaxation-time (MRT) LBM, that suggests to perform the collision in a space of raw (or absolute) moments~\cite{d2002multiple}. While second-order ones are related to the flow physics and relax with a frequency directly linked to the fluid kinematic viscosity, higher-order moments are related to non-hydrodynamic modes~\cite{Latt2006165}, and their relaxation rates can be considered as free parameters. The MRT formulation allows to directly tune the relaxation frequencies related to these high-order contributions, eventually leading to an improved stability of the resulting LBM. In addition, the original MRT-LBM further increases the dissipation of acoustic waves, hence reducing their impact on the surrounding fluid dynamics~\cite{marie2009comparison}. Usually, both BGK and MRT LBMs relax to equilibrium states derived through a second-order truncated Taylor expansion in the local Mach number of the above-mentioned continuous Maxwellian distribution. As a consequence, these are Galilean invariant only up to the second order. Unfortunately, the lack of Galilean invariance at higher orders leads to numerical instability in the low viscous regime~\cite{0295-5075-81-3-34005, COREIXAS_PhD_2018}.\\
\indent To cope with this problem, Geier \textit{et al.}~\cite{geier2006cascaded} proposed LBMs based on the concept of central moments (CMs), the latter being obtained by shifting the lattice directions by the local fluid velocity. In the actual practice, it is more convenient to operate in terms of raw moments, thus requiring a transformation between the two types of moments. This is achieved by using binomial formulas, that generate central moments of a certain order as a function of lower order raw moments only. Consequently, the collision in the CM space shows a pyramidal hierarchical topology, where the post-collision state of CMs is constructed starting from the lowest order, and then proceeding in ascending sequence, hence the name ``cascaded'' lattice Boltzmann method (CLBM). Undoubtedly, the CLBM drastically outperforms both BGK and MRT in terms of stability~\cite{geier2006cascaded, geier2007properties, geier2008aliasing, asinari2008generalized, geier2009factorized, geier2015cumulant, DeRosis_central2016, geier2017parametrization, geier2017parametrization2, geier2018fourth, fei2017consistent, fei2018three, shah2017cascaded, kumar2017numerical, sharma2017new, fei2018cascaded, fei2018cascaded_2, safari2018lattice, premnath2009incorporating, premnath2011three, FLD:FLD4208, hajabdollahi2017improving}. 
{The latter point can partially be explained by the \emph{positive} hyperviscosity naturally introduced by CM-LBMs, and that can further be adjusted, through fine tuning of relaxation times related to high-order CMs, in order to find the best compromise between stability and accuracy}~\cite{CHAVEZMODENA_CF_172_2018}.\\
An important step in the design of the CLBM is the match between continuous and discrete central moments, that enforces the Galilean invariance at all orders. In other words, CMs are forced to relax to the equilibrium state of the continuous Maxwellian distribution, rather than to its discrete counterpart. Consequently, this methodology assumes that equilibrium CMs are unchanged by the velocity discretization of the Boltzmann equation, which might not be true depending on both equilibrium CMs of interest and on the considered lattice of discrete velocities. 

More recently, we approached central moments from a different viewpoint~\cite{derosis2016epl_d2q9, de2017nonorthogonal, PhysRevE.95.023311, 0295-5075-117-3-34003}. Given a certain lattice, our methodology consists of building a transformation matrix allowing us to move from the space of populations to the one of central moments and vice versa. The resultant algorithm loses the above-mentioned pyramidal cascaded structure and, as a consequence, it can be interpreted as a ``non-cascaded'' way to apply the collision step in CM space~\cite{de2017preconditioned}. In addition, this methodology is based on CMs computed from the polynomial expansion of populations instead of those derived from the continuous ones. Thus, one does not need anymore to choose which continuous CMs should be kept for the collision step. Consequently, the proposed approach allows the derivation of CM-LBMs for any lattice of discrete velocities in a straightforward manner.
This was thoroughly demonstrated by successfully recovering different sets of governing equations with this approach, hence allowing the simulation of a rich variety of physics problems such as shallow waters~\cite{de2017central_shallow}, magnetohydrodynamic~\cite{de2018advanced}, and multicomponent flows~\cite{saito2018color} among others.

Nonetheless, it is important to understand that the above attempt to derive a CMs-based scheme in the D3Q27 space was relying on the classical second-order truncated equilibrium state~\cite{de2017nonorthogonal}. Even if the stability of the algorithm was greatly improved, it involved a huge computational cost~\cite{fei2018three}. This is mainly due to the use of an improper (or incomplete) equilibrium state, which eventually leads to a non-negligible number of non-zero velocity-dependent equilibrium CMs.
Indeed, as pointed out by Malaspinas~\cite{malaspinas2015increasing} and Coreixas \emph{et al.}~\cite{coreixas2017recursive, COREIXAS_PhD_2018}, the full potential of the D3Q27 velocity space can only be achieved by employing the correct (or complete) equilibrium populations accounting for the whole set of 27 Hermite polynomials. By including these high-order polynomials, the methodology outlined in~\cite{de2017nonorthogonal} was proven to lead to Galilean invariant CMs that were used for the derivation of a force-free D3Q27-CM-LBM~\cite{derosisHermite}. This formulation was further shown to recover the behavior of the original cascaded LBM by only relying on the correct set of Hermite polynomials. Since this family of polynomials is tightly linked to the velocity discretization of the Boltzmann equation~\cite{grad1949kinetic,SHAN_PRL_80_1998,shan2006kinetic}, the correct set of Hermite polynomials is known in an \emph{a priori} way, hence ensuring its consistency for any kind of lattices.

When it comes to external or internal forces, it is well known that  they are ubiquitous in many fluid systems. Indeed, gravity, Coriolis or Lorentz forces are just few examples of the fields that a fluid might undergo. It is then of paramount importance to properly take into account these forces in the lattice Boltzmann framework. To meet this need, a seminal contribution has been proposed by Guo \textit{et al.}~\cite{guo2002discrete} in 2002, where the authors analyzed the discrete lattice effects on the forcing term in the BGK-LBM. Later, Guo $\&$ Zheng~\cite{guo2008analysis} extended the approach to the MRT-LBM. Concerning the adoption of CMs, Premnath $\&$ Banerjee derived forcing schemes based on the cascaded approach for two-and three-dimensional simulations~\cite{premnath2009incorporating,premnath2011three}. In particular, they approximated the forcing term according to the formula in the pioneering work by He \textit{et al.}~\cite{he1998novel}. In 2017, this forcing term has been adopted by Fei $\&$ Luo~\cite{fei2017consistent} within a framework where equilibrium CMs were computed from the continuous Maxwellian distributions. The same force treatment has been adopted in~\cite{PhysRevE.95.023311}, where equilibrium CMs were derived from the classical second-order-truncated expansion. More recently, it has been demonstrated that the forcing term can be written as a Hermite polynomial expansion up to the fourth order in the D2Q9 space~\cite{Huang2018}, where the well-known formula by Guo \textit{et al.}~\cite{guo2002discrete} is recovered if Hermite polynomials of order higher than two are neglected.

In this paper, we derive a three dimensional CMs-based model with external forcing in an \emph{a priori} manner. Following the idea originally proposed by Shan \emph{et al.}~\cite{shan2006kinetic}, and further adopted for the D2Q9 space by Huang \textit{et al.}~\cite{Huang2018}, the forcing term is written as an expansion of Hermite polynomials. By adopting the very same set of Hermite polynomials for both the equilibrium state and the forcing term, it is thoroughly shown that corresponding CMs become Galilean invariant, i.e., all remaining error terms depending on the mean flow vanish. As a consequence, the number of non-zero CMs of the forcing term drops to nine, which drastically simplify its implementation. 

The rest of the paper is organized as follows. The adopted methodology is first devised in Sec.~\ref{Sec2}, and the impact of high-order Hermite polynomials on CMs of the forcing term is thoroughly quantified. In Sec.~\ref{Sec3}, several numerical experiments of increasing complexity highlight the consistency, accuracy {and universality} of the proposed model. A comparative study with six other forcing schemes 
is also provided to further evaluate the numerical properties of the present approach. Some conclusions are finally drawn in Section \ref{Sec4}. For the sake of completeness, details regarding Hermite polynomials, the D2Q9-CM-LBM with external forcing, the lattice Boltzmann method for the magnetic field, {and the CMs-based color-gradient method} are reported in Apps.~\ref{sec:AppHermite},~\ref{sec:AppD2Q9},~\ref{sec:AppMHD}, and ~\ref{sec:AppCG}, respectively.

\section{Lattice Boltzmann schemes with higher-order Hermite polynomials}
\label{Sec2}

Let us consider an Eulerian basis $\bm{x}=[x,y,z]$ and the D3Q27 space~\cite{SucciBook}. The lattice Boltzmann equation predicts the space and time evolution of the particle distribution functions $\displaystyle | f_{i}\rangle = \left[ f_0,\, \ldots ,\, f_i,\, \ldots,\, f_{26}    \right]^{\top}$ that collide and stream along the generic link $i=0 \ldots 26$ with the discrete veloctities $\displaystyle \bm{c}_i=[| c_{ix}\rangle ,\, | c_{iy}\rangle ,\, | c_{iz}\rangle]$ defined as
\begin{widetext}
\begin{align}
%
 | c_{ix}\rangle &=  [0,\phantom{-}1, -1, \phantom{-}0, \phantom{-}0, \phantom{-}0, \phantom{-}0, \phantom{-}1, -1, \phantom{-}1, -1, \phantom{-}1, -1, \phantom{-}1, -1, \phantom{-}0, \phantom{-}0, \phantom{-}0, \phantom{-}0, \phantom{-}1, -1, \phantom{-}1, -1, \phantom{-}1, -1, \phantom{-}1, -1 ]^{\top}, \nonumber\\
| c_{iy}\rangle &= [0, \phantom{-}0, \phantom{-}0, \phantom{-}1, -1, \phantom{-}0, \phantom{-}0, \phantom{-}1, \phantom{-}1, -1, -1, \phantom{-}0, \phantom{-}0, \phantom{-}0, \phantom{-}0, \phantom{-}1, -1, \phantom{-}1, -1, \phantom{-}1, \phantom{-}1, -1, -1, \phantom{-}1, \phantom{-}1, -1, -1 ]^{\top}, \nonumber\\
| c_{iz}\rangle &= [0, \phantom{-}0, \phantom{-}0, \phantom{-}0, \phantom{-}0, \phantom{-}1, -1, \phantom{-}0, \phantom{-}0, \phantom{-}0, \phantom{-}0, \phantom{-}1, \phantom{-}1, -1, -1, \phantom{-}1, \phantom{-}1, -1, -1, \phantom{-}1, \phantom{-}1, \phantom{-}1, \phantom{-}1, -1, -1, -1, -1 ]^{\top}.
\end{align}
\end{widetext}

Let us employ the symbols $| \bullet \rangle$ and $\top$ to denote a column vector and the transpose operator, respectively. Within the BGK approximation~\cite{bhatnagar1954model}, the numerical discretization of the lattice Boltzmann equation with external forcing reads as follows:
\begin{align}
f_i(\bm{x}+\bm{c}_i, t+1) = f_i(\bm{x}, t) &+ \omega \left[ f_i^{eq}(\bm{x}, t) -f_i(\bm{x}, t)   \right] \nonumber\\
&+ \left(1-\dfrac{\omega}{2}\right)\mathcal{F}_i(\bm{x}, t),\label{lbe}
\end{align}
where the lattice unit system was assumed. As usual, this numerical scheme can be divided into two parts, i.e., collision:
\begin{align}
f_i^{\star}(\bm{x}, t) = f_i(\bm{x}, t) &+ \omega \left[  f_i^{eq}(\bm{x}, t) - f_i(\bm{x}, t)\right] \nonumber\\
&+ \left(1-\dfrac{\omega}{2}\right)\mathcal{F}_i(\bm{x}, t),\label{collision}
\end{align}
and streaming:
\begin{equation}\label{streaming}
f_i(\bm{x}+\bm{c}_i, t+1) = f_i^{\star}(\bm{x}, t),
\end{equation}
where the superscript $\star$ denotes post-collision quantities here and henceforth. If not otherwize stated, the dependence on the space and the time will be tacitly assumed in the rest of the paper. The term $\mathcal{F}_i$ accounts for external body forces $\bm{F} = [F_x, \, F_y, \, F_z]$ and it will be discussed later. The fluid density $\rho$ and velocity $\bm{u} = [u_x, \, u_y, \, u_z]$ are computed as 
\begin{align}
\rho &= \sum_i{f_i}, \\
\rho \bm{u} &= \sum_i{f_i {\bm c}_i} + \frac{{\bm F}}{2}, \label{macroscopic}
\end{align}	
respectively. The second term of the right-hand side of Eq.~(\ref{collision}) is the BGK collision operator, that forces all the populations to relax to a discrete local equilibrium $ f_i^{eq}$ with the same rate $\omega = \left(\nu/c_s^2+1/2 \right)^{-1}$, $\nu$ being the fluid kinematic viscosity. Following the works by Malaspinas~\cite{malaspinas2015increasing} and Coreixas~\cite{coreixas2017recursive, COREIXAS_PhD_2018}, the full potential of the D3Q27 lattice can only be achieved by correctly expanding the equilibrium distribution onto the \emph{complete} basis composed of 27 Hermite polynomials
\begin{widetext}
\begin{align}
f_i^{eq} &= w_i \rho \bigg\{1+ \frac{\bm{c}_i \cdot \bm{u}}{c_s^2} + \frac{1}{2 c_s^4} \bigg[\mathcal{H}_{ixx}^{(2)}u_x^2+\mathcal{H}_{iyy}^{(2)}u_y^2+\mathcal{H}_{izz}^{(2)}u_z^2+ 2\bigg(\mathcal{H}_{ixy}^{(2)}u_xu_y+\mathcal{H}_{ixz}^{(2)}u_xu_z+\mathcal{H}_{iyz}^{(2)}u_yu_z\bigg)\bigg]\nonumber\\
&\quad +\frac{1}{2 c_s^6}\bigg[\mathcal{H}_{ixxy}^{(3)} u_x^2 u_y + \mathcal{H}_{ixxz}^{(3)} u_x^2 u_z + \mathcal{H}_{ixyy}^{(3)} u_x u_y^2 + \mathcal{H}_{ixzz}^{(3)} u_x u_z^2  +\mathcal{H}_{iyzz}^{(3)} u_y u_z^2 + \mathcal{H}_{iyyz}^{(3)} u_y^2 u_z + 2 \mathcal{H}_{ixyz}^{(3)} u_x u_y u_z \bigg] \nonumber\\
&\quad +\frac{1}{4 c_s^8} \bigg[ \mathcal{H}_{ixxyy}^{(4)}u_x^2 u_y^2 + \mathcal{H}_{ixxzz}^{(4)}u_x^2 u_z^2 + \mathcal{H}_{iyyzz}^{(4)}u_y^2 u_z^2 + 2 \bigg(\mathcal{H}_{ixyzz}^{(4)} u_x u_y u_z^2 + \mathcal{H}_{ixyyz}^{(4)} u_x u_y^2 u_z + \mathcal{H}_{ixxyz}^{(4)}u_x^2 u_y u_z\bigg) \bigg] \nonumber\\
&\quad +\frac{1}{4 c_s^{10}} \bigg[ \mathcal{H}_{ixxyzz}^{(5)} u_x^2 u_y u_z^2 + \mathcal{H}_{ixxyyz}^{(5)} u_x^2 u_y^2 u_z + \mathcal{H}_{ixyyzz}^{(5)} u_x u_y^2 u_z^2 \bigg]\nonumber\\
&\quad +\frac{1}{8 c_s^{12}} \mathcal{H}_{ixxyyzz}^{(6)} u_x^2 u_y^2 u_z^2\bigg\}, \label{eq:truncation}
\end{align}
\end{widetext}
where the weights are $w_0=8/27$, $w_{1 \ldots 6} = 2/27$, $w_{7 \ldots 18}=1/54$, $w_{19 \ldots 26}=1/216$ and $c_s=1/\sqrt{3}$ is the lattice sound speed. $\mathcal{H}_i^{(n)}$ denotes the $n$th-order Hermite polynomial tensor. Coefficients before these tensors are $(n_x!n_y!n_z! c_s^{2n})^{-1}$ where $n_x$, $n_y$ and $n_z$ are the number of occurrences of $x$, $y$ and $z$ respectively. 
Full expressions of the Hermite polynomials are compiled in App.~\ref{sec:AppHermite}. 

Before going any further, it is important to understand that in the present context the sole purpose of the extended equilibrium state~(\ref{eq:truncation}) is to derive Galilean invariant equilibrium and forcing CMs. Hence, only these CMs are required for the implementation of the D3Q27-CM-LBM with or without external forcing. In addition, one can note that Eq.~(\ref{eq:truncation}) recovers the classical second-order truncated equilibrium if $\mathcal{H}_i^{(3)}$, $\mathcal{H}_i^{(4)}$, $\mathcal{H}_i^{(5)}$ and $\mathcal{H}_i^{(6)}$ are neglected.

Regarding the forcing term, in the continuous Boltzmann equation it is expressed as $\mathcal{F} = \displaystyle - \bm{F} \cdot \boldsymbol \nabla_{\bm{c}} f$, where $\nabla_{\bm{c}}$ is the derivative with respect to the mesoscopic velocity $\bm{c}$~\cite{shan2006kinetic}. Based on the Chapman-Enskog expansion, the Navier-Stokes-Fourier equations can be recovered if the force term is approximated as $\displaystyle - \bm{F} \cdot \boldsymbol \nabla_{\bm{c}} f^{eq}$. Then, it is possible to write an Hermite expansion of this forcing term based on the expansion of the equilibrium state~\cite{shan2006kinetic}
\begin{align}
\mathcal{F} &= - \bm{F} \cdot \nabla_{\bm{c}} f^{eq}\notag\\ 
&=- \bm{F} \cdot \nabla_{\bm{c}} \bigg( w(\bm{c}) \sum_{n=0}^{\infty} \frac{1}{n!} \bm{a}_{eq}^{(n)} \cdot \mathcal{H}^{(n)} \bigg)\notag\\
&=- \bm{F} \cdot \sum_{n=0}^{\infty} \frac{1}{n!} \bm{a}_{eq}^{(n)} \cdot  \nabla_{\bm{c}}\bigg(w(\bm{c})\mathcal{H}^{(n)}\bigg)\notag \\
&=\omega(\bm{c})\sum_{n=1}^{\infty} \frac{1}{n!} \rho\left[\bm{F} \bm{u}^{(n-1)}\right] \cdot \mathcal{H}^{(n)}
\label{generalForcing}
\end{align}
where both Rodrigues' formula and $\bm{a}_{eq}^{(n-1)}=\rho \bm{u}^{(n-1)}$ have been used~\cite{malaspinas2015increasing}. $\bm{u}^{(n-1)}$ is the velocity tensor of rank $(n-1)$. The square bracket $[\bullet ]$ stands for cyclic permutations, e.g., for $n=3$, $$\displaystyle [\bm{F} \bm{u}  \bm{u} ] = \bm{F} \bm{u}  \bm{u} + \bm{u} \bm{F}  \bm{u}  + \bm{u} \bm{u}  \bm{F}.$$
By adopting the extended equilibrium state related to the D3Q27 lattice~(\ref{eq:truncation}), the expansion of the \emph{complete} forcing term reads as
\begin{widetext}
\begin{align}
\mathcal{F}_i =& w_i \rho \bigg\{\frac{\bm{F} \cdot \bm{c}_i}{c_s^2} + \frac{1}{2 c_s^4} \bigg[\mathcal{H}_{ixx}^{(2)}(2u_xF_x)+\mathcal{H}_{iyy}^{(2)}(2u_yF_y)+\mathcal{H}_{izz}^{(2)}(2u_zF_z)+ 2\mathcal{H}_{ixy}^{(2)}(u_xF_y+u_yF_x) + 2\mathcal{H}_{ixz}^{(2)}(u_xF_z+u_zF_x)\nonumber\\
&+ 2\mathcal{H}_{iyz}^{(2)}(u_yF_z+u_zF_y)\bigg] +\frac{1}{2 c_s^6}\bigg[\mathcal{H}_{ixxy}^{(3)} (2u_xu_yF_x +u_x^2F_y) + \mathcal{H}_{ixxz}^{(3)} (2u_xu_zF_x +u_x^2F_z) + \mathcal{H}_{ixyy}^{(3)} (2u_xu_yF_y +u_y^2F_x) \nonumber\\
&+ \mathcal{H}_{ixzz}^{(3)} (2u_xu_zF_z +u_z^2F_x)  +\mathcal{H}_{iyzz}^{(3)} (2u_yu_zF_z +u_z^2F_y) + \mathcal{H}_{iyyz}^{(3)} (2u_yu_zF_y +u_y^2F_z) + 2 \mathcal{H}_{ixyz}^{(3)} (u_xu_yF_z +u_xF_yu_z \nonumber\\
& +F_xu_yu_z) \bigg] +\frac{1}{4 c_s^8} \bigg[ \mathcal{H}_{ixxyy}^{(4)} (2 u_x^2 u_y F_y + 2 u_x u_y^2 F_x)+ \mathcal{H}_{ixxzz}^{(4)} (2 u_x^2 u_z F_z + 2 u_x u_z^2 F_x) + \mathcal{H}_{iyyzz}^{(4)} (2 u_y^2 u_z F_z + 2 u_y u_z^2 F_y) \nonumber\\
& + 2 \mathcal{H}_{ixyzz}^{(4)} (2 u_x u_y u_z F_z + u_x F_y u_z^2 + F_x u_y u_z^2) + 2 \mathcal{H}_{ixyyz}^{(4)} (2 u_x u_y u_z F_y + u_x u_y^2 F_z + F_x u_y^2 u_z) \nonumber\\
& + 2 \mathcal{H}_{ixxyz}^{(4)} (2 u_x u_y u_z F_x + u_y u_x^2 F_z + u_x^2 u_z F_y) \bigg] +\frac{1}{4 c_s^{10}} \bigg[ \mathcal{H}_{ixxyyz}^{(5)} (2 u_x u_y^2 u_z F_x + 2 u_x^2 u_y u_z F_y + u_x^2 u_y^2 F_z)\nonumber\\
& + \mathcal{H}_{ixxyzz}^{(5)} (2 u_x u_y u_z^2 F_x+ u_x^2 u_z^2 F_y + 2 u_x^2 u_y u_z F_z) + \mathcal{H}_{ixyyzz}^{(5)} (u_y^2 u_z^2 F_x + 2 u_x u_y u_z^2 F_y + 2 u_x u_y^2 u_z F_z) \bigg]\nonumber\\
& +\frac{1}{8 c_s^{12}} \mathcal{H}_{ixxyyzz}^{(6)} (2 u_x u_y^2 u_z^2 F_x + 2 u_x^2 u_y u_z^2 F_y + 2 u_x^2 u_y^2 u_z F_z)\bigg\},\label{forcing}
\end{align}
\end{widetext}
where $1/c_s^{2n}$ terms appear due to the use of \emph{discrete} Hermite polynomials $\mathcal{H}_i^{(n)}$. Once again, the full form~(\ref{forcing}) is only used to derive Galilean invariant CMs of the forcing term. As a consequence, only these CMs are required for the implementation of the D3Q27-CM-LBM with external forcing, while the extended equilibrium state only serves a theoretical purpose in the present context. Regarding the forcing term itself,
one can notice that the popular formula by Guo \textit{et al.}~\cite{guo2002discrete} is recovered if terms proportional to $\mathcal{H}_i^{(3)}$, $\mathcal{H}_i^{(4)}$, $\mathcal{H}_i^{(5)}$ and $\mathcal{H}_i^{(6)}$ are neglected.

Now, let us focus on central moments. The pillar to design any CMs-based collision operator is to shift the lattice directions by the local fluid velocity~\cite{geier2006cascaded}. Therefore, it is possible to define $\displaystyle \bar{\bm{c}}_i=[| \bar{c}_{ix}\rangle ,\, | \bar{c}_{iy}\rangle ,\, | \bar{c}_{iz}\rangle]$, where 
\begin{eqnarray}
| \bar{c}_{ix}\rangle &=& |c_{ix}-u_x \rangle, \nonumber \\
| \bar{c}_{iy}\rangle &=& |c_{iy}-u_y \rangle, \nonumber \\
| \bar{c}_{iz}\rangle &=& |c_{iz}-u_z \rangle. \label{central_directions}
\end{eqnarray}
One possibility is to adopt the basis proposed in Ref.~\cite{de2017nonorthogonal}, which is $\bar{\mathcal{T}} = \left[  \bar{T}_0,\, \ldots, \, \bar{T}_i,\, \ldots, \, \bar{T}_{26}  \right]$ with
\begin{eqnarray}\label{basis}
| \bar{T}_0\rangle &=& |1, \, \ldots ,\, 1\rangle,\nonumber \\
| \bar{T}_1 \rangle &=& | \bar{c}_{ix}\rangle ,\nonumber \\ 
| \bar{T}_2 \rangle &=& | \bar{c}_{iy}\rangle ,\nonumber \\
| \bar{T}_3 \rangle &=& | \bar{c}_{iz}\rangle ,\nonumber \\  
| \bar{T}_4 \rangle &=& | \bar{c}_{ix} \bar{c}_{iy}\rangle ,\nonumber \\
| \bar{T}_5 \rangle &=& | \bar{c}_{ix} \bar{c}_{iz} \rangle,\nonumber \\ 
| \bar{T}_6 \rangle &=& | \bar{c}_{iy} \bar{c}_{iz} \rangle,\nonumber \\
| \bar{T}_7 \rangle &=& | \bar{c}_{ix}^2- \bar{c}_{iy}^2 \rangle,\nonumber \\ 
| \bar{T}_8 \rangle &=& | \bar{c}_{ix}^2- \bar{c}_{iz}^2 \rangle,\nonumber \\
| \bar{T}_9 \rangle &=& | \bar{c}_{ix}^2+ \bar{c}_{iy}^2 + \bar{c}_{iz}^2 \rangle,\nonumber \\
| \bar{T}_{10} \rangle &=& | \bar{c}_{ix}\bar{c}_{iy}^2+ \bar{c}_{ix}\bar{c}_{iz}^2 \rangle,\nonumber \\ 
| \bar{T}_{11} \rangle &=& | \bar{c}_{ix}^2\bar{c}_{iy}+ \bar{c}_{iy}\bar{c}_{iz}^2 \rangle,\nonumber \\
| \bar{T}_{12} \rangle &=& | \bar{c}_{ix}^2\bar{c}_{iz}+ \bar{c}_{iy}^2\bar{c}_{iz} \rangle, \nonumber \\
| \bar{T}_{13} \rangle &=& | \bar{c}_{ix}\bar{c}_{iy}^2- \bar{c}_{ix}\bar{c}_{iz}^2 \rangle,\nonumber \\
| \bar{T}_{14} \rangle &=& | \bar{c}_{ix}^2\bar{c}_{iy}- \bar{c}_{iy}\bar{c}_{iz}^2 \rangle,\nonumber \\
| \bar{T}_{15} \rangle &=& | \bar{c}_{ix}^2\bar{c}_{iz}- \bar{c}_{iy}^2\bar{c}_{iz} \rangle,\nonumber \\
| \bar{T}_{16} \rangle &=& | \bar{c}_{ix} \bar{c}_{iy} \bar{c}_{iz} \rangle,\nonumber \\
| \bar{T}_{17} \rangle &=& | \bar{c}_{ix}^2\bar{c}_{iy}^2+ \bar{c}_{ix}^2\bar{c}_{iz}^2+ \bar{c}_{iy}^2\bar{c}_{iz}^2 \rangle,\nonumber \\
| \bar{T}_{18} \rangle &=& | \bar{c}_{ix}^2\bar{c}_{iy}^2+ \bar{c}_{ix}^2\bar{c}_{iz}^2- \bar{c}_{iy}^2\bar{c}_{iz}^2 \rangle,\nonumber \\
| \bar{T}_{19} \rangle &=& | \bar{c}_{ix}^2\bar{c}_{iy}^2- \bar{c}_{ix}^2\bar{c}_{iz}^2 \rangle,\nonumber \\
| \bar{T}_{20} \rangle &=& | \bar{c}_{ix}^2 \bar{c}_{iy} \bar{c}_{iz} \rangle,\nonumber \\
| \bar{T}_{21} \rangle &=& | \bar{c}_{ix} \bar{c}_{iy}^2 \bar{c}_{iz} \rangle,\nonumber \\
| \bar{T}_{22} \rangle &=& | \bar{c}_{ix} \bar{c}_{iy} \bar{c}_{iz}^2 \rangle,\nonumber \\
| \bar{T}_{23} \rangle &=& | \bar{c}_{ix} \bar{c}_{iy}^2 \bar{c}_{iz}^2 \rangle,\nonumber \\
| \bar{T}_{24} \rangle &=& | \bar{c}_{ix}^2 \bar{c}_{iy} \bar{c}_{iz}^2 \rangle,\nonumber \\ 
| \bar{T}_{25} \rangle &=& | \bar{c}_{ix}^2 \bar{c}_{iy}^2 \bar{c}_{iz} \rangle,\nonumber \\ 
| \bar{T}_{26} \rangle &=& | \bar{c}_{ix}^2 \bar{c}_{iy}^2 \bar{c}_{iz}^2 \rangle.
\end{eqnarray}

Let us first recall important results concerning the force-free CM-LBM. Pre-collision, equilibrium and post-collision CMs are defined as
\begin{eqnarray}
| k_i \rangle &=& \left[ k_0,\, \ldots, \, k_i,\, \ldots, \, k_{26}   \right]^{\top},\nonumber \\
| k_i^{eq} \rangle &=& \left[ k_0^{eq},\, \ldots, \, k_i^{eq},\, \ldots, \, k_{26}^{eq}   \right]^{\top},\nonumber \\
| k_i^{\star} \rangle &=& \left[ k_0^{\star},\, \ldots, \, k_i^{\star},\, \ldots, \, k_{26}^{\star}   \right]^{\top},
\end{eqnarray}
respectively. The first two quantities are evaluated by applying the transformation matrix $\bar{\mathcal{T}}$ to the corresponding distributions, i.e,
\begin{eqnarray}
| k_i \rangle &=& \bar{\mathcal{T}}^{\top} |f_i \rangle, \nonumber\\
| k_i^{eq} \rangle &=& \bar{\mathcal{T}}^{\top} |f_i^{eq} \rangle,
\end{eqnarray}
where $|f_i^{eq} \rangle = [f_0^{eq} ,\, \ldots f_i^{eq}, \, \ldots f_{26}^{eq} ]^{\top}$. By adopting a Hermite polynomial expansion of the equilibrium state up to $\mathcal{H}_i^{(6)}$ (Eq.~(\ref{eq:truncation})), equilibrium CMs read as follows
\begin{eqnarray}
k_0^{eq} &=& \rho,\nonumber \\
k_9^{eq} &=& 3\rho c_s^2,\nonumber \\
k_{17}^{eq} &=& \rho c_s^2,\nonumber \\
k_{18}^{eq} &=& \rho c_s^4,\nonumber \\
k_{26}^{eq} &=& \rho c_s^6, \label{equil_central_moments6}
\end{eqnarray}
with $k_{1 \ldots 8}^{eq} = k_{10 \ldots 16}^{eq} = k_{19 \ldots 25}^{eq} = 0$. Notably, only five equilibrium CMs assume non-zero values, and they have the same form as those of the continuous Maxwellian. In other words, Galilean invariant equilibrium CMs are obtained when the full set of Hermite polynomials is considered, as originally demonstrated in Ref.~\cite{derosisHermite}.

The post-collision CMs can be written as
\begin{equation}
| k_i^{\star} \rangle = \left( \mathbf{I}-\boldsymbol \Lambda \right) \bar{\mathcal{T}}^{\top} |f_i \rangle + \boldsymbol \Lambda \bar{\mathcal{T}}^{\top} |f_i^{eq} \rangle,
\end{equation}
or
\begin{equation}
| k_i^{\star} \rangle = \left( \mathbf{I}-\boldsymbol \Lambda \right) |k_i \rangle + \boldsymbol \Lambda |k_i^{eq} \rangle,
\end{equation}
where $\mathbf{I}$ is the $27 \times 27$ unit tensor and 
\begin{equation}
\displaystyle \boldsymbol \Lambda = \mathrm{diag}\left[1,\,1,\,1,\,1, \, \omega,\, \omega,\, \omega,\, \omega,\, \omega,\,1,\,\ldots, \, 1    \right]
\label{eq:relaxationMatrix}
\end{equation}
is the $27 \times 27$ relaxation matrix adopted in the present work. The latter allows us to increase the numerical stability through (1) the overdissipation of acoustic waves~\cite{d2002multiple}, and (2) the equilibriation of high-order moments~\cite{Latt2006165}. The post-collision state of CMs then becomes
\begin{eqnarray}
k_0^{\star} &=& \rho, \nonumber \\
k_4^{\star} &=& \left(1-\omega \right)k_4, \nonumber \\
k_5^{\star} &=& \left(1-\omega \right)k_5, \nonumber \\
k_6^{\star} &=& \left(1-\omega \right)k_6 \nonumber \\
k_7^{\star} &=& \left(1-\omega \right)k_7, \nonumber \\
k_8^{\star} &=& \left(1-\omega \right)k_8, \nonumber \\
k_9^{\star} &=& 3 \rho c_s^2,\nonumber \\
k_{17}^{\star} &=& \rho c_s^2,\nonumber \\
k_{18}^{\star} &=& \rho c_s^4,\nonumber \\
k_{26}^{\star} &=& \rho c_s^6, \label{post_collision_central_moments}
\end{eqnarray}
where
\begin{eqnarray} \label{pre_collision}
k_4 &=& \sum_i f_i \bar{c}_{ix} \bar{c}_{iy},\nonumber \\
k_5 &=& \sum_i f_i \bar{c}_{ix} \bar{c}_{iz},\nonumber \\
k_6 &=& \sum_i f_i \bar{c}_{iy} \bar{c}_{iz},\nonumber \\
k_7 &=& \sum_i f_i ( \bar{c}_{ix}^2 - \bar{c}_{iy}^2 ),\nonumber \\
k_8 &=& \sum_i f_i ( \bar{c}_{ix}^2 - \bar{c}_{iz}^2 ),
\end{eqnarray}
and $\displaystyle k_{1 \ldots 3}^{\star} = k_{10 \ldots 16}^{\star} = k_{19 \ldots 25}^{\star} = 0$. Eventually, the post-collision populations are reconstructed as
\begin{equation}\label{system}
 | f_i^{\star} \rangle = \left(\bar{\mathcal{T}}^{\top} \right)^{-1} | k_i^{\star} \rangle,
\end{equation}
with $|f_i^{\star} \rangle = [f_0^{\star} ,\, \ldots f_i^{\star}, \, \ldots f_{26}^{\star} ]^{\top}$, before being streamed (see Eq.~(\ref{streaming})). One can notice that the computation of only five pre-collision CMs and ten post-collision CMs is required in the present work. Hence, not only the two previous choices (equilibrium~(\ref{eq:truncation}) and relaxation matrix~(\ref{eq:relaxationMatrix})) lead to a robust and accurate CM-LBM~\cite{derosisHermite}, but they also lead to a very concise formulation. The CPU time required for the computation of the remaining CMs can further be reduced by adopting the fast CM transform~\cite{geier2015cumulant}.

In presence of an external force, the post-collision CMs can be rewritten as:
\begin{equation}
| k_i^{\star} \rangle = \left( \mathbf{I}-\boldsymbol \Lambda \right) \bar{\mathcal{T}}^{\top} |f_i \rangle + \boldsymbol \Lambda \bar{\mathcal{T}}^{\top} |f_i^{eq} \rangle + \left( \mathbf{I}- \frac{\boldsymbol \Lambda}{2} \right) \bar{\mathcal{T}}^{\top} |\mathcal{F}_i \rangle,
\end{equation}
or
\begin{equation}
| k_i^{\star} \rangle = \left( \mathbf{I}-\boldsymbol \Lambda \right) |k_i \rangle + \boldsymbol \Lambda |k_i^{eq} \rangle + \left( \mathbf{I}- \frac{\boldsymbol \Lambda}{2} \right)  |R_i \rangle,
\end{equation}
where the CMs of the forcing term are
\begin{equation}
|R_i \rangle= \bar{\mathcal{T}}^{\top} |\mathcal{F}_i \rangle.
\end{equation}
Now, let us proceed by assuming the presence of Hermite polynomials of progressively higher order. If only $\mathcal{H}_i^{(1)}$ and $\mathcal{H}_i^{(2)}$ are considered, Eq.~(\ref{generalForcing}) leads to the following CMs for the forcing term:
\begin{widetext}
\begin{eqnarray}
R_1 &=& F_x, \nonumber \\
R_2 &=& F_y, \nonumber \\
R_3 &=& F_z, \nonumber \\
R_{10} &=& 2F_x c_s^2 - F_x u_y^2 - 2 F_y u_x u_y - F_x u_z^2 - 2 F_z u_x u_z, \nonumber \\
R_{11} &=& 2F_y c_s^2 - F_y u_x^2 - 2 F_x u_y u_x - F_y u_z^2 - 2 F_z u_y u_z, \nonumber \\
R_{12} &=& 2F_z c_s^2 - F_z u_x^2 - 2 F_x u_z u_x - F_z u_y^2 - 2 F_y u_z u_y, \nonumber \\
R_{13} &=& -F_x u_y^2 - 2 F_y u_x u_y + F_x u_z^2 + 2 F_z u_x u_z, \nonumber \\
R_{14} &=& -F_y u_x^2 - 2 F_x u_y u_x + F_y u_z^2 + 2 F_z u_y u_z, \nonumber \\
R_{15} &=& - F_z u_x^2 - 2 F_x u_z u_x + F_z u_y^2 + 2 F_y u_z u_y, \nonumber \\
R_{16} &=& - F_z u_x u_y - F_y u_x u_z - F_x u_y u_z, \nonumber \\
R_{17} &=& 4 F_y u_x^2 u_y + 4 F_z u_x^2 u_z + 4 F_x u_x u_y^2 + 4 F_x u_x u_z^2 + 4 F_z u_y^2 u_z + 4 F_y u_y u_z^2, \nonumber \\
R_{18} &=& 4 F_y u_x^2 u_y + 4 F_z u_x^2 u_z + 4 F_x u_x u_y^2 + 4 F_x u_x u_z^2  - 4 F_z u_y^2 u_z - 4 F_y u_y u_z^2, \nonumber \\
R_{19} &=&  4 u_x (F_x u_y^2 + F_y u_x u_y - F_x u_z^2 - F_z u_x u_z), \nonumber \\
R_{20} &=&  2 u_x (F_z u_x u_y + F_y u_x u_z + 2 F_x u_y u_z), \nonumber \\
R_{21} &=& 2 u_y (F_z u_x u_y + 2 F_y u_x u_z + F_x u_y u_z), \nonumber \\
R_{22} &=& 2 u_z (2 F_z u_x u_y + F_y u_x u_z + F_x u_y u_z), \nonumber \\
R_{23} &=& F_x c_s^4- 3 F_x u_y^2 u_z^2 - 6 F_z u_x u_y^2 u_z - F_x u_y^2/3 - 6 F_y u_x u_y u_z^2  - 2 F_y u_x u_y/3 - F_x u_z^2/3 - 2 F_z u_x u_z/3, \nonumber \\
R_{24} &=&  F_y c_s^4 - 3 F_y u_x^2 u_z^2 - 6 F_z u_y u_x^2 u_z - F_y u_x^2/3 - 6 F_x u_y u_x u_z^2 - 2 F_x u_y u_x/3 - F_y u_z^2/3 - 2 F_z u_y u_z/3, \nonumber \\
R_{25} &=& F_z c_s^4 - 3 F_z u_x^2 u_y^2 - 6 F_y u_z u_x^2 u_y - F_z u_x^2/3 - 6 F_x u_z u_x u_y^2 - 2 F_x u_z u_x/3 - F_z u_y^2/3 - 2 F_y u_z u_y/3, \nonumber \\
R_{26} &=& 8 (F_z u_x^2 u_y^2 u_z + F_y u_x^2 u_y u_z^2 + F_x u_x u_y^2 u_z^2)+ 4c_s^2( F_y u_x^2 u_y  + F_z u_x^2 u_z  + F_x u_x u_y^2 + F_x u_x u_z^2 + F_z u_y^2 u_z + F_y u_y u_z^2),
\end{eqnarray}
\end{widetext}
with $R_0 = R_{4 \ldots 9} = 0$. One can notice that most of these CMs are not Galilean invariant since they contain velocity-dependent terms. By further taking into account third order Hermite polynomials, terms proportional to the square of the velocity vanish, and the corresponding CMs become
\begin{eqnarray}
R_1 &=& F_x, \nonumber \\
R_2 &=& F_y, \nonumber \\
R_3 &=& F_z, \nonumber \\
R_{10} &=& 2F_x c_s^2, \nonumber \\
R_{11} &=& 2F_y c_s^2 , \nonumber \\
R_{12} &=& 2F_z c_s^2, \nonumber \\
R_{17} &=&  - 2 F_y u_x^2 u_y - 2 F_z u_x^2 u_z - 2 F_x u_x u_y^2 - 2 F_x u_x u_z^2 \nonumber \\
&\quad& - 2 F_z u_y^2 u_z - 2 F_y u_y u_z^2, \nonumber \\
R_{18} &=& - 2 F_y u_x^2 u_y - 2 F_z u_x^2 u_z - 2 F_x u_x u_y^2 - 2 F_x u_x u_z^2 \nonumber \\
&\quad& + 2 F_z u_y^2 u_z + 2 F_y u_y u_z^2, \nonumber \\
R_{19} &=&  -2 u_x (F_x u_y^2 + F_y u_x u_y - F_x u_z^2 - F_z u_x u_z), \nonumber \\
R_{20} &=& -u_x (F_z u_x u_y + F_y u_x u_z + 2 F_x u_y u_z), \nonumber \\
R_{21} &=& -u_y (F_z u_x u_y + 2 F_y u_x u_z + F_x u_y u_z), \nonumber \\
R_{22} &=& -u_z (2 F_z u_x u_y + F_y u_x u_z + F_x u_y u_z), \nonumber \\
R_{23} &=& F_x c_s^4 + 3 F_x u_y^2 u_z^2 + 6 F_z u_x u_y^2 u_z + 6 F_y u_x u_y u_z^2, \nonumber \\
R_{24} &=&  F_y c_s^4 + 3 F_y u_x^2 u_z^2 + 6 F_z u_y u_x^2 u_z + 6 F_x u_y u_x u_z^2, \nonumber \\
R_{25} &=& F_z c_s^4 + 3 F_z u_x^2 u_y^2 + 6 F_y u_z u_x^2 u_y + 6 F_x u_z u_x u_y^2, \nonumber \\
R_{26} &=& - 12(F_z u_x^2 u_y^2 u_z - F_y u_x^2 u_y u_z^2 - F_x u_x u_y^2 u_z^2)\nonumber \\
&\quad&+ 2 c_s^2( F_y u_x^2 u_y -  F_z u_x^2 u_z -  F_x u_x u_y^2 - F_x u_x u_z^2\nonumber \\
&\quad& -  F_z u_y^2 u_z - F_y u_y u_z^2),
\end{eqnarray}
with now $R_0 = R_{4 \ldots 9}  = R_{13 \ldots 16} = 0$. If $\mathcal{H}_i^{(4)}$ polynomials are further considered in the definition of the forcing term~(\ref{generalForcing}), we obtain
\begin{eqnarray}
R_1 &=& F_x, \nonumber \\
R_2 &=& F_y, \nonumber \\
R_3 &=& F_z, \nonumber \\
R_{10} &=& 2F_x c_s^2, \nonumber \\
R_{11} &=& 2F_y c_s^2 , \nonumber \\
R_{12} &=& 2F_z c_s^2, \nonumber \\
R_{23} &=& F_x c_s^4 - F_x u_y^2 u_z^2 - 2 F_z u_x u_y^2 u_z - 2 F_y u_x u_y u_z^2, \nonumber \\
R_{24} &=&  F_y c_s^4 - F_y u_x^2 u_z^2 - 2 F_z u_y u_x^2 u_z - 2 F_x u_y u_x u_z^2, \nonumber \\
R_{25} &=& F_z c_s^4 - F_z u_x^2 u_y^2 - 2 F_y u_z u_x^2 u_y - 2 F_x u_z u_x u_y^2, \nonumber \\
R_{26} &=& 8 u_x u_y u_z (F_z u_x u_y + F_y u_x u_z + F_x u_y u_z),
\end{eqnarray}
with $R_0 = R_{4 \ldots 9}  = R_{13 \ldots 22} = 0$. By adding $\mathcal{H}_i^{(5)}$, CMs are
\begin{eqnarray}
R_1 &=& F_x, \nonumber \\
R_2 &=& F_y, \nonumber \\
R_3 &=& F_z, \nonumber \\
R_{10} &=& 2F_x c_s^2, \nonumber \\
R_{11} &=& 2F_y c_s^2 , \nonumber \\
R_{12} &=& 2F_z c_s^2, \nonumber \\
R_{23} &=& F_x c_s^4 , \nonumber \\
R_{24} &=&  F_y c_s^4 , \nonumber \\
R_{25} &=& F_z c_s^4,  \nonumber \\
R_{26} &=& -2 u_x u_y u_z (F_z u_x u_y + F_y u_x u_z + F_x u_y u_z).
\end{eqnarray}
Finally by matching the Hermite polynomial expansion of the forcing term to the one of the extended equilibrium state~(\ref{eq:truncation}), the following expressions are derived:
\begin{eqnarray}
R_1 &=& F_x, \nonumber \\
R_2 &=& F_y, \nonumber \\
R_3 &=& F_z, \nonumber \\
R_{10} &=& 2F_x c_s^2, \nonumber \\
R_{11} &=& 2F_y c_s^2 , \nonumber \\
R_{12} &=& 2F_z c_s^2, \nonumber \\
R_{23} &=& F_x c_s^4 , \nonumber \\
R_{24} &=&  F_y c_s^4 , \nonumber \\
R_{25} &=& F_z c_s^4, \label{force_cms_sixth}
\end{eqnarray}
with $R_0 = R_{4 \ldots 9}  = R_{13 \ldots 22} = R_{26} = 0$, and where all the velocity-dependent terms have vanished, eventually leading to a fully Galilean invariant forcing term. 

Before going any further, let us point out some interesting properties of the present approach. First, the same expressions of $|\mathcal{R}_i \rangle$ in Eqs.~(\ref{force_cms_sixth}) can be achieved when CMs are computed from the continuous Maxwellian distribution~\cite{fei2018three}. This is consistent with a recent paper by De Rosis $\&$ Luo~\cite{derosisHermite}, where it is shown that the CMs of the discrete equilibrium distribution reduce to those of the continuous counterpart when the full set of Hermite polynomials is considered. Hence the proposed approach consistently lead to Galilean invariant forcing terms in accordance with the velocity discretization of interest. This methodology can then be extended to any kind of lattices in a \emph{straightforward} and \emph{a priori} manner, i.e., without having to rely on the CMs of the Maxwellian equilibrium state. Second, as an alternative to the present Hermite expansion of the forcing term, the latter could have been expressed by the formula $\displaystyle \bm{F} \cdot (\bm{c}_i-\bm{u}) f_i^{eq}/\left( \rho c_s^2 \right)$ in He \textit{et al.}~\cite{he1998novel}. 
However, we find that, in this case, the CMs of the force show a dependence on the third power of the velocity. This is due to the fact that the forcing term $\displaystyle - \bm{F} \cdot \boldsymbol \nabla_{\bm{c}_i} f^{eq}$ is only approximated as $\displaystyle - \bm{F} \cdot \frac{\bm{c}_i-\bm{u}}{c_s^2} f^{eq}$ in the pioneering work by He \textit{et al.}~\cite{he1998novel}. This is consistent with the analysis presented in~\cite{Huang2018} for the D2Q9 space, and such a problem is avoided in the present work thanks to the extended Hermite polynomial expansion.

Eventually, post-collision CMs accounting for sixth-order Hermite polynomials in both equilibrium populations and forcing term are:
\begin{eqnarray}
k_0^{\star} &=& \rho, \nonumber \\
k_1^{\star} &=& F_x /2, \nonumber \\
k_2^{\star} &=& F_y /2, \nonumber \\
k_3^{\star} &=& F_z /2, \nonumber \\
k_4^{\star} &=& \left(1-\omega \right)k_4, \nonumber \\
k_5^{\star} &=& \left(1-\omega \right)k_5, \nonumber \\
k_6^{\star} &=& \left(1-\omega \right)k_6 \nonumber \\
k_7^{\star} &=& \left(1-\omega \right)k_7, \nonumber \\
k_8^{\star} &=& \left(1-\omega \right)k_8, \nonumber \\
k_9^{\star} &=& 3\rho c_s^2,\nonumber \\
k_{10}^{\star} &=& F_x c_s^2, \nonumber \\
k_{11}^{\star} &=& F_y c_s^2 , \nonumber \\
k_{12}^{\star} &=& F_z c_s^2, \nonumber \\
k_{17}^{\star} &=& \rho c_s^2,\nonumber \\
k_{18}^{\star} &=& \rho c_s^4,\nonumber \\
k_{23}^{\star} &=& F_x c_s^4/2 , \nonumber \\
k_{24}^{\star} &=& F_y c_s^4/2 , \nonumber \\
k_{25}^{\star} &=& F_z c_s^4/2,\nonumber \\
k_{26}^{\star} &=& \rho c_s^6, \label{post_collision_central_moments_forcing}
\end{eqnarray}
with $k_{13 \ldots 16}^{\star} = k_{19 \ldots 22}^{\star} = 0$. Once again, only five pre-collision CMs need to be computed in the present work. Nevertheless, the inclusion of the forcing term now requires the computation of nine more post-collision CMs, and this brings the number of non-zero post-collision CMs to nineteen instead of ten for the force-free case~(\ref{post_collision_central_moments}).

To conclude, a time iteration of the present algorithm can be summarized as follows:
\begin{enumerate}
\item Computation of macroscopic variables~(\ref{macroscopic});
\item Evaluation of pre-collision CMs $k_{4 \ldots 8}$~(\ref{pre_collision});
\item Calculation of post-collision CMs~(\ref{post_collision_central_moments_forcing});
\item Reconstruction of post-collision populations~(\ref{system});
\item Streaming~(\ref{streaming}).
\end{enumerate}
For the sake of completeness, a script is provided to allow the reader to perform all the symbolic manipulations required for the implementation of the present approach~\footnote{See Supplemental Material at [\protect \url{D3Q27_CentralMomentsWithForcingScheme.m}] for performing all the symbolic manipulations to easily compute all formulas required for the implementation (the transformation matrix, the post-collision CMs, etc).}.
\section{Numerical tests}
\label{Sec3}
In this Section, the accuracy, stability {and universality} of the present approach are discussed using several numerical testcases. Unless differently specified, the density field is initialized as $\rho(\bm{x}, t=0) = \rho_0$, with $\rho_0 = 1$, and the fluid is initially at rest, i.e., $\bm{u} (\bm{x},t=0 )=0$. Velocity boundary conditions are assigned through the regularized technique~\cite{latt2008straight}. Six test cases of increasing complexity are investigated: (1) four-rolls mill, (2) Hartmann flow, (3) two-, (4) three-dimensional Orszag-Tang vortex, (5) static droplet, and (6) the Rayleigh-Taylor instability.

The first three tests are two-dimensional (i.e, only one point is used for the spatial discretization in the $z$ direction), while the last one is a fully three-dimensional case. In the first scenario, the effect of a force driving the flow is simulated. In the second, third and fourth ones, an electrically conductive fluid is considered and the proposed treatment of the forcing scheme is applied to the Lorentz force. A self-consistent internal force is considered in order to simulate a two-components flow in the fifth scenario. Finally, the flexibility of the present approach to deal with more sophisticated LB models is shown through simulations of the Rayleigh-Taylor instability by means of the color-gradient method. 

In all the tests, a diffusive scaling is adopted for the computation of the time step. In addition, all quantities are expressed in lattice units hereafter.

To further evaluate the robustness and the accuracy of the present method (referred to as $\mathrm{M_1}$), a comparative study with six other forcing schemes is conducted:
\begin{itemize}
\item[--] $\mathrm{M_2}$: a recent effort by Fei $\&$ Luo~\cite{fei2017consistent}, where equilibrium CMs are derived from the continuous Maxwellian distribution (equivalent to the six-order truncated expansion of the discrete equilibrium state), and CMs of the forcing terms are computed from the approximation by He \textit{et al.}~\cite{he1998novel} ; 
\item[--] $\mathrm{M_3}$: the methodology discussed in~\cite{PhysRevE.95.023311}, where only a second-order truncated expansion is used for the discrete equilibrium populations, while the forcing scheme is the same as for $\mathrm{M_2}$ (i.e., He \textit{et al.}~\cite{he1998novel});
\item[--] $\mathrm{M_4}$: a model using sixth-order truncated expansion for the discrete equilibrium populations and a second-order force treatment (i.e., Guo \textit{et al.}~\cite{guo2002discrete});
\item[--] $\mathrm{M_5}$: a model using the second-order truncated expansion for both the discrete equilibrium populations and the forcing term (i.e., Guo \textit{et al.}~\cite{guo2002discrete});
\item[--] $\mathrm{M_6}$: the cascaded approach proposed by Premnath $\&$ Banerjee~\cite{premnath2009incorporating};
\item[--] $\mathrm{M_7}$: the exact difference method by Kupershtokh $\&$ Medvedev~\cite{KUPERSHTOKH2006581}.
\end{itemize}
Comparisons between methods $\mathrm{M_1}$, $\mathrm{M_4}$ and $\mathrm{M_5}$ are proposed to better understand the contribution of each high-order modification (equilibrium or forcing term) in the context of the Hermite polynomial expansion. $\mathrm{M_2}$ and $\mathrm{M_4}$, as well as $\mathrm{M_3}$ and $\mathrm{M_5}$, further highlight the impact of the forcing scheme since they are based on the same equilibrium CMs. On the contrary, comparing $\mathrm{M_2}$ and $\mathrm{M_3}$ (or $\mathrm{M_4}$ and $\mathrm{M_5}$) gives more information about the influence of equilibrium CMs. Eventually, both $\mathrm{M_6}$ and $\mathrm{M_7}$ were also considered for the present comparative study since the former was derived in the context of CMs, whereas the latter is a common forcing scheme used in many applications of the lattice Boltzmann framework. 

\subsection{Two-dimensional four-rolls mill}
\begin{figure}[bp!]
\centering
\includegraphics[width=0.3\textwidth]{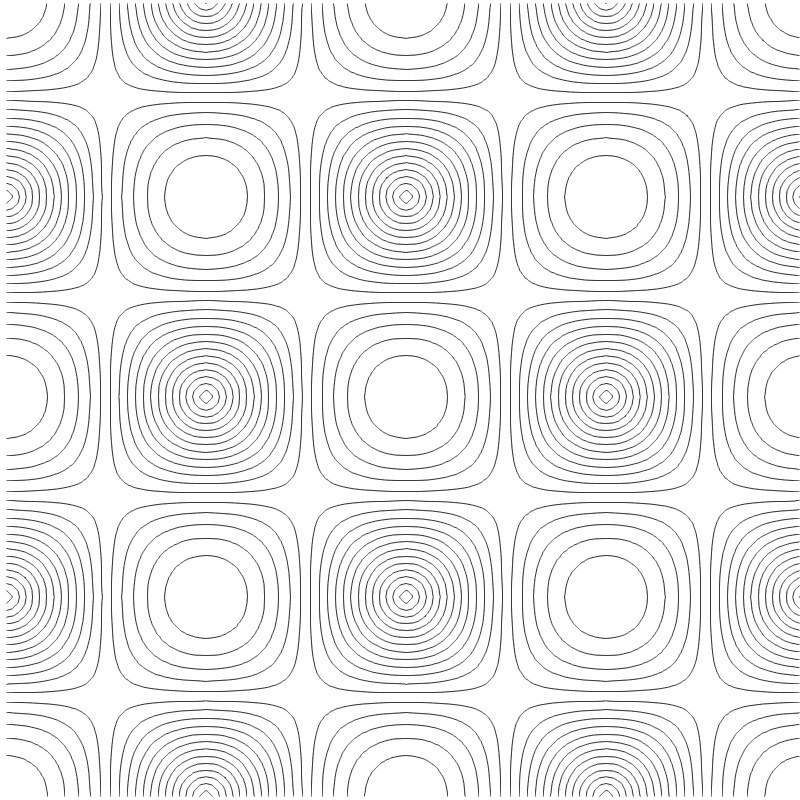}
\caption{Four-rolls mill: contour lines of the velocity magnitude.}
\label{Figure1a}
\end{figure}

Let us consider a square periodic box of size $[x,y] \in [0:N]^2$, $N$ being the number of points in each spatial direction. This test simulates the effect of four rotating cylinders that are replaced by a constant body force driving the flow. It reads as
\begin{equation}
\bm{F}(\bm{x}) =  2\nu u_0 \psi^2\left[\sin(\psi x) \sin(\psi y),\cos(\psi x) \cos(\psi y)   \right].
\end{equation}
This test is known as four-rolls mill~\cite{malaspinas2010lattice}, and admits a steady state. When the latter is reached, the following velocity field is recovered
\begin{equation}
\bm{u} \left(\bm{x} \right) = u_0 \left[\sin \left(\psi x \right) \sin \left(\psi y \right) , \,  \cos \left(\psi x \right) \cos \left(\psi y \right) \right], \label{tg_vel}
\end{equation}
where $u_0=10^{-3}$ and $\psi=2\pi/N$. A contour plot of the velocity magnitude is depicted in FIG.~\ref{Figure1a}, exhibiting the pattern that is typical of this kind of flow.

\begin{figure}[tbp!]
\centering
\includegraphics[scale=1]{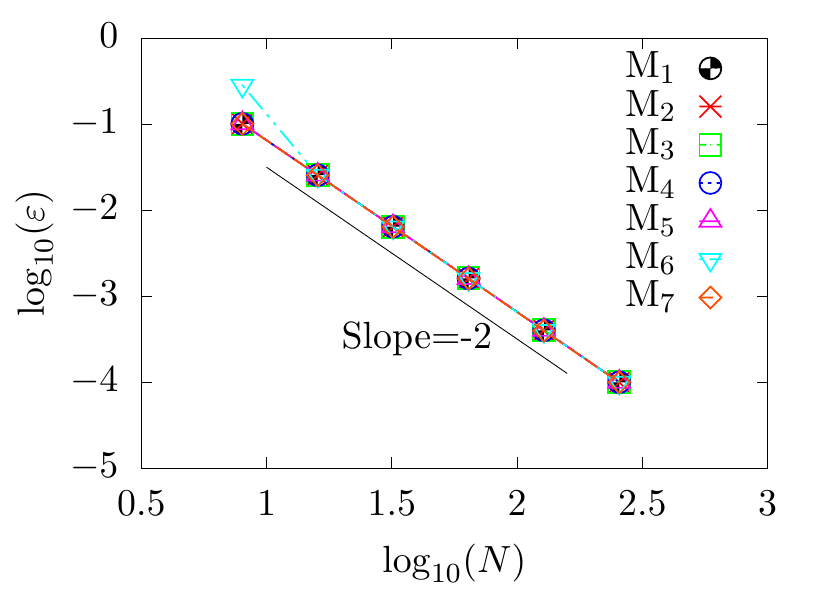}
\caption{Four-rolls mill at $u_0=10^{-3}$: results of a convergence analysis by adopting different lattice Boltzmann models. The black continuous line corresponds to the expected second-order accuracy.}
\label{Figure1}
\end{figure}

\begin{figure}[tp!]
\centering
\includegraphics[scale=1]{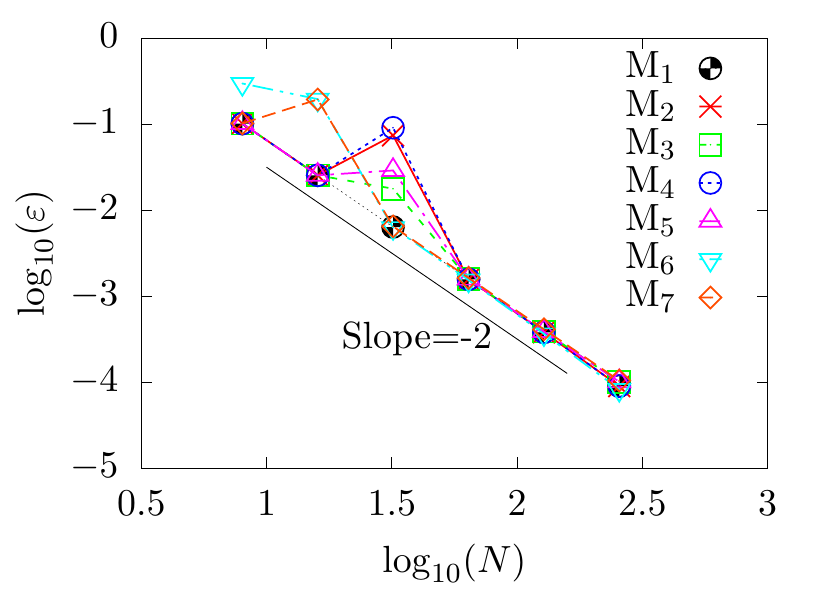}
\caption{Four-rolls mill at $u_0=10^{-2}$: results of a convergence analysis by adopting different lattice Boltzmann models. The black continuous line corresponds to the expected second-order accuracy. Only the present approach ($\mathrm{M_1}$) recovers the proper behavior for all considered grid meshes.}
\label{Figure1b}
\end{figure}

In the present study, a Reynolds number $\displaystyle \mathrm{Re} = u_0 N/\nu$ of 100 is considered. To assert the accuracy of the present numerical scheme, a convergence analysis is carried out by varying $N$ as $N = \left[ 8,\, 16,\, 32,\, 64,\, 128,\, 256 \right]$. Let us collect the velocity field from Eq.~(\ref{tg_vel}) and the one from our numerical experiments in the vectors $\mathbf{r}_{\mathrm{an}}$ and $\mathbf{r}_{\mathrm{num}}$, respectively. By doing so, it is then possible to compute the $L_2$-norm of the relative error $\varepsilon$ between present findings and the analytical solution as
\begin{equation}\label{error}
\varepsilon = \frac{\| \mathbf{r}_{\mathrm{an}}- \mathbf{r}_{\mathrm{num}}  \|}{ \| \mathbf{r}_{\mathrm{an}} \|}.
\end{equation}
As shown in FIG.~\ref{Figure1}, the expected second-order accuracy property of the present method ($\mathrm{M_1}$) is successfully recovered, as an excellent convergence rate equal to 1.998 is found.
The approach $\mathrm{M_6}$ manifests a poor performance for the coarset mesh grid ($N=8$), while other models show the same accuracy. The latter result can simply be explained by the fact that discrepancies between these models are proportional to powers of the velocity.\\ 
\indent By increasing the mean velocity from $u_0=10^{-3}$ to $10^{-2}$, these discrepancies can be exerted, as illustrated in FIG.~\ref{Figure1b}.
Generally speaking, the proposed approach shows a desirable convergence rate $\sim 2$ for each value of $N$, while the other methodologies require finer grid meshes ($N \geq 64$) to recover the expected behavior. By looking at the results more closely, it is clear that $\mathrm{M_6}$ generates large errors for the lowest values of $N$. The exact difference method $\mathrm{M_7}$ deviates from the ideal second-order of convergence when $N=16$. Non-negligible errors are experienced for $N=32$ by adopting $\mathrm{M_2} - \mathrm{M_5}$. Surprisingly, $\mathrm{M_3}$ and $\mathrm{M_5}$ show the lowest mismatch while they are only second-order approaches regarding both the equilibrium state and the forcing scheme. In fact, this can be explained by the fact that $\mathrm{M_2}$ and $\mathrm{M_4}$ are inconsistent models since they do not share the same order for both the equilibrium and the forcing terms. Consequently, one can easily show that high-order CMs of these terms contain velocity-dependent terms that violate the principle of Galilean invariance.

\subsection{Two-dimensional Hartmann flow}
Let us now consider the flow of an electrically conductive fluid of magnetic resistivity $\eta=\nu$, and which is subjected to a magnetic field $\bm{b}$. The interest reader shall refer to App.~\ref{sec:AppMHD} for further details about the computation of the magnetic field $\bm{b}$ in the lattice Boltzmann framework. Let us assume a rectangular domain of length $L_x$ and height $L_y$. Initial conditions consist of $$\bm{b}(\bm{x}, t=0) = [0, b_{y0}, 0].$$
We reproduce the so-called Hartmann flow, that it is analogous to the Poiseuille flow, but in the former case: (1) the fluid is assumed to be conductive, and (2) a constant uniform vertical magnetic field ($b_{y0}$) is enforced at the bottom and top walls, where the no-slip condition is enforced too. At the other sides, the magnetic field is assumed to be periodic. Magnetic boundary conditions are assigned according to~\cite{dellar2013moment}. The analytical solution of the Hartmann flow reads as follows:
\begin{equation} \label{imposed_hartman}
u_x(\bm{x},t) =\frac{4 \nu u_0}{L_y b_{y0} \tanh \left(\mathrm{Ha} \right)} \left[ 1-\frac{\cosh \left(\mathrm{Ha} y^{\prime}/L_y \right)}{\cosh \left(\mathrm{Ha} \right)}    \right],
\end{equation}
where $y^{\prime}=2y-L_y$ and the Hartmann number is defined as $\mathrm{Ha} = b_{y0} L_y/\sqrt{4\rho_0 \nu \eta}$~\cite{dellar2002lattice}. Moreover, Eq.~(\ref{imposed_hartman}) is prescribed at the inlet section, while an outflow boundary condition is applied at the opposite side of the domain as $\bm{n}^{\ddagger} \cdot \nabla \bm{u} = 0$, $\bm{n}^{\ddagger}$ being the unit vector normal to the boundary. The presence of the magnetic field gives rise to the Lorentz force $\bm{F}= \bm{j} \times \bm{b}$, $\bm{j}$ being the electric current that is computed directly from the populations~\cite{pattison2008progress}.

\begin{figure}[h!]
\centering
\includegraphics[scale=1]{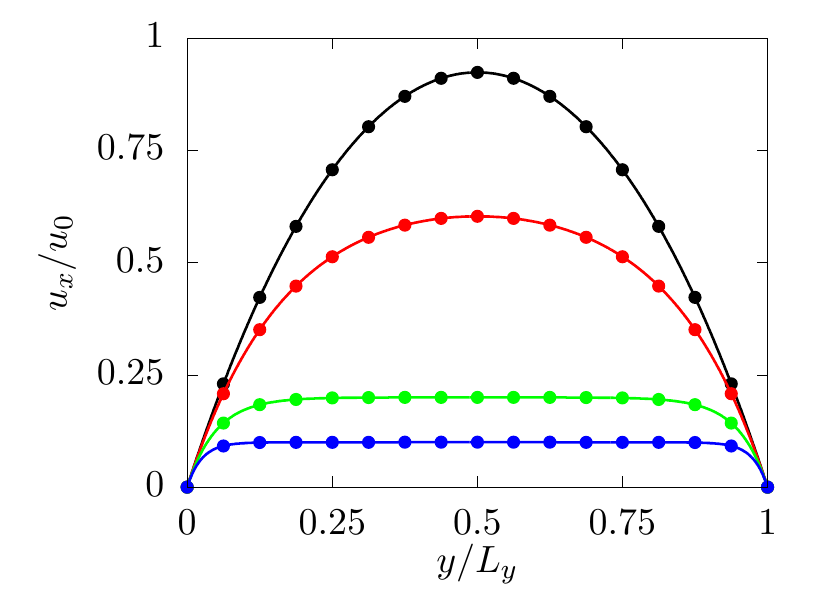}
\caption{Hartmann flow: normalized axial velocity at $\mathrm{Ha}=1$ (black), 3 (red), 10 (green) and 20 (blue). Continuous lines and circles respectively account for the analytical predictions and the numerical solutions obtained with $\mathrm{M_1}$ using $L_y=1025$.}
\label{Figure2a}
\end{figure}

\begin{figure}[h!]
\centering
\includegraphics[scale=1]{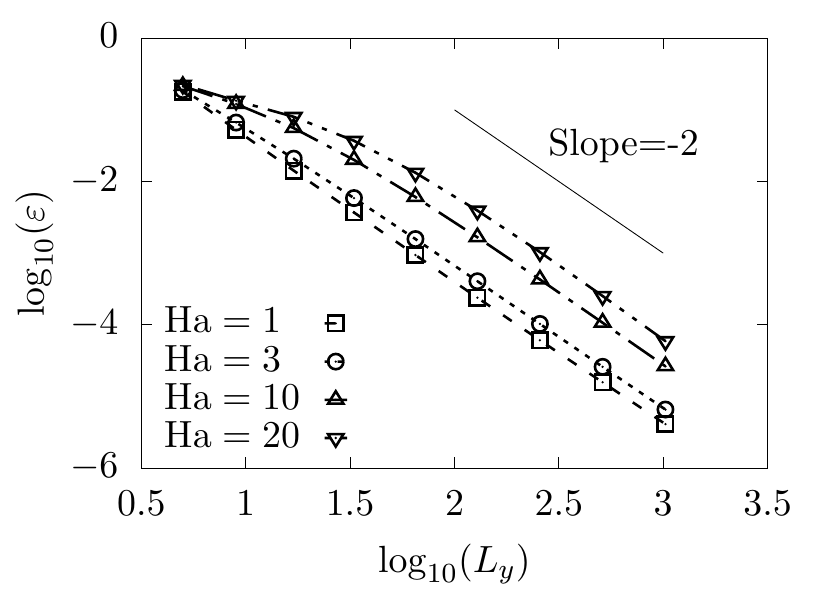}
\caption{Hartmann flow simulation using $\mathrm{M_1}$: convergence analysis for $\mathrm{Ha}=1$ (squares, dashed line), 3 (circles, dotted line), 10 (triangles, dot-dashed line) and 20 (inverted triangles, dot-dot-dashed line). The continuous line denotes a convergence rate equal to 2.}
\label{Figure2}
\end{figure}

\begin{figure}[h!]
\centering
\includegraphics[scale=1]{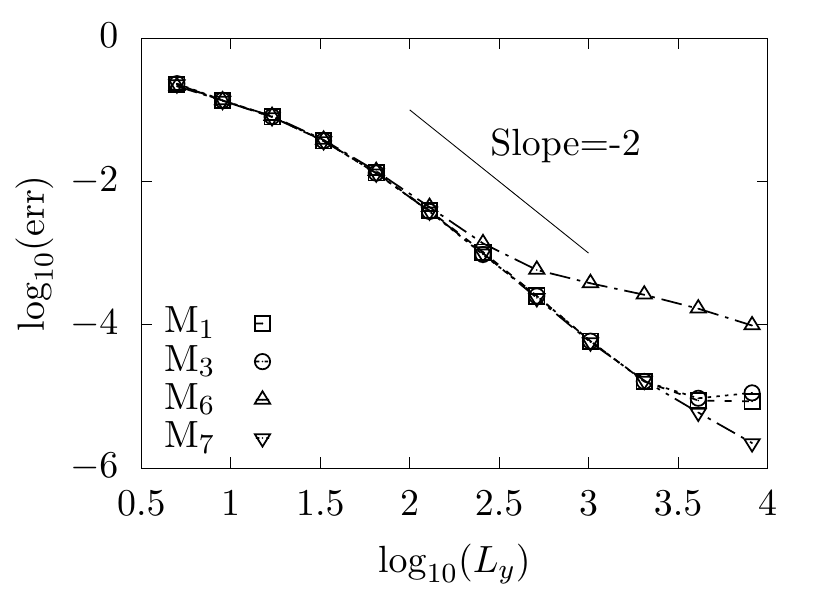}
\caption{Hartmann flow: further investigation of the convergence order of models $\mathrm{M}_1$ (squares), $\mathrm{M}_3$ (circles), $\mathrm{M}_6$ (triangles) and $\mathrm{M}_7$ (inverted triangles), for the most complicated configuration ($\mathrm{Ha}=20$).}
\label{Figure2c}
\end{figure}

In FIG.~\ref{Figure2a}, results obtained with the present approach are compared to the analytical solution~(\ref{imposed_hartman}) for four values of the Hartmann number, i.e., $\mathrm{Ha} = 1 ,\, 3 ,\, 10 ,\, 20$. From them, it is clear that $\mathrm{M_1}$ is able to correctly recover the physics of the Hartmann flow when the grid mesh is fine enough. 
The impact of the grid mesh, as well as, the convergence order of the present method is further studied by evaluating the relative error~(\ref{error}) for different values of $L_y \in [5:1025]$, and using the analytical solution~(\ref{imposed_hartman}).
In FIG.~\ref{Figure2}, the relative error is plotted against the number of points required to discretize the vertical dimension. Interestingly, a poor convergence is experienced for low values of $L_y$, especially for high values of $\mathrm{Ha}$. This behavior can be explained by the presence of progressively thinner Hartmann layers, the latter requiring a larger number of grid points to be successfully and accurately captured. 
\\
\indent This test is further considered to compare the behavior of the proposed approach with other forcing schemes. Before moving to the results, one should note that the Hartmann flow is a purely one-dimensional problem (i.e., $u_y = u_z=0$). Hence,  only terms properly to $u_x$ remain in Eqs.~(\ref{eq:truncation}, \ref{forcing})
. As a consequence, it is expected that $\mathrm{M}_1$ and $\mathrm{M}_4$ reduce to $\mathrm{M}_5$, whereas $\mathrm{M}_2$ should recover the behavior of $\mathrm{M}_3$. Indeed, for any combination of $\mathrm{Ha}$ and $L_y$, it has been confirmed that high-order approaches lead exactly to the same results, in terms of relative errors and velocity profiles, as their low-order versions. In the following, $\mathrm{M_1}$, $\mathrm{M_3}$, $\mathrm{M_6}$ and $\mathrm{M_7}$ are then compared against each other for the most difficult configuration ($\mathrm{Ha}=20$). In addition, the maximal number of points in the vertical direction is increased ($L_y^{\mathrm{max}}=4097$) in order to exert discrepancies between all the models. 
\\
\indent Results compiled in FIG.~\ref{Figure2c} show that the exact difference method exhibits the best performance by keeping the second-order accuracy even for the finest grid mesh. On the contrary, a deterioration of the convergence rate is experienced by all the other methods. More precisely, $\mathrm{M}_1$  and  $\mathrm{M}_3$ show non-negligible errors for $L_y\geq 1025$. Eventually, the cascaded model $\mathrm{M}_6$ starts deviating from the expected second-order convergence for $L_y \geq 129$, which confirms the poor behavior of this forcing scheme, as already pointed out in the four-mill test case.

\subsection{Two-dimensional Orszag-Tang vortex\label{sec:Orszag2D}}
The next test case consists in the simulation of the Orszag-Tang vortex problem~\cite{dellar2002lattice,orszag1979small}. Its simulation represents a hard task since smaller and smaller structures appear in the domain as the Reynolds number grows. This test case is then a perfect candidate to assess (1) the accuracy of the present method to capture the existence of fine flow features, as well as (2) its ability to deal with high local gradients that arise during the simulation~\cite{de2018advanced}.

\begin{figure}[htbp]
\centering
\includegraphics[scale=1]{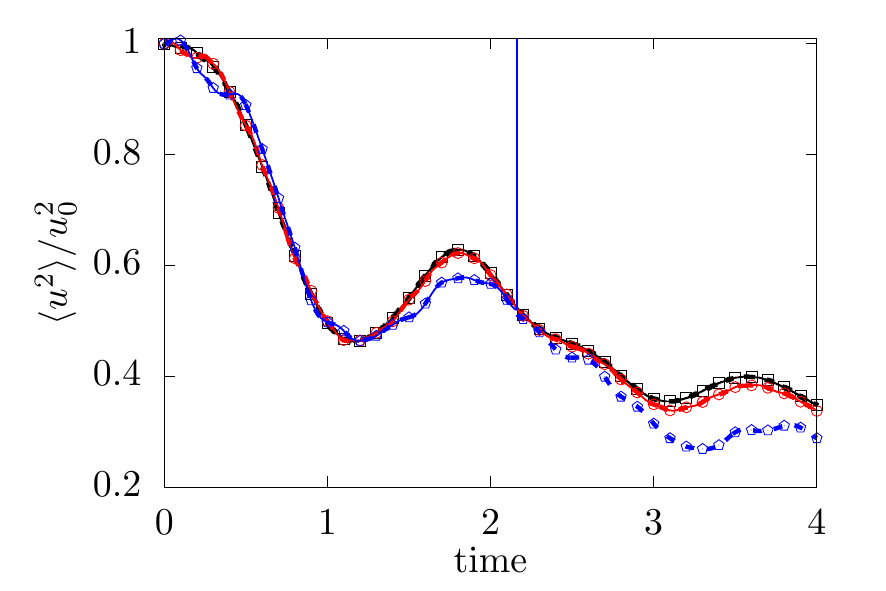}
\caption{Orszag-Tang vortex in two dimensions:  time history of the kinetic energy normalized by its initial value for different Mach numbers, i.e., $\mathrm{Ma}=0.07$ (black), 0.14 (red) and 0.28 (blue). Continuous and dashed lines account for simulations carried out by using Hermite polynomials (used for both equilibrium populations and forcing term) up to the second ($M_5$) and sixth ($M_1$) orders, respectively. The fluid domain consists of 256 points in each direction. The accuracy of these results is assessed by repeating the experiments with a finer grid ($512^2$ lattice points) and by depicting the corresponding solution with symbols (squares for $\mathrm{Ma}=0.07$, circles for $\mathrm{Ma}=0.14$ and pentagons for $\mathrm{Ma}=0.28$). From them, it is clear that a mesh convergence has been achieved.}
\label{Figure3}
\end{figure}

In what follows, a square periodic box of length $2\pi$ is considered with a Reynolds number of $\mathrm{Re} = 1600$, and initial conditions are set to
\begin{eqnarray}
\bm{u}(\bm{x},t=0) &=& u_0 \left[ - \sin (\psi y),\,  \sin (\psi x)\right], \nonumber \\
\bm{b}(\bm{x},t=0) &=& b_0 \left[  -\sin (\psi y) , \,  \sin (2\psi x)  \right].
\end{eqnarray}
In FIG.~\ref{Figure3}, the mean kinetic energy $\langle u^2 \rangle$  (normalized by its initial value $\langle u_0^2 \rangle$) is reported for different values of the Mach number, i.e., $\mathrm{Ma}=u_0/c_s =[0.07, \, 0.14,\, 0.28]$ with $c_s = 1/\sqrt{3}$. Moreover, it is set  $b_0=u_0$. Experiments are carried out by adopting Hermite polynomials up to the second and sixth orders for both equilibrium populations and forcing term, corresponding to the models $M_1$ and $M_5$, respectively. Here, the stability properties of the proposed algorithm clearly manifest. In fact, the two strategies are both stable for $\mathrm{Ma}=0.07$ and 0.14. By further increasing the Mach number, $M_5$ becomes unstable whereas the present approach, $M_1$, leads to the correct behavior.

We also use this scenario to evaluate the accuracy of different LB approaches. In TABLE~\ref{Table2}, the peak values of the current $j_{\mathrm{max}}$ and vorticity $\zeta_{\mathrm{max}}$ are compared to reference high-resolution spectral values in~\cite{dellar2002lattice}. All the models show very good accuracy, with the relative errors lower than $0.22 \%$. To further evaluate their robustness, the Mach number is increased up to $\mathrm{Ma}=0.28$ for all the involved forcing schemes. From this, it is found that only $\mathrm{M_5}$ manifests stability issues.

Now that the good numerical properties of our approach have been rigorously demonstrated in the 2D case, its validation will further be carried out considering 3D flows. Due to the non-negligible CPU time required to run these simulations, $\mathrm{M_1}$ will only be compared to the second-order formulations $\mathrm{M_4}$ and $\mathrm{M_5}$, with the exception of the Rayleigh-Taylor instability test case, where data are already available in the literature.  

\begin{table*}[!htbp]
\centering
\begin{tabular}{ C{1cm}  C{1cm}  C{1cm}  C{1.4cm}  C{1.4cm}   C{1.4cm}  C{1.4cm}  C{1.4cm}   C{1.4cm}    C{1.4cm} }
\hline\hline
\multicolumn{1}{c }{}  & $t$ &~\cite{dellar2002lattice} & $\mathrm{M_1}$ & $\mathrm{M_2}$ & $\mathrm{M_3}$& $\mathrm{M_4}$ & $\mathrm{M_5}$ & $\mathrm{M_6}$ & $\mathrm{M_7}$\\
\hline
\multirow{2}{*}{$j_{\mathrm{max}}$} & 0.5 & 18.24 & 18.26 & 18.26 & 18.26 & 18.26 & 18.26 & 18.27 & 18.26\\
& 1 & 46.59 & 46.66 & 46.66 & 46.66 & 46.66 & 46.66 & 46.69  & 46.66 \\
\hline
\multirow{2}{*}{$\zeta_{\mathrm{max}}$} & 0.5  & 6.758 & 6.756 & 6.756 & 6.756  & 6.756  & 6.756  & 6.755 & 6.756 \\
& 1 & 14.20 & 14.10 & 14.10 & 14.11 & 14.10 & 14.10 & 14.10 & 14.10\\
\hline\hline
\end{tabular}
\caption{Orszag-Tang vortex: findings from different LB schemes and reference spectral values from~\cite{dellar2002lattice} in terms of peak value of the current, $j_{\mathrm{max}}$, and vorticity, $\zeta_{\mathrm{max}}$, at two representative time instants.}
\label{Table2}
\end{table*}

\subsection{Three-dimensional Orszag-Tang vortex}
This test involves an Orszag-Tang vortex that develops in a three-dimensional cubic periodic box of length $2 \pi$ with the initial conditions defined according to~\cite{mininni2006small}, i.e.,
\begin{eqnarray}
\bm{u}(\bm{x},t=0) &=& u_0 \left[ 2  \sin (y),\, 2  \sin (x),\, 0\right],  \nonumber \\
\bm{b}(\bm{x},t=0) &=& 0.8 b_0 [-2\sin (2 y) + \sin (z), \nonumber \\
		&\qquad& 2 \sin (x)+\sin (z) , \sin (x)+\sin (y) ],
\end{eqnarray}
with $u_0=b_0 = 0.0203$. Similarly to the two-dimensional case, here the flow is characterized by the presence of intermittent small-scale structures. However, the topology of the reconnection of the magnetic field is considerably more complicated in three dimensions~\cite{greene1988geometrical}.
Firstly, the accuracy of the method is assessed by running a simulation at $\mathrm{Re=100}$ with three different grid sizes, i.e., $N=\left[32, \, 64,\, 128 \right]$.
\begin{figure}[bp]
\centering
\includegraphics[scale=1]{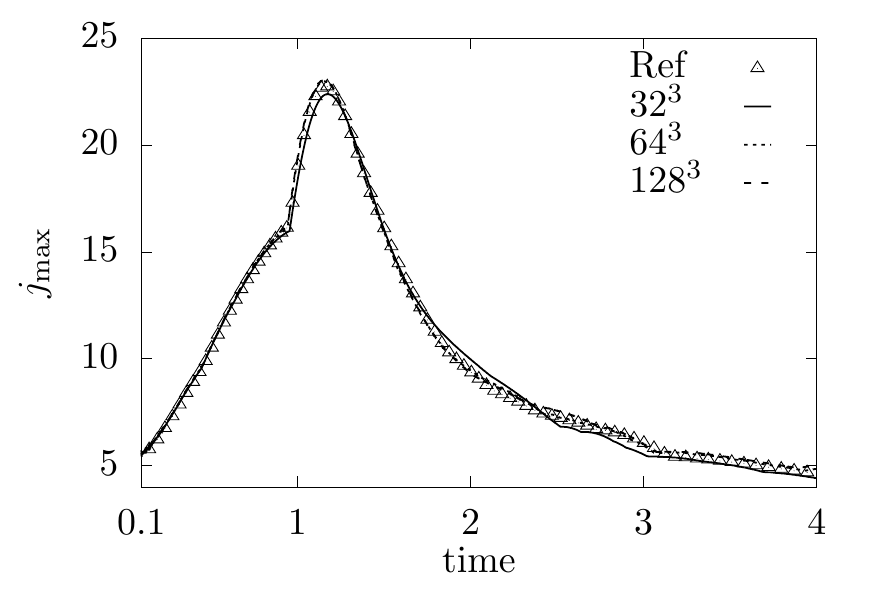}
\caption{Orszag-Tang vortex in three dimensions: the time evolutions of the peak value of the current magnitude, $j_{\mathrm{max}}$, for different grid dimensions, i.e., $32^3$ (continuous line), $64^3$ (dotted line), $128^3$ (dashed line), generated by the present method are compared to those from a high-resolution pseudo-spectral simulation from~\cite{de2017nonorthogonal} (triangles, Ref).}
\label{Figure4}
\end{figure}
In FIG. \ref{Figure4}, the time evolution of the peak value of the current magnitude ($j_{\mathrm{max}}$) is depicted for the different grid dimensions. As $N$ increases, the numerical solution gets progressively closer to the reference one obtained from a high-resolution pseudo-spectral simulation~\cite{de2017nonorthogonal}. This is further quantified by defining the vectors $\mathbf{r}_{\mathrm{ref}}$ and $\mathbf{r}_{\mathrm{num}}$, gathering findings from the reference pseudo-spectral analysis and our results, respectively. Then, we introduce $\varepsilon_{\infty}$ as the $L_{\infty}$-norm of the relative discrepancy, i.e.,
 $$ \displaystyle \varepsilon_{\infty} = \frac{\mathrm{max} |\mathbf{r}_{\mathrm{ref}} -\mathbf{r}_{\mathrm{num}} |    }{\mathrm{max} |\mathbf{r}_{\mathrm{ref}} |,    }$$ 
that is of order $10^{-2}$ for the finest grid, thus highlighting the accuracy of the method. Then, in order to evaluate the stability of the algorithm, higher values of the Reynolds number are considered, i.e., $\mathrm{Re} = [570, \, 1040, \, 3040, \, 5600]$. In FIG.~(\ref{Figure5}), the time evolution of $j_{\mathrm{max}}$ is reported for all these Reynolds numbers. From this, it is clear that the present approach ($\mathrm{M_1}$) remains stable even for the highest value of the Reynolds number. To further highlight the robustness of $\mathrm{M_1}$, the case $\mathrm{Re} = 570$ (the one that is expected to be the least prone to the onset of instability) is also simulated using this time the BGK collision operator~\cite{pattison2008progress}. One can notice that, for the latter collision model, the run blows-up at $t \sim 0.43$. Conversely, the adoption of CMs (either the sixth- or the second-order one, i.e. $\mathrm{M_1}$, $\mathrm{M_4}$ and $\mathrm{M_5}$) overcome this problem, allowing us to simulate a larger time span for the whole desired set of $\mathrm{Re}$. This is consistent with the observations in~\cite{de2018advanced} regarding two-dimensional magnetohydrodynamic flows simulated by a CMs-based D2Q9 model. Nonetheless, by increasing the Mach number, $\mathrm{M_5}$ becomes unstable as it was already the case in Sec.~\ref{sec:Orszag2D}. It should be noted that this paper delivers the first tangible solution of the three-dimensional high-Reynolds Orszag-Tang vortex within the framework of the lattice Boltzmann method.
\begin{figure}[tp]
\centering
\includegraphics[scale=1]{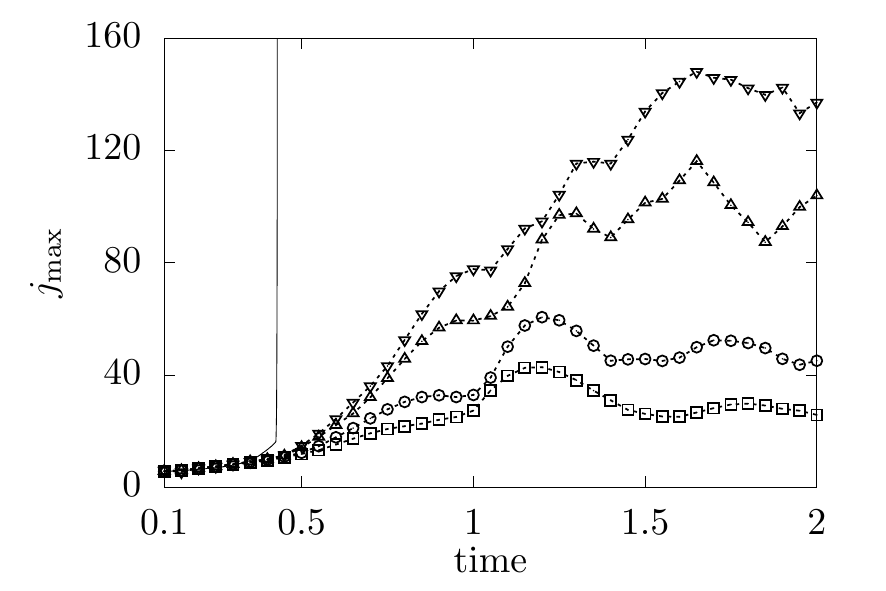}
\caption{Orszag-Tang vortex in three dimensions: time evolution of the peak value of the magnitude of the current for different Reynolds numbers. Symbols denote results from simulations by CMs at $\mathrm{Re} = 570$ (squares), $\mathrm{Re} = 1040$ (circles), $\mathrm{Re} = 3040$ (triangles) and $\mathrm{Re} = 5600$ (inverted triangles). The BGK run at $\mathrm{Re} = 570$ (continuous line) shows an unstable behavior, while CMs-runs always lead to a stable solution. Every run is conducted with $N=128$.}
\label{Figure5}
\end{figure}
Finally, a contour plot of the current magnitude is sketched in FIG.~(\ref{Figure6}) at different time instants for the configuration $\mathrm{Re} = 570$ with $N=128$. One can appreciate that the present method is able to capture the existence of fine flow features, as well as the presence of strong gradients.
\begin{figure}[htbp!]
\centering
\subfigure{\includegraphics[width=0.22\textwidth]{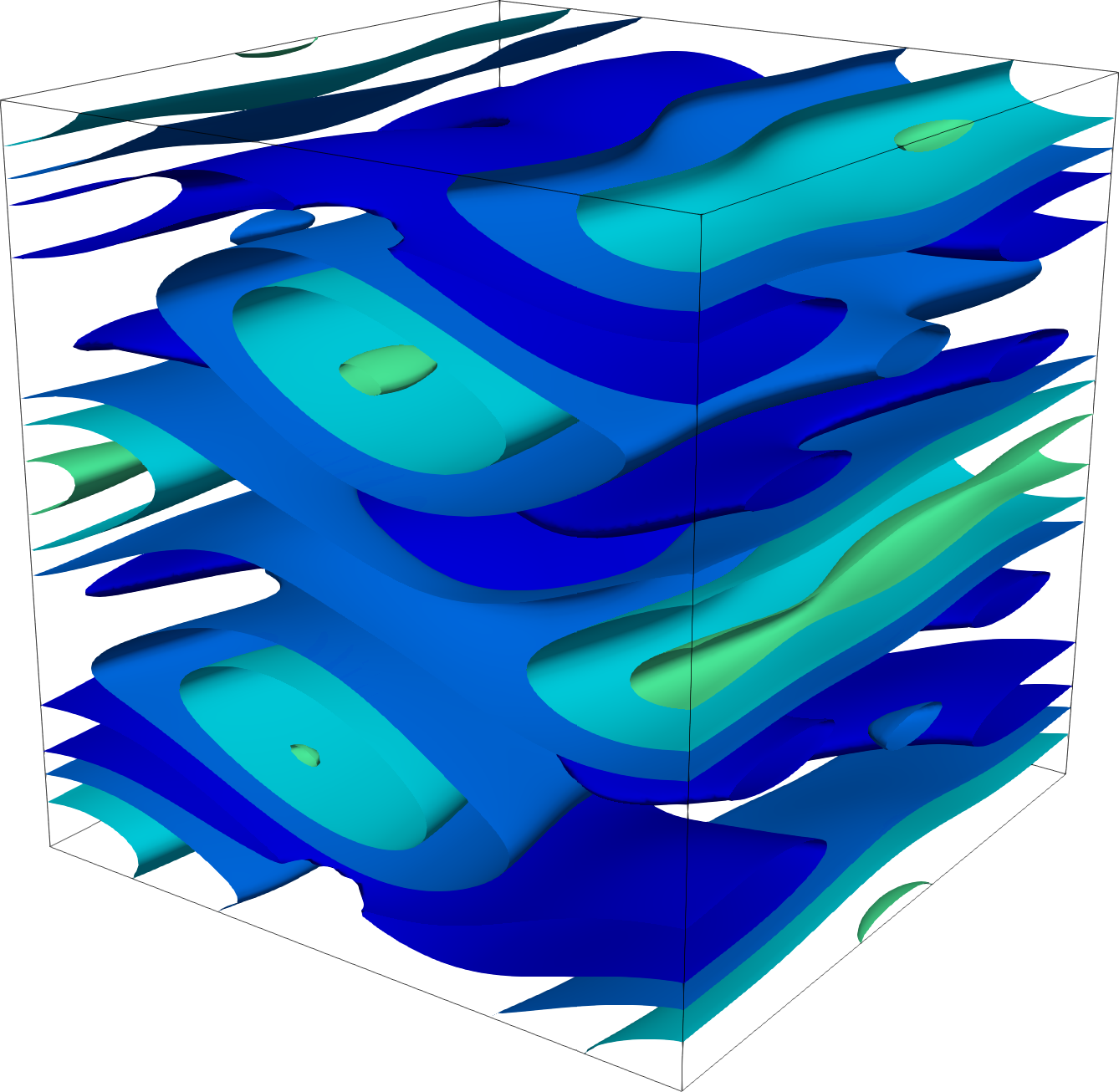}}
\subfigure{\includegraphics[width=0.22\textwidth]{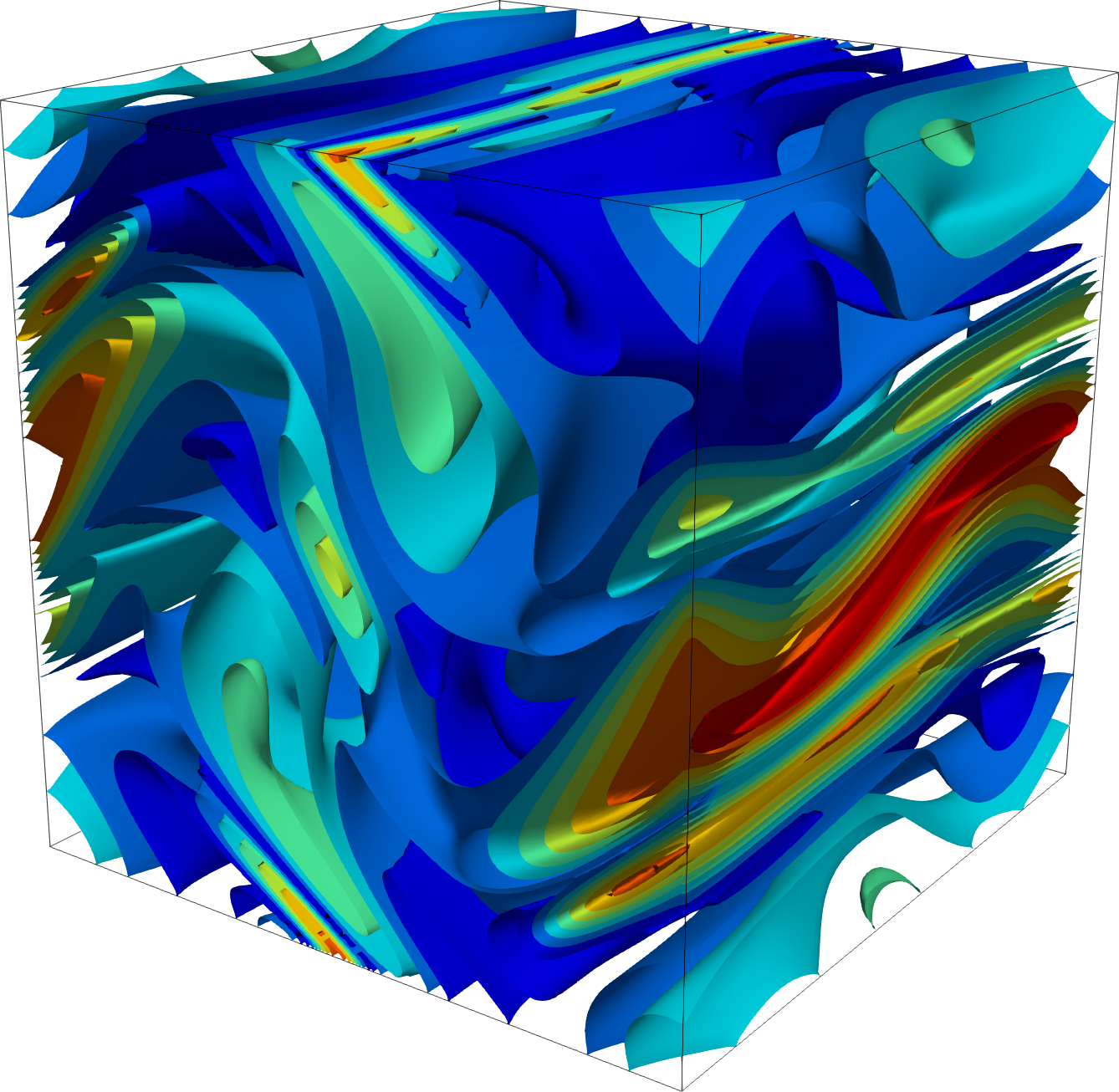}}
\subfigure{\includegraphics[width=0.22\textwidth]{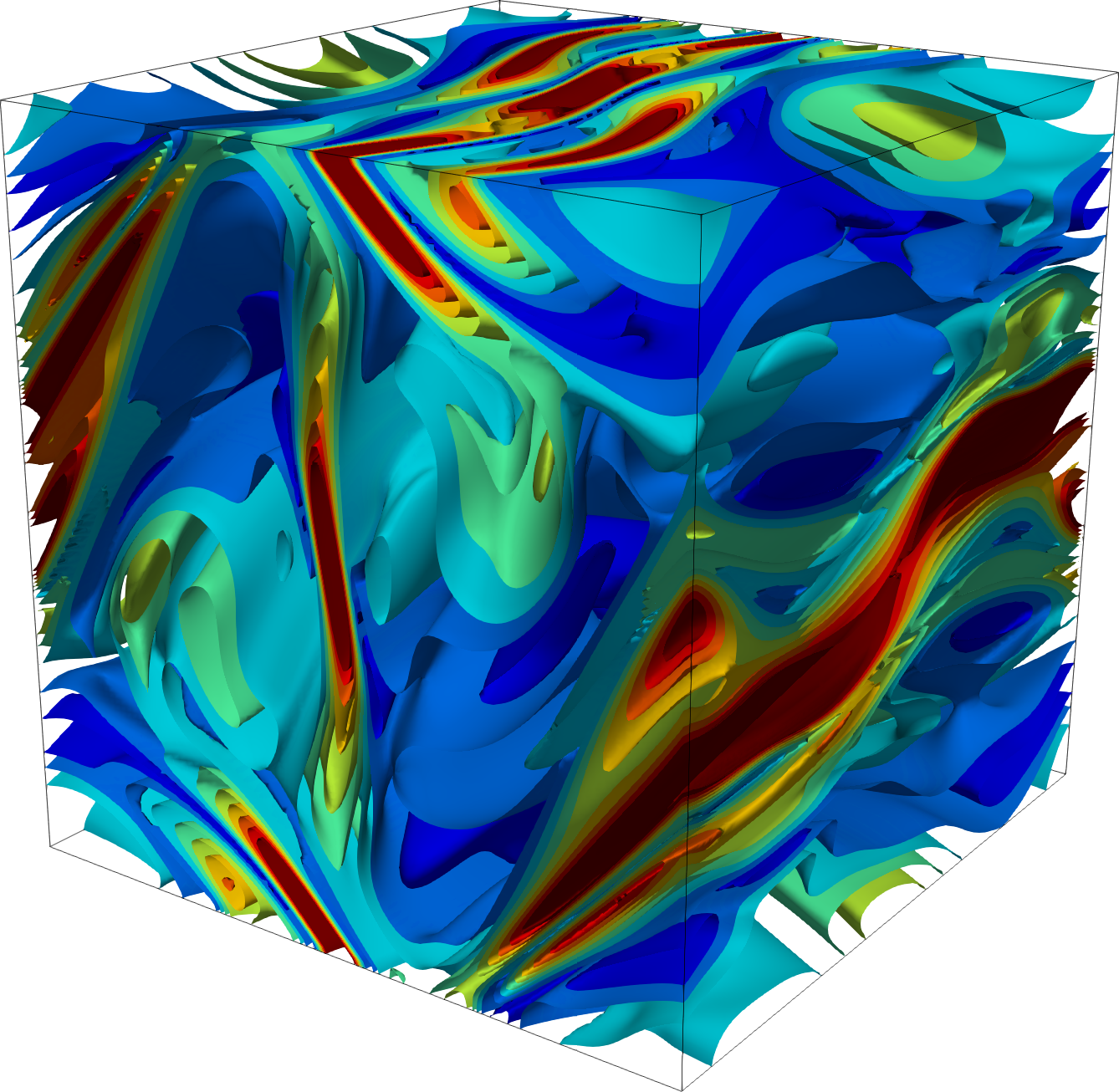}}
\subfigure{\includegraphics[width=0.22\textwidth]{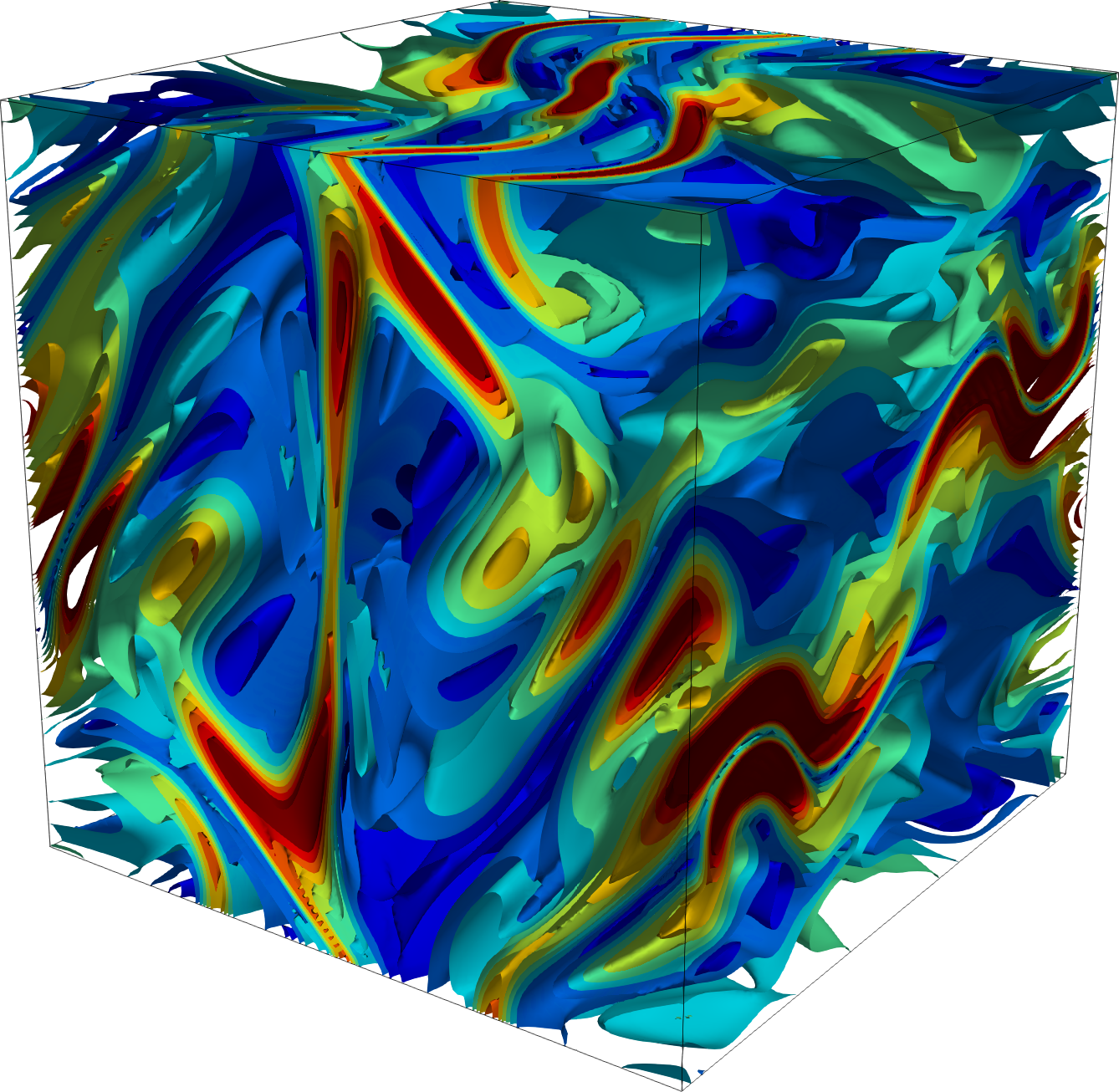}}
\subfigure{\includegraphics[width=0.22\textwidth]{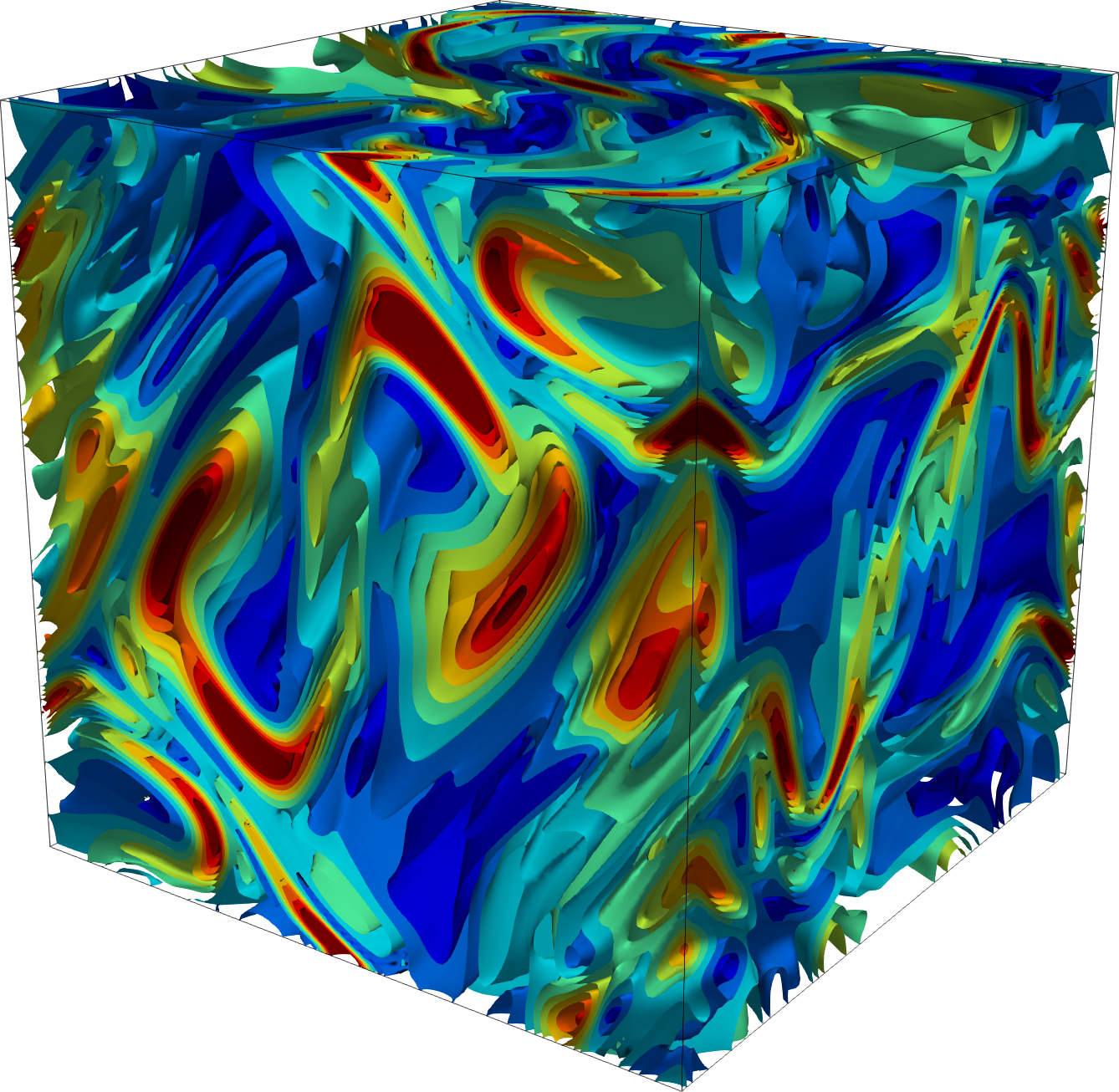}}
\caption{Simulation of the Orszag-Tang vortex in three dimensions using $\mathrm{M_1}$: time evolution of the (isocontours of the) current magnitude for $\mathrm{Re} = 570$ and $N=128$. From top left to bottom, $t=0.1$, 0.5, 1.0, 1.5 and 2.0.}
\label{Figure6}
\end{figure}
%
%
%
\subsection{Static droplet}
In the following, the behavior of the present model is further investigated in the context of a multiphase flow simulated through the popular Shan-Chen model~\cite{shan1993lattice}. Specifically, the latter introduces an interaction force
\begin{equation}
\bm{F} (\bm{x},t) = -G \psi(\bm{x},t) \sum_i w_i \bm{c}_i \psi(\bm{x}+\bm{c}_i,t),
\end{equation}
that mimics the molecular interactions leading to phase segregation. $\psi = 1-e^{-\rho}$ is the so-called pseudo-potential and $G$ is a parameter controlling the strength of the interaction. 

\begin{figure}[tp]
\centering
\includegraphics[scale=1]{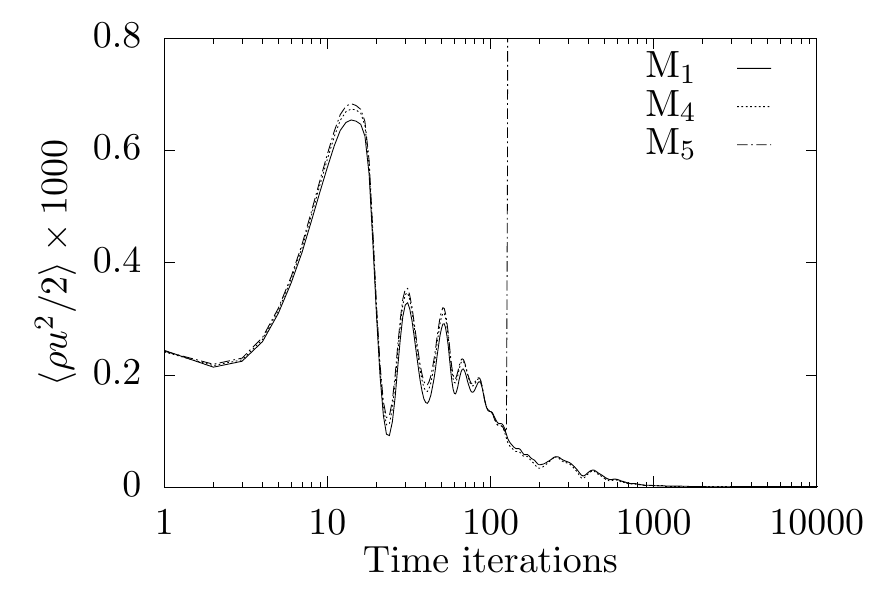}
\caption{Static droplet: time evolution of the mean kinetic energy by adopting Hermite polynomials up to the second (solid line) and sixth (dotted line) orders.}
\label{Figure7}
\end{figure}

Let us consider a periodic domain consisting of $200$ lattice points in each direction. A droplet of radius $R$ and density equal to $\rho_0$ is placed in the center of the domain, while the density is set to $\rho_0/8$ elsewhere. The parameter $G$ is set equal to -6. The pseudo-potential LBM is known to be prone to numerical instability due to the presence of spurious velocity currents. In FIG.~\ref{Figure7}, the time evolution of the mean kinetic energy is reported for numerical experiments carried out by using the models $M_1$, $M_4$ and $M_5$. Here, the beneficial effect of using sixth-order Hermite polynomials for the forcing term are clearly visible. Indeed, the run corresponding to the poorest model (i.e., $M_5$) blows-up after $\sim 120$ time iterations. Conversely, the adoption of sixth-order Hermite polynomials allows us to simulate a considerably larger time span, with the kinetic energy that tends to vanish after initial large-amplitude oscillations. This behavior is slightly impacted by the form of the forcing term, as both $M_1$ and $M_4$ lead to stable simulations. A small mismatch between the two forcing schemes is found in the early stage of the simulation, where larger velocity magnitude are encountered. They emphasize the impact of higher-order Hermite polynomials on the forcing term, and suggest a slightly better behavior of $M_1$, as compared to $M_4$, since the former leads to a lower maximal kinetic energy.

\subsection{Rayleigh-Taylor instability}
\begin{figure}[htbp!]
\centering
\subfigure{\includegraphics[width=0.14\textwidth]{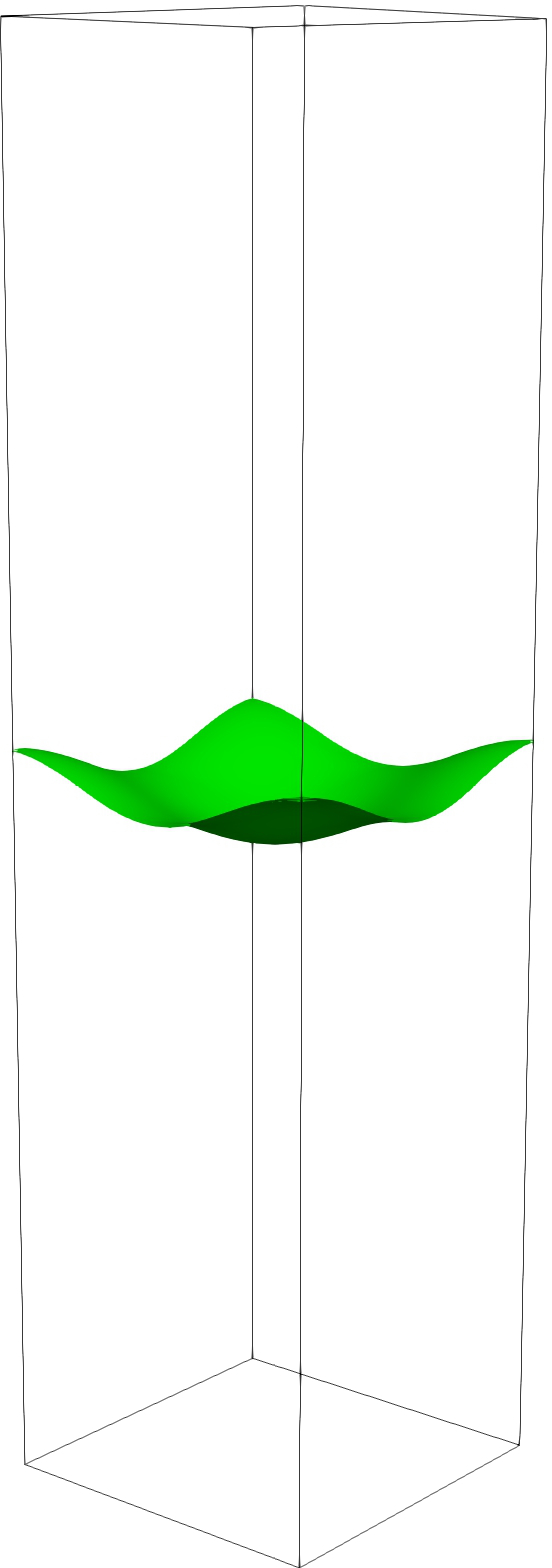}}
\subfigure{\includegraphics[width=0.14\textwidth]{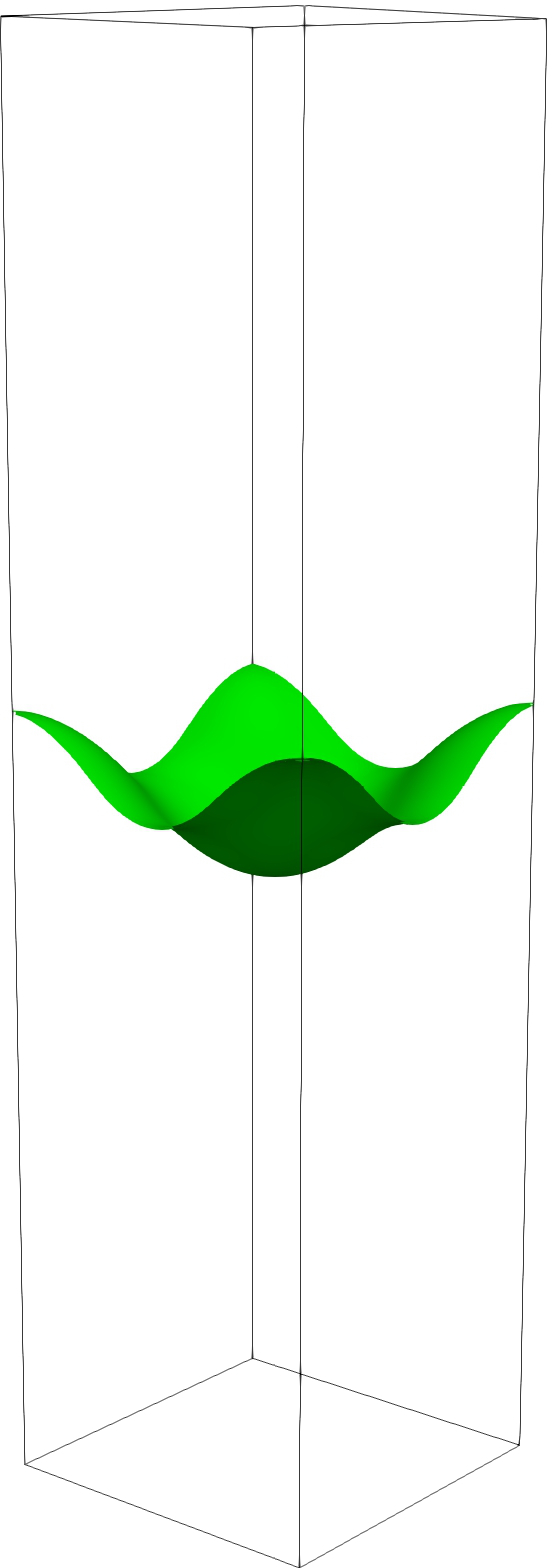}}
\subfigure{\includegraphics[width=0.14\textwidth]{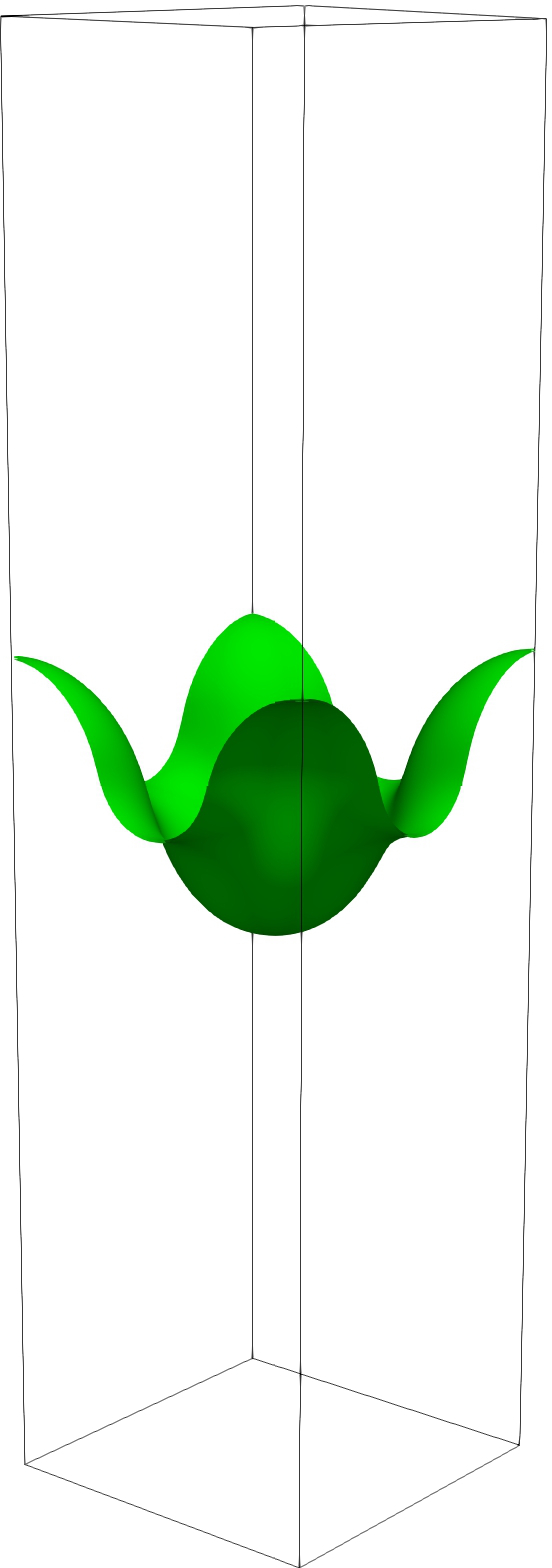}}
\subfigure{\includegraphics[width=0.14\textwidth]{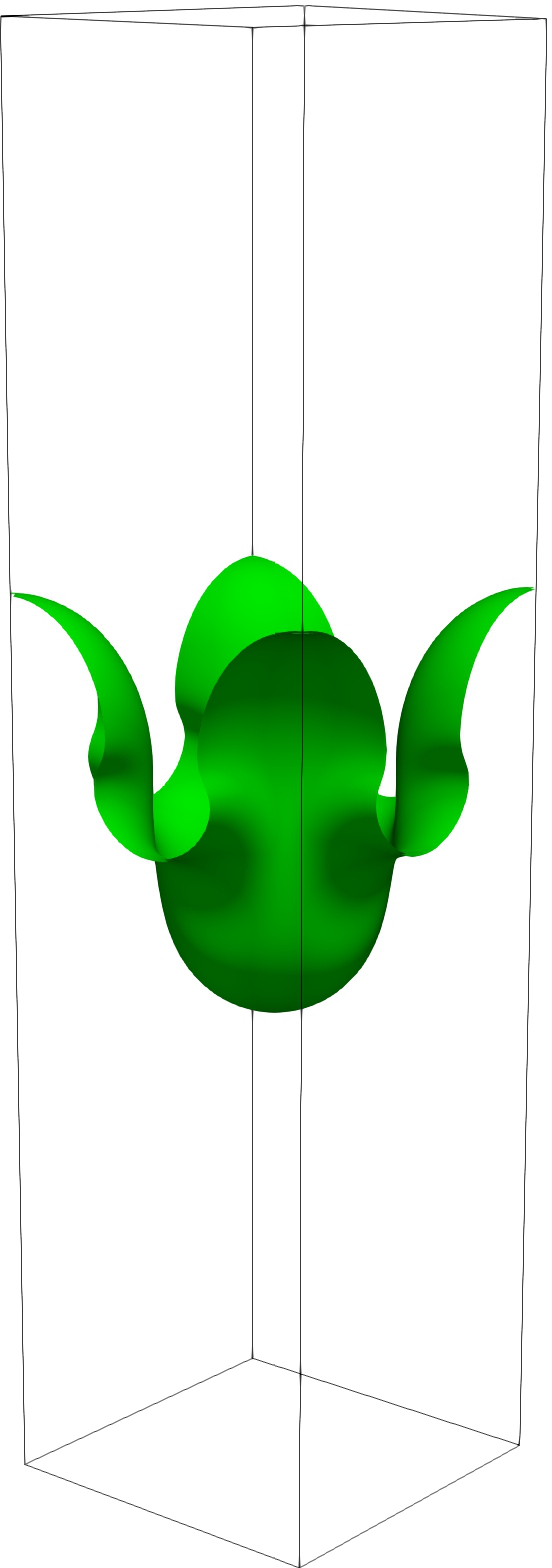}}
\subfigure{\includegraphics[width=0.14\textwidth]{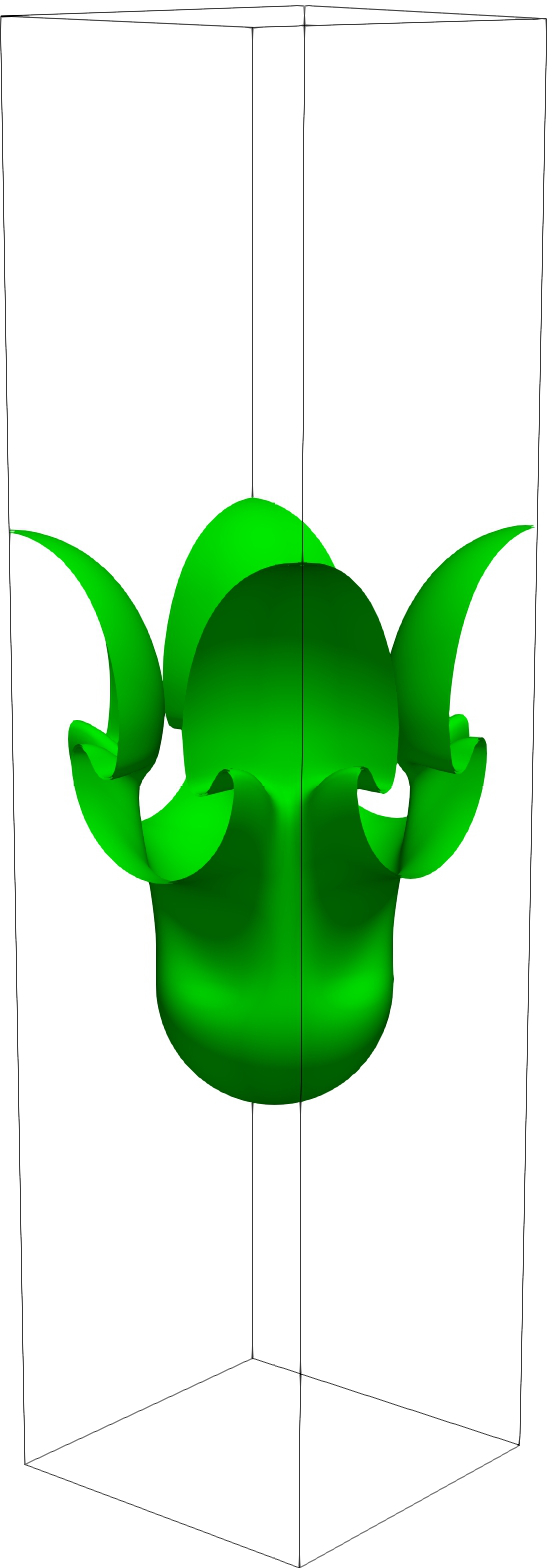}}
\subfigure{\includegraphics[width=0.14\textwidth]{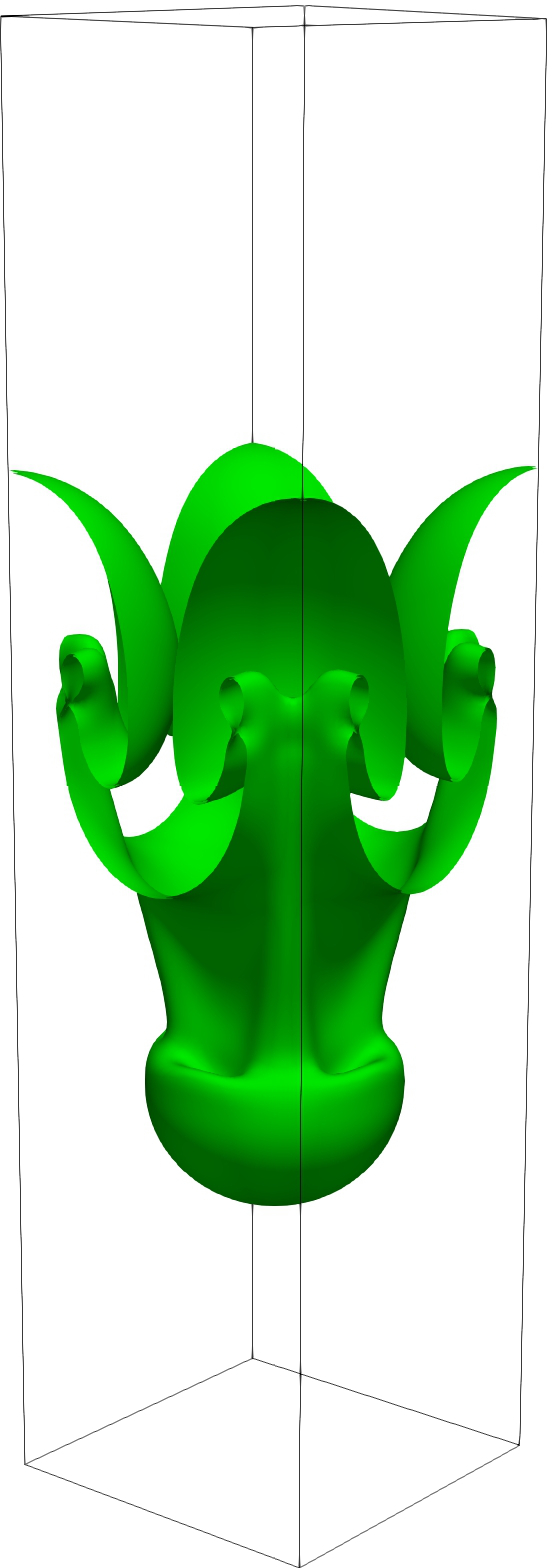}}
\caption{Rayleigh-Taylor instability: interface evolution at salient time instants, i.e., $t/t_0=0.5$ (top left), 1.0 (top center), 1.5 (top right), 2.0 (bottom left), 2.5 (bottom center) and 3.0 (bottom right).}
\label{Figure8}
\end{figure}

\begin{figure}[tp]
\centering
\includegraphics[scale=1]{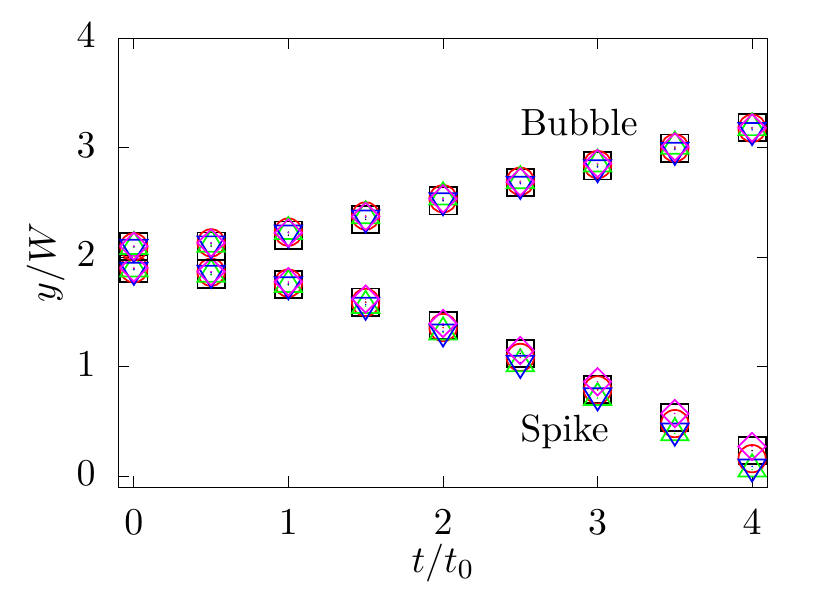}
\caption{Rayleigh-Taylor instability: time evolution of the position of the spike and the bubble of the interface at salient time instants. Present results (black squares) are plotted together with those from (i) a MRT LB study~\cite{saito2017lattice} (red circles), (ii) a BGK LB model for multiphase flows~\cite{he1999three} (green triangles), (iii) a phase-field MRT LB scheme~\cite{WANG2016340} (blue inverted triangles), and (iv) a solution of the coupled Navier-Stokes-Cahn-Hilliard equations~\cite{LEE20131466} (magenta diamonds).}
\label{Figure9}
\end{figure}

To conclude the numerical investigation of the present method, the simulation of the Rayleigh-Taylor instability is carried out. The fundamental mechanism, upon which this kind of instability relies, can be summarized as follows. Let us consider a heavy fluid on top of a lighter one, both being immiscible fluids that are separated by an interface. Assuming that they are subject to a gravitational field, their interface will not be able to stay in its equilibrium state. As a consequence, plumes (spikes) will start forming and flowing downward (upward).\\ 
\indent This Rayleigh-Taylor instability mechanism is simulated hereafter by adopting the color-gradient method~\cite{PhysRevE.71.056702, Reis_2007, LECLAIRE20122237}.  Recently, Saito \textit{et al.}~\cite{saito2018color} have proposed a central-moments-based color-gradient model by using third- and second-order terms for the equilibrium and forcing, respectively. Here, the approach is recasted within the present framework, hence adopting sixth-order Hermite polynomials for both the equilibrium and the forcing term. This problem, as well as its solution procedure by the color-gradient method, clearly shows the universality of the present approach. The CM-based color-gradient method is briefly outlined in App.~\ref{sec:AppCG}.\\
\indent Let us consider a three-dimensional domain of size $W \times 4W \times W$, with $W=64$. A fluid of density $\rho_h$ is placed over a lighter one of density $\rho_l=1$. The initial flow field is perturbed as
\begin{eqnarray}
\rho(\bm{x},0) &=& \rho_h, \, \mathrm{if} \, y>2W + 0.05W \left[ \cos \left( 2 \pi x \right)+\cos \left( 2 \pi z \right)  \right], \nonumber \\
\rho(\bm{x},0) &=& \rho_l, \, \mathrm{otherwise.}
\end{eqnarray}
No-slip walls are enforced at the top and bottom sections, while periodic boundaries are assumed at the other sides of the domain. A gravitational body force is considered as
\begin{equation}
\bm{F} = - \left[ \rho(\bm{x},t) - \frac{\rho_h+\rho_l}{2}  \right] \bm{g},
\end{equation}
with $\bm{g}=(0, \, -g, \, 0)$, and $g$ chosen so that $\sqrt{gW} = 0.04$~\cite{PhysRevE.71.056702}. The problem is governed by two dimensionless parameters: the Reynolds number 
\begin{equation}
\mathrm{Re} = \frac{W\sqrt{gW} }{\nu},
\end{equation}
and the Atwood number
\begin{equation}
\mathrm{At} = \frac{\rho_h-\rho_l}{\rho_h+\rho_l},
\end{equation}
that are set here to $\mathrm{Re} =512$ and $\mathrm{At}=0.5$, respectively. In addition, the characteristic time used for the time evolution of the Rayleigh-Taylor instability is defined as
\begin{equation}
t_0 = \frac{t}{\sqrt{Wg}}.
\end{equation}
In the following, let us denote the bottom and top points of the interface as spike and bubble, respectively. In FIG.~\ref{Figure8}, the evolution of the interface between the two fluids is sketched at salient time instants. They show that the edge of the spike rolls up at $t/t_0=3$. This is consistent with another LB study~\cite{saito2017lattice}, where the Rayleigh-Taylor instability has been investigated by means of the multiple-relaxation-time kernel. 
\\
\indent More quantitative results, regarding the time evolution of the position of the two reference points, are reported in TABLE~\ref{Table3}, and depicted in FIG.~\ref{Figure9}. Present findings are further compared to several models to assess its accuracy: (1) a MRT LB study~\cite{saito2017lattice}, (2) a BGK LB model for multiphase flows~\cite{he1999three}, (3) a phase-field MRT LB scheme~\cite{WANG2016340} and (4) a solution of the coupled Navier-Stokes-Cahn-Hilliard equations~\cite{LEE20131466}. From this, it is clear that the present method shows a pretty good agreement with data from the literature. This confirms the good numerical properties of the proposed approach, as well as, its universality.
\begin{table*}[!htbp]
\centering
\begin{tabular}{ C{1cm} | C{1.3cm}  C{1.3cm}  C{1.3cm}  C{1.3cm}   C{1.3cm}  |  C{1.3cm}  C{1.3cm}  C{1.3cm}  C{1.3cm}   C{1.3cm} }
\hline\hline
 & \multicolumn{5}{c|}{Spike} & \multicolumn{5}{c}{Bubble}\\
\hline
$t / t_0$ & Present  &  \cite{saito2017lattice} & \cite{he1999three} & \cite{WANG2016340} &  \cite{LEE20131466}  & Present  &  \cite{saito2017lattice} & \cite{he1999three} & \cite{WANG2016340} &  \cite{LEE20131466} \\
\hline
0.0 & 1.897 & 1.895 & 1.887 & 1.888 & 1.904 & 2.100 & 2.095 & 2.092 & 2.096 & 2.101\\
0.5 & 1.845 & 1.864 & 1.839 & 1.860 & 1.869 & 2.101 & 2.131 & 2.113 & 2.129 & 2.131\\
1.0 & 1.753 & 1.763 & 1.744 & 1.755 & 1.776 & 2.202 & 2.230 & 2.229 & 2.228 & 2.224\\
1.5 & 1.591 & 1.587 & 1.555 & 1.569 & 1.618 & 2.345 & 2.377 & 2.372 & 2.364 & 2.372\\
2.0 & 1.378 & 1.357 & 1.312 & 1.325 & 1.396 & 2.516 & 2.535 & 2.545 & 2.524 & 2.535\\
2.5 & 1.121 & 1.085 & 1.022 & 1.037 & 1.149 & 2.682 & 2.695 & 2.693 & 2.672 & 2.688\\
3.0 & 0.791 & 0.788 & 0.712 & 0.740 & 0.863 & 2.834 & 2.847 & 2.846 & 2.824 & 2.856\\
3.5 & 0.537 & 0.481 & 0.390 & 0.419 & 0.572 & 2.997 & 2.999 & 3.009 & 2.984 & 3.009\\
4.0 & 0.233 & 0.160 & 0.060 & 0.090 & 0.271 & 3.184 & 3.179 & 3.178 & 3.164 & 3.181\\
\hline\hline
\end{tabular}
\caption{Rayleigh-Taylor instability: time evolution of the position of the spike and the bubble of the interface at salient time instants. Present results are compared those from (i) a MRT LB study~\cite{saito2017lattice}, (ii) a BGK LB model for multiphase flows~\cite{he1999three}, (iii) a phase-field MRT LB scheme~\cite{WANG2016340}, and (iv) a solution of the coupled Navier-Stokes-Cahn-Hilliard equations~\cite{LEE20131466}.}
\label{Table3}
\end{table*}

\section{Conclusions}
\label{Sec4}
In the present work, a \emph{systematic} way to derive {CM-LBMs} with external forcing has been proposed. It naturally flows from a previous work based on the D3Q27 velocity discretization~\cite{derosisHermite}, where Galilean invariant CMs were derived from the discrete equilibrium state instead of the continuous Maxwell Boltzmann distribution. Here, by relying on the very same set of 27 Hermite polynomials, Galilean invariant CMs of the forcing term are also obtained but without any assumption on these CMs. By further equilibrating acoustic and high-order CMs, the present method only requires the computation of five pre-collision CMs. Both points eventually lead to a very concise algorithm.\\
\indent Interestingly, it is also shown that the present method recovers the behavior of the cascaded LBM~\cite{fei2018three}. But differently from the latter, the proposed methodology allows the derivation of forcing terms in an \emph{a priori} way thanks to the tight link existing between the Hermite polynomial expansion and the lattice of discrete velocities. As a consequence, its extension to any kind of velocity discretizations can be done in a \emph{systematic} and \emph{straightforward} manner.

The simulation of the four-rolls mill, the Hartmann flow, the two- and three-dimensional Orszag-Tang vortex, a static droplet and the Rayleigh-Taylor instability were conducted to evaluate both the numerical behavior as well as the universality of the present method. By further comparing the latter with six other forcing schemes on several of these testcases, its good accuracy and robustness properties were properly demonstrated.\\
\indent All in all, the consistency with the cascaded LBM, the compactness of the algorithm as well as its excellent numerical properties  lead the present scheme to be a very good candidate to perform multiphysics simulations with (or without) the presence of external forces. Regarding future works, the extension to both high-order lattices and to other kinds of collision space (raw moment, Hermite moment, central Hermite moment and cumulant~\cite{COREIXAS_ARXIV_1904_2019}) is currently under investigation. Corresponding results shall be presented in the near future.

\section*{Acknowledgments}
This article is based upon work from COST Action MP1305, supported by COST (European Cooperation in Science and Technology). The authors would like to express their gratitude to the reviewers for their relevant remarks and insightful questions. A.D.R. is grateful to Dr. Shimpei Saito for very valuable hints and discussions related to the setup of the simulations of the Rayleigh-Taylor instability.

\appendix
\section{Hermite polynomials \label{sec:AppHermite}}
Discrete Hermite polynomials are defined as~\cite{malaspinas2015increasing,coreixas2017recursive}
\begin{equation}
\boldsymbol{\mathcal{H}}_i^{(n)} = \dfrac{(-c_s^2)^n}{\omega(\bm{c}_i)}\boldsymbol\nabla^n_{\bm{c}_i}\omega(\bm{c}_i),
\label{eq:RodriguesFormula}
\end{equation}
with the corresponding Gaussian weight function
\begin{equation}
w(\bm{c}_i) = \dfrac{1}{(2\pi c_s^2)^{D/2}}\exp \left( -\frac{c_i^2}{2 c_s^2} \right),
\label{eq:GaussianWeight}
\end{equation}
where $D$ is the number of physical dimensions and $c_s$ is the lattice constant.
In the 3D case, discrete Hermite tensors can be computed using tensor products of their 1D formulation,
\begin{widetext}
\begin{align}
\mathcal{H}_{ix\ldots xy\ldots yz\ldots z}^{(nx+ny+nz)} &= \dfrac{(-c_s^2)^{nx+ny+nz}}{w(\bm{c}_i)}\bm\nabla^{nx+ny+nz}_{\bm{c}_i}w(\bm{c}_i)\nonumber\\
	&=\left[\dfrac{(-c_s^2)^{nx}}{w'(c_{ix})}\nabla^{nx}_{c_{ix}}w'(c_{ix})\right] \left[\dfrac{(-c_s^2)^{ny}}{w'(c_{iy})}\nabla^{ny}_{c_{iy}}w'(c_{iy})\right] \left[\dfrac{(-c_s^2)^{nz}}{w'(c_{iz})}\nabla^{nz}_{c_{iz}}w'(c_{iz})\right]\nonumber\\
	&= \mathcal{H}_{ix\ldots x}^{(nx)}\mathcal{H}_{iy\ldots y}^{(ny)}\mathcal{H}_{iz\ldots z}^{(nz)},
	\label{eq:HermiteTensorProduct}
\end{align}
\end{widetext}
where $w'$ is the one-dimensional version of the Gaussian weight~(\ref{eq:GaussianWeight}), and $(nx,ny,nz)$ are the number of occurrences of indexes $(x,y,z)$. Since the D3Q27 lattice is a tensor product of three D1Q3 lattices, then only 1D Hermite polynomials of degree up to $n=2$ can be used for the construction of their 3D counterparts. This means that the degree of each Hermite polynomial will be at most \emph{two per direction}. Indeed, 
\begin{align*}
\mathcal{H}^{(0)}_{i}&=1,\\
\mathcal{H}^{(1)}_{ix}&=c_{ix},\\
\mathcal{H}^{(1)}_{iy}&=c_{iy},\\
\mathcal{H}^{(1)}_{iz}&=c_{iz},\\
\mathcal{H}^{(2)}_{ixx}&=c_{ix}^2-c_s^2,\\
\mathcal{H}^{(2)}_{iyy}&=c_{iy}^2-c_s^2,\\
\mathcal{H}^{(2)}_{izz}&=c_{iz}^2-c_s^2,\\
\mathcal{H}^{(2)}_{ixy}&=\mathcal{H}^{(1)}_{ix}\mathcal{H}^{(1)}_{iy},\\
\mathcal{H}^{(2)}_{ixz}&=\mathcal{H}^{(1)}_{ix}\mathcal{H}^{(1)}_{iz},\\
\mathcal{H}^{(2)}_{iyz}&=\mathcal{H}^{(1)}_{iy}\mathcal{H}^{(1)}_{iz},\\
\mathcal{H}^{(3)}_{ixxy}&=\mathcal{H}^{(2)}_{ixx}\mathcal{H}^{(1)}_{iy},\\
\mathcal{H}^{(3)}_{ixxz}&=\mathcal{H}^{(2)}_{ixx}\mathcal{H}^{(1)}_{iz},\\
\mathcal{H}^{(3)}_{iyyz}&=\mathcal{H}^{(2)}_{iyy}\mathcal{H}^{(1)}_{iz},\\
\mathcal{H}^{(3)}_{ixyy}&=\mathcal{H}^{(1)}_{ix}\mathcal{H}^{(2)}_{iyy},\\
\mathcal{H}^{(3)}_{ixzz}&=\mathcal{H}^{(1)}_{ix}\mathcal{H}^{(2)}_{izz},\\
\mathcal{H}^{(3)}_{iyzz}&=\mathcal{H}^{(1)}_{iy}\mathcal{H}^{(2)}_{izz},\\
\mathcal{H}^{(3)}_{ixyz}&=\mathcal{H}^{(1)}_{ix}\mathcal{H}^{(1)}_{iy}\mathcal{H}^{(1)}_{iz},\\
\mathcal{H}^{(4)}_{ixxyy}&=\mathcal{H}^{(2)}_{ixx}\mathcal{H}^{(2)}_{iyy},\\
\mathcal{H}^{(4)}_{ixxzz}&=\mathcal{H}^{(2)}_{ixx}\mathcal{H}^{(2)}_{izz},\\
\mathcal{H}^{(4)}_{iyyzz}&=\mathcal{H}^{(2)}_{iyy}\mathcal{H}^{(2)}_{izz},\\
\mathcal{H}^{(4)}_{ixxyz}&=\mathcal{H}^{(2)}_{ixx}\mathcal{H}^{(1)}_{iy}\mathcal{H}^{(1)}_{iz},\\
\mathcal{H}^{(4)}_{ixyyz}&=\mathcal{H}^{(1)}_{ix}\mathcal{H}^{(2)}_{iyy}\mathcal{H}^{(1)}_{iz},\\
\mathcal{H}^{(4)}_{ixyzz}&=\mathcal{H}^{(1)}_{ix}\mathcal{H}^{(1)}_{iy}\mathcal{H}^{(2)}_{izz},\\
\mathcal{H}^{(5)}_{ixxyyz}&=\mathcal{H}^{(2)}_{ixx}\mathcal{H}^{(2)}_{iyy}\mathcal{H}^{(1)}_{iz},\\
\mathcal{H}^{(5)}_{ixxyzz}&=\mathcal{H}^{(2)}_{ixx}\mathcal{H}^{(1)}_{iy}\mathcal{H}^{(2)}_{izz},\\
\mathcal{H}^{(5)}_{ixyyzz}&=\mathcal{H}^{(1)}_{ix}\mathcal{H}^{(2)}_{iyy}\mathcal{H}^{(2)}_{izz},\\
\mathcal{H}^{(6)}_{ixxyyzz}&=\mathcal{H}^{(2)}_{ixx}\mathcal{H}^{(2)}_{iyy}\mathcal{H}^{(2)}_{izz}.
\end{align*}

\section{Two-dimensional model \label{sec:AppD2Q9}}
Let us consider an Eulerian basis $\bm{x}=[x,y]$. Here, we derive the two-dimensional LB model by means of the D2Q9 velocity discretization, where lattice directions $\displaystyle \bm{c}_i=[|c_{ix}\rangle ,\, |c_{iy}\rangle]$ are 
\begin{eqnarray}
	\begin{split}
	|c_{ix} \rangle &= [0, \phantom{-}1,-1,\phantom{-} 0,\phantom{-}0,\phantom{-}1,-1,\phantom{-}1, -1]^{\top}, \\
	|c_{iy} \rangle &= [0,\phantom{-}0,\phantom{-}0,\phantom{-}1,-1,\phantom{-}1,-1,-1,\phantom{-}1]^{\top},
	\label{eq:velVec}
	\end{split}
\end{eqnarray}
with $i=0 \dots 8$. Let us define the particle distribution functions $\displaystyle |f_i \rangle = [f_0 ,\, \ldots f_i, \, \ldots f_{8} ]^{\top}$ and the velocity vector $\bm{u} = [u_x, u_y]$. In two dimensions, the equilibrium distributions $|f_i^{eq} \rangle = [f_0^{eq} ,\, \ldots f_i^{eq}, \, \ldots f_{8}^{eq} ]^{\top}$ can be expanded onto a basis of Hermite polynomials $\mathcal{H}^{(n)}$ up to the fourth order~\cite{malaspinas2015increasing, coreixas2017recursive, COREIXAS_PhD_2018}, i.e
\begin{widetext}
\begin{equation} \label{eq:D2Q9}
f_i^{eq} = w_i \rho \bigg(1+ \frac{\bm{c}_i \cdot \bm{u}}{c_s^2} + \frac{1}{2 c_s^4} \mathcal{H}_i^{(2)} : \bm{u}\bm{u} +\frac{1}{2 c_s^6}( \mathcal{H}_{ixxy}^{(3)} u_x^2 u_y + \mathcal{H}_{ixyy}^{(3)} u_x u_y^2)  +\frac{1}{4 c_s^8} \mathcal{H}_{ixxyy}^{(4)}u_x^2 u_y^2 \bigg),
\end{equation}
\end{widetext}
where $w_0=4/9$, $w_{1 \ldots 4} = 1/9$, $w_{5 \ldots 8}=1/36$ and $c_s=1/\sqrt{3}$ is the lattice sound speed ~\cite{SucciBook}. Again, the development of a central-moments-based collision operator begins by shifting the lattice directions by the local fluid velocity, i.e., $\displaystyle \bar{\bm{c}}_i=[| \bar{c}_{ix}\rangle ,\, | \bar{c}_{iy}\rangle]$, where
\begin{eqnarray}
| \bar{c}_{ix}\rangle &=& |c_{ix}- u_x \rangle, \nonumber \\
| \bar{c}_{iy}\rangle &=& |c_{iy}- u_y \rangle.
\end{eqnarray}
Let us choose the basis $\bar{\mathcal{T}} = \left[  \bar{T}_0,\, \ldots, \, \bar{T}_i,\, \ldots, \, \bar{T}_8  \right]$ which components are
\begin{eqnarray}
|\bar{T}_0\rangle &=& |1, \, \ldots ,\, 1\rangle,\nonumber \\
|\bar{T}_1\rangle &=& | \bar{c}_{ix}\rangle,\nonumber \\
|\bar{T}_2\rangle &=& | \bar{c}_{iy}\rangle ,\nonumber \\
|\bar{T}_3\rangle &=& | \bar{c}_{ix}^2+ \bar{c}_{iy}^2\rangle ,\nonumber \\
|\bar{T}_4\rangle &=& | \bar{c}_{ix}^2- \bar{c}_{iy}^2\rangle ,\nonumber \\
|\bar{T}_5\rangle &=& | \bar{c}_{ix} \bar{c}_{iy}\rangle, \nonumber \\
|\bar{T}_6\rangle &=& | \bar{c}_{ix}^2 \bar{c}_{iy} \rangle,\nonumber \\
|\bar{T}_7\rangle &=& | \bar{c}_{ix} \bar{c}_{ii}^2 \rangle, \nonumber \\
|\bar{T}_8\rangle &=& | \bar{c}_{ix}^2 \bar{c}_{iy}^2 \rangle.
\end{eqnarray}
Let us collect pre-collision, equilibrium and post-collision CMs as
\begin{eqnarray}
| k_i \rangle &=& \left[ k_0,\, \ldots, \, k_i,\, \ldots, \, k_{8}   \right]^{\top},\nonumber \\
| k_i^{eq} \rangle &=& \left[ k_0^{eq},\, \ldots, \, k_i^{eq},\, \ldots, \, k_{8}^{eq}   \right]^{\top},\nonumber \\
| k_i^{\star} \rangle &=& \left[ k_0^{\star},\, \ldots, \, k_i^{\star},\, \ldots, \, k_{8}^{\star}   \right]^{\top},
\end{eqnarray}
respectively. The first two quantities are evaluated by applying the transformation matrix $\bar{\mathcal{T}}$ to the corresponding distribution, that is
\begin{eqnarray}
| k_i \rangle &=& \bar{\mathcal{T}}^{\top} |f_i \rangle, \nonumber\\
| k_i^{eq} \rangle &=& \bar{\mathcal{T}}^{\top} |f_i^{eq} \rangle.
\end{eqnarray}
By adopting equilibrium populations with the full set of Hermite polynomials (see Eq.~(\ref{eq:D2Q9})), equilibrium CMs are
\begin{eqnarray}
k_0^{eq} &=& \rho,\nonumber \\
k_3^{eq} &=& 2 \rho c_s^2 \nonumber \\
k_8^{eq} &=& \rho c_s^4.
\end{eqnarray}
and $k_{1,2,4,5,6,7}^{eq}=0$. In analogy with the 3D case, it should be noted that (1) the equilibrium state is fully Galilean invariant since it does not show any dependence on the fluid velocity, and (2) the discrete equilibrium CMs have the same form of the continuous counterpart when the full set of Hermite polynomials is considered.\\
\indent Again, the post-collision CMs can be written as
\begin{equation}
| k_i^{\star} \rangle = \left( \mathbf{I}-\boldsymbol \Lambda \right) \bar{\mathcal{T}}^{\top} |f_i \rangle + \boldsymbol \Lambda \bar{\mathcal{T}}^{\top} |f_i^{eq} \rangle + \left( \mathbf{I}- \frac{\boldsymbol \Lambda}{2} \right) \bar{\mathcal{T}}^{\top} |\mathcal{F}_i \rangle,
\end{equation}
or
\begin{equation}
| k_i^{\star} \rangle = \left( \mathbf{I}-\boldsymbol \Lambda \right) |k_i \rangle + \boldsymbol \Lambda |k_i^{eq} \rangle + \left( \mathbf{I}- \frac{\boldsymbol \Lambda}{2} \right)  |R_i \rangle,
\end{equation}
where $\mathbf{I}$ is the ($9 \times 9$) unit tensor, $\displaystyle \boldsymbol \Lambda = \mathrm{diag}\left[1,\,1,\,1,\,1, \, \omega,\, \omega,\,1,\,1, \, 1    \right]$ is a ($9 \times 9$) relaxation matrix.\\
\indent Now, let us define the two-dimensional forcing term $\mathcal{F}_i$. Specifically, we employ the expression adopted by Huang \textit{et al.}~\cite{Huang2018}:
\begin{widetext}
\begin{equation}
\mathcal{F}_i = w_i \left( \frac{\bm{F} \cdot \bm{c}_i}{c_s} \cdot  \mathcal{H}_i^{(1)} + \frac{[\bm{F} \bm{u}]}{2 c_s^2}  \cdot  \mathcal{H}_i^{(2)} + \frac{[\bm{F} \bm{u} \bm{u}]}{6 c_s^3}  \cdot  \mathcal{H}_{i,[xyy],[xxy]}^{(3)} + \frac{[\bm{F} \bm{u} \bm{u} \bm{u}]}{24 c_s^4}  \cdot  \mathcal{H}_{i,[xxyy]}^{(4)} \right),
\end{equation}
\end{widetext}
where the square bracket in Hermite coefficient denotes permutations. Notice that the popular formula by Guo \textit{et al.}~\cite{guo2002discrete} is recovered if $\mathcal{H}_i^{(3)}$ and $\mathcal{H}_i^{(4)}$ are neglected. The CMs of the forcing term can be computed as 
\begin{equation}
|R_i \rangle = \bar{\mathcal{T}}^{\top} |\mathcal{F}_i \rangle
\end{equation}
and read as follows:
\begin{eqnarray}
R_1 &=& F_x, \nonumber \\
R_2 &=& F_y, \nonumber \\
R_6 &=& F_y c_s^2, \nonumber \\
R_7 &=& F_x c_s^2,
\end{eqnarray}
and $R_{0,3,4,5,8}=0$. The resultant expressions of the post-collision CMs are:
\begin{eqnarray}
k_0^{\star} &=& \rho, \nonumber \\
k_1^{\star} &=& F_x/2, \nonumber \\
k_2^{\star} &=& F_y/2, \nonumber \\
k_3^{\star} &=& 2\rho c_s^2, \nonumber \\
k_4^{\star} &=& (1-\omega) k_4, \nonumber \\
k_5^{\star} &=& (1-\omega) k_5, \nonumber \\
k_6^{\star} &=& F_y c_s^2/2, \nonumber \\
k_7^{\star} &=& F_x c_s^2/2, \nonumber \\
k_8^{\star} &=& \rho c_s^4.
\end{eqnarray}
Again, one can appreciate that the post-collision state in terms of CMs show simple expressions. Then, the post-collision populations are reconstructed as usual:
\begin{equation}\label{system2}
| f_i^{\star} \rangle = \left(\bar{\mathcal{T}}^{\top} \right)^{-1} | k_i^{\star} \rangle,
\end{equation}
with $|f_i^{\star} \rangle = [f_0^{\star} ,\, \ldots f_i^{\star}, \, \ldots f_{8}^{\star} ]^{\top}$, that are eventually streamed.
%
%
%


\section{Lattice Boltzmann method for the evolution of the magnetic field \label{sec:AppMHD}}
The evolution of the magnetic field $\bm{b}$ is predicted by a second LBM based on the D3Q7 velocity space, that is
$\forall l \in \llbracket 0,6 \rrbracket$,
\begin{equation}
\bm{h}_l(\bm{x}+ \boldsymbol \xi_l , t+1)  = \bm{h}_l(\bm{x}, t) + \omega_m \left[\bm{h}_l^{eq}(\bm{x}, t)-\bm{h}_l(\bm{x}, t)    \right].
\end{equation}
with $\omega_m$ the relaxation frequency that is related to the magnetic resistivity as $\displaystyle \eta = c_s^2\left(\frac{1}{\omega_m} - \frac{1}{2} \right)$, where $c_s=1/2$ for the D3Q7 lattice. For this particular lattice, discrete velocities $\displaystyle \boldsymbol \xi_l=[| \xi_{lx}\rangle ,\, | \xi_{ly}\rangle ,\, | \xi_{lz}\rangle]$ are defined as
\begin{eqnarray}
| \xi_{lx}\rangle &=& [0,\phantom{-}1, -1,\phantom{-}0,\phantom{-}0,\phantom{-}0,\phantom{-}0 ]^{\top}, \nonumber\\
| \xi_{ly}\rangle &=& [0,\phantom{-} 0,\phantom{-} 0,\phantom{-} 1, -1,\phantom{-} 0,\phantom{-} 0 ]^{\top}, \nonumber\\
| \xi_{lz}\rangle &=& [0,\phantom{-} 0,\phantom{-} 0,\phantom{-} 0,\phantom{-} 0,\phantom{-} 1, -1]^{\top}.
\end{eqnarray}
Vector-valued populations $\bm{h}_l$ are necessary due to the anti-symmetry of the electric tensor~\cite{dellar2002lattice}. Their equilibrium state is defined as~\cite{pattison2008progress}
\begin{equation}
h_{l\alpha }^{eq} = w_{l} \left[  b_{\alpha} + 4 \xi_{l\beta } \left( u_{\beta} b_{\alpha} - b_{\beta} u_{\alpha}    \right)    \right],
\end{equation}
where $w_{0}=1/4$, $w_{1 \ldots 6}=1/8$ and $(\alpha, \beta) \in \{x,y,z\}^2$. Regarding magnetic boundary conditions, we follow the approach in~\cite{dellar2013moment} that is built on the fact that the magnetic field can be computed as
\begin{equation}
\bm{b} = \sum_{l=0}^6 \bm{h}_l.
\end{equation}
Specifically, let us assume that one wants to enforce a certain magnetic field $\bm{b}^{\dagger} = [b_x^{\dagger}, \, b_y^{\dagger}, \, b_z^{\dagger}]$ to the bottom section of the considered domain. In this case, the only population to be assigned is $\bm{h}_2$, that is evaluated as
\begin{eqnarray}
h_{2x} &=& b_x^{\dagger} - \sum_{l \neq 2} h_{lx}, \nonumber\\
h_{2y} &=& b_y^{\dagger} - \sum_{l \neq 2} h_{ly}, \nonumber\\
h_{2z} &=& b_z^{\dagger} - \sum_{l \neq 2} h_{lz}.
\end{eqnarray}

\section{Central-moments-based color-gradient method \label{sec:AppCG}}
Let us consider two immiscible fluids, namely, red and blue. The evolution of populations $f_i^k$ is
\begin{equation}
f_i^k(\bm{x}+\bm{c}_i, t+1) = f_i^k(\bm{x}, t) + \Omega_i^k(\bm{x}, t),
\end{equation} 
where $k=r$ for the red fluid, and $k=b$ for the blue one. Moreover, it is possible to define the total distribution functions as $\displaystyle f_i = f_i^r + f_i^b$. The collision process $\Omega_i^k$ is composed of three sub-stages:
\begin{equation}
\Omega_i^k = \big(\Omega_i^k \big)^{(3)} \left[  \left(  \Omega_i^k \right)^{(1)} + \left(  \Omega_i^k \right)^{(2)}   \right],
\end{equation}
where $ \left( \Omega_i^k \right)^{(1)}$, $ \left( \Omega_i^k \right)^{(2)}$ and $ \left( \Omega_i^k \right)^{(3)}$ are the single-phase, perturbation and recoloring operators, respectively. Within the CM-based framework~\cite{saito2018color}, the single-phase collision operator can be written as
\begin{equation}
 \left(\Omega_i^k \right)^{(1)} =\left(  \bar{\mathcal{T}}^{\top} \right)^{-1}  \boldsymbol \Lambda \bar{\mathcal{T}}^{\top} \left( |f_i^{eq} \rangle  - |f_i \rangle \right) + \left( \mathbf{I}- \frac{\boldsymbol \Lambda}{2} \right) \bar{\mathcal{T}}^{\top} |\mathcal{F}_i \rangle,
\end{equation}
with $|\mathcal{F}_i \rangle$, $\bar{\mathcal{T}}^{\top}$ and $\Lambda$ that obey Eqs.~(\ref{forcing}),~(\ref{basis}) and~(\ref{eq:relaxationMatrix}), respectively. Macroscopic variables are given by
\begin{eqnarray}
\rho_k &=& \sum_i f_i^k, \nonumber \\
\rho &=& \sum_k \rho_k,  \nonumber \\
\rho \bm{u} &=& \sum_i f_i  \bm{c}_i + \frac{1}{2}\bm{F},
\end{eqnarray}
where $\rho_k$ is the density of the fluid $k$, $\rho$ is the total mass density, $\bm{u}$ is the total momentum and $\bm{F}$ is a body force. Distributions relax to an enhanced local equilibrium~\cite{LECLAIRE2013159, saito2017lattice} that, by adopting sixth-order Hermite polynomials, read as follows:
\begin{widetext}
\begin{align}
f_i^{eq}\left( \rho, \bm{u} \right) &= \rho \bigg\{\varphi_i + w_i\bigg[ \frac{\bm{c}_i \cdot \bm{u}}{c_s^2} + \frac{1}{2 c_s^4} \bigg[\mathcal{H}_{ixx}^{(2)}u_x^2+\mathcal{H}_{iyy}^{(2)}u_y^2+\mathcal{H}_{izz}^{(2)}u_z^2+ 2\bigg(\mathcal{H}_{ixy}^{(2)}u_xu_y+\mathcal{H}_{ixz}^{(2)}u_xu_z+\mathcal{H}_{iyz}^{(2)}u_yu_z\bigg)\bigg]\nonumber\\
&\quad +\frac{1}{2 c_s^6}\bigg[\mathcal{H}_{ixxy}^{(3)} u_x^2 u_y + \mathcal{H}_{ixxz}^{(3)} u_x^2 u_z + \mathcal{H}_{ixyy}^{(3)} u_x u_y^2 + \mathcal{H}_{ixzz}^{(3)} u_x u_z^2  +\mathcal{H}_{iyzz}^{(3)} u_y u_z^2 + \mathcal{H}_{iyyz}^{(3)} u_y^2 u_z + 2 \mathcal{H}_{ixyz}^{(3)} u_x u_y u_z \bigg] \nonumber\\
&\quad +\frac{1}{4 c_s^8} \bigg[ \mathcal{H}_{ixxyy}^{(4)}u_x^2 u_y^2 + \mathcal{H}_{ixxzz}^{(4)}u_x^2 u_z^2 + \mathcal{H}_{iyyzz}^{(4)}u_y^2 u_z^2 + 2 \bigg(\mathcal{H}_{ixyzz}^{(4)} u_x u_y u_z^2 + \mathcal{H}_{ixyyz}^{(4)} u_x u_y^2 u_z + \mathcal{H}_{ixxyz}^{(4)}u_x^2 u_y u_z\bigg) \bigg] \nonumber\\
&\quad +\frac{1}{4 c_s^{10}} \bigg[ \mathcal{H}_{ixxyzz}^{(5)} u_x^2 u_y u_z^2 + \mathcal{H}_{ixxyyz}^{(5)} u_x^2 u_y^2 u_z + \mathcal{H}_{ixyyzz}^{(5)} u_x u_y^2 u_z^2 \bigg]\nonumber\\
&\quad +\frac{1}{8 c_s^{12}} \mathcal{H}_{ixxyyzz}^{(6)} u_x^2 u_y^2 u_z^2 \bigg] \bigg\} + \Phi_i.
\end{align}
\end{widetext}
Here, quantities $\varphi_i$ and $\Phi_i$ are
\begin{equation}
	\varphi_i = 
	\begin{cases}
		~\bar{\alpha}, & |\bm{c}_i|^2=0,\\
		~2(1-\bar{\alpha})/19, & |\bm{c}_i|^2=1, \\
		~(1-\bar{\alpha})/38, & |\bm{c}_i|^2=2, \\
		~(1-\bar{\alpha})/152, & |\bm{c}_i|^2=3,
		\label{eq:varphi_ik}
\end{cases}
\end{equation}
and 
\begin{equation}
	\Phi_i = 
	\begin{cases}
		~-3 \bar{\nu}({\bf u}\cdot \nabla \rho), & |\bm{c}_i|^2=0,\\
		~+16\bar{\nu}(\mathbf{G} : \bm{c}_i \otimes \bm{c}_i), & |\bm{c}_i|^2=1, \\
		~+4 \bar{\nu}(\mathbf{G} : \bm{c}_i \otimes \bm{c}_i), & |\bm{c}_i|^2=2, \\
		~+1 \bar{\nu}(\mathbf{G} : \bm{c}_i \otimes \bm{c}_i), & |\bm{c}_i|^2=3, \label{eq:Phi}
\end{cases}
\end{equation}
where $\otimes$ is the tensor (outer) product, and : stands for the contraction of index (Frobenius inner product). The tensor $\mathbf{G}$ is defined as
\begin{equation}
	\mathbf{G} = \frac{1}{48} \left[{\bf u}\otimes \nabla \rho
	+ ({\bf u}\otimes \nabla \rho)^{\top} \right],
\end{equation}
where gradients are computed by a second-order isotropic central scheme~\cite{PhysRevE.85.046309}. $\bar{\nu}$ is the kinematic viscosity interpolated by~\cite{LECLAIRE20122237}
\begin{equation}
	\frac{1}{\bar{\nu}} = \frac{1+\phi}{2}\frac{1}{\nu_r} + \frac{1-\phi}{2}\frac{1}{\nu_b}.
\end{equation}	
To distinguish the two components, the order parameter $\phi$ is introduced, that is
\begin{equation}
	\phi = \left(\frac{\rho_r}{\rho_r^0}-\frac{\rho_b}{\rho_b^0} \right) \bigg/ \left(\frac{\rho_r}{\rho_r^0}+\frac{\rho_b}{\rho_b^0} \right).
	\label{eq:order}
\end{equation}
The values $\phi=1,-1$, and $0$ correspond to a purely red fluid, a purely blue fluid, and the interface, respectively. To obtain a stable interface, the density ratio between the fluids must be taken into account as follows to obtain a stable interface~\cite{grunau1993lattice}:
\begin{equation}
\frac{\rho_r^0}{\rho_b^0} = \frac{1-\alpha_b}{1-\alpha_r},
\end{equation}
where the superscript ``0'' indicates the initial value of the density at the beginning of the simulation~\cite{PhysRevE.95.033306}. The pressure of the fluid is given as an isothermal equation of state:
\begin{equation}
p = \rho \left(c_s^k \right)^2 = \rho_k \frac{9(1-\bar{\alpha})}{19},
\label{eq:press}
\end{equation}
for the D3Q27 lattice, where $c_s^k$ is the speed of sound of the fluid $k$~\cite{LECLAIRE20122237}, $\bar{\alpha}$ is interpolated by
\begin{equation}
\bar{\alpha} = \frac{1 + \phi}{2} \alpha_r + \frac{1 - \phi}{2} \alpha_b,
\end{equation}
with $\alpha_b = 8/27$ and $c_s^b=1/\sqrt{3}$~\cite{saito2018color}.\\
\indent Following~\cite{BRACKBILL1992335, Reis_2007, PhysRevE.85.046309}, the interfacial tension is modeled by the so-called perturbation operator:
\begin{equation}
	\left(\Omega_i \right)^{(2)} = \frac{A}{2} |\nabla \phi| 
	\left[w_i \frac{(\bm{c}_i \cdot \nabla \phi)}{|\nabla \phi|^2} - B_i \right],
	\label{eq:perturb}
\end{equation}
where
\begin{equation}
	B_i = 
	\begin{cases}
		~-10/27, & |\bm{c}_i|^2=0,\\
		~+2/27, & |\bm{c}_i|^2=1, \\
		~+1/54, & |\bm{c}_i|^2=2, \\
		~+1/216, & |\bm{c}_i|^2=3. \label{eq:B_i}
\end{cases}
\end{equation} 
\indent Eventually, the following recoloring operator is applied to promote phase segregation and maintain the interface:
\begin{eqnarray}
	(\Omega_i^r)^{(3)} = \frac{\rho_r}{\rho}f_i 
	+ \gamma \frac{\rho_r \rho_b}{\rho^2} \cos(\theta_i)
	f_i^{eq}(\rho,\bf{0}), \label{eq:op3r}
\\
	(\Omega_i^b)^{(3)} = \frac{\rho_b}{\rho}f_i 
	- \gamma \frac{\rho_r \rho_b}{\rho^2} \cos(\theta_i)
	f_i^{eq}(\rho,\bf{0}), \label{eq:op3b}
\end{eqnarray}
where $\gamma=0.7$~\cite{LECLAIRE20122237, saito2017lattice, saito2018color}, $\theta_i$ is the angle between $\nabla\phi$ and $\bm{c}_i$, which is defined by
\begin{equation}
	\cos(\theta_i) = \frac{\bm{c}_i \cdot \nabla \phi}{|\bm{c}_i|  |\nabla \phi|}.
\end{equation}

\bibliographystyle{apsrev4-1}
\bibliography{bibliography}

\begin{thebibliography}{77}%
\makeatletter
\providecommand \@ifxundefined [1]{%
 \@ifx{#1\undefined}
}%
\providecommand \@ifnum [1]{%
 \ifnum #1\expandafter \@firstoftwo
 \else \expandafter \@secondoftwo
 \fi
}%
\providecommand \@ifx [1]{%
 \ifx #1\expandafter \@firstoftwo
 \else \expandafter \@secondoftwo
 \fi
}%
\providecommand \natexlab [1]{#1}%
\providecommand \enquote  [1]{``#1''}%
\providecommand \bibnamefont  [1]{#1}%
\providecommand \bibfnamefont [1]{#1}%
\providecommand \citenamefont [1]{#1}%
\providecommand \href@noop [0]{\@secondoftwo}%
\providecommand \href [0]{\begingroup \@sanitize@url \@href}%
\providecommand \@href[1]{\@@startlink{#1}\@@href}%
\providecommand \@@href[1]{\endgroup#1\@@endlink}%
\providecommand \@sanitize@url [0]{\catcode `\\12\catcode `\$12\catcode
  `\&12\catcode `\#12\catcode `\^12\catcode `\_12\catcode `\%12\relax}%
\providecommand \@@startlink[1]{}%
\providecommand \@@endlink[0]{}%
\providecommand \url  [0]{\begingroup\@sanitize@url \@url }%
\providecommand \@url [1]{\endgroup\@href {#1}{\urlprefix }}%
\providecommand \urlprefix  [0]{URL }%
\providecommand \Eprint [0]{\href }%
\providecommand \doibase [0]{http://dx.doi.org/}%
\providecommand \selectlanguage [0]{\@gobble}%
\providecommand \bibinfo  [0]{\@secondoftwo}%
\providecommand \bibfield  [0]{\@secondoftwo}%
\providecommand \translation [1]{[#1]}%
\providecommand \BibitemOpen [0]{}%
\providecommand \bibitemStop [0]{}%
\providecommand \bibitemNoStop [0]{.\EOS\space}%
\providecommand \EOS [0]{\spacefactor3000\relax}%
\providecommand \BibitemShut  [1]{\csname bibitem#1\endcsname}%
\let\auto@bib@innerbib\@empty
\bibitem [{\citenamefont {Benzi}\ \emph {et~al.}(1992)\citenamefont {Benzi},
  \citenamefont {Succi},\ and\ \citenamefont {Vergassola}}]{benzi1992lattice}%
  \BibitemOpen
  \bibfield  {author} {\bibinfo {author} {\bibfnamefont {R.}~\bibnamefont
  {Benzi}}, \bibinfo {author} {\bibfnamefont {S.}~\bibnamefont {Succi}}, \ and\
  \bibinfo {author} {\bibfnamefont {M.}~\bibnamefont {Vergassola}},\ }\href
  {\doibase 10.1016/0370-1573(92)90090-M} {\bibfield  {journal} {\bibinfo
  {journal} {Phys. Rep.}\ }\textbf {\bibinfo {volume} {222}},\ \bibinfo {pages}
  {145} (\bibinfo {year} {1992})}\BibitemShut {NoStop}%
\bibitem [{\citenamefont {Succi}(2001)}]{SucciBook}%
  \BibitemOpen
  \bibfield  {author} {\bibinfo {author} {\bibfnamefont {S.}~\bibnamefont
  {Succi}},\ }\href
  {https://global.oup.com/academic/product/the-lattice-boltzmann-equation-9780198503989?cc=nl&lang=en&}
  {\emph {\bibinfo {title} {The Lattice {B}oltzmann Equation for Fluid Dynamics
  and Beyond}}}\ (\bibinfo  {publisher} {Clarendon},\ \bibinfo {year}
  {2001})\BibitemShut {NoStop}%
\bibitem [{\citenamefont {Succi}(2015)}]{succi2015lattice}%
  \BibitemOpen
  \bibfield  {author} {\bibinfo {author} {\bibfnamefont {S.}~\bibnamefont
  {Succi}},\ }\href {\doibase 10.1209/0295-5075/109/50001} {\bibfield
  {journal} {\bibinfo  {journal} {Europhys. Lett.}\ }\textbf {\bibinfo {volume}
  {109}},\ \bibinfo {pages} {50001} (\bibinfo {year} {2015})}\BibitemShut
  {NoStop}%
\bibitem [{\citenamefont {Succi}(2016)}]{succi2016chimaera}%
  \BibitemOpen
  \bibfield  {author} {\bibinfo {author} {\bibfnamefont {S.}~\bibnamefont
  {Succi}},\ }\href
  {https://royalsocietypublishing.org/doi/10.1098/rsta.2016.0151} {\bibfield
  {journal} {\bibinfo  {journal} {Phil. Trans. R. Soc. A}\ }\textbf {\bibinfo
  {volume} {374}} (\bibinfo {year} {2016})}\BibitemShut {NoStop}%
\bibitem [{\citenamefont {Kr{\"u}ger}\ \emph {et~al.}(2016)\citenamefont
  {Kr{\"u}ger}, \citenamefont {Kusumaatmaja}, \citenamefont {Kuzmin},
  \citenamefont {Shardt}, \citenamefont {Silva},\ and\ \citenamefont
  {Viggen}}]{kruger2016lattice}%
  \BibitemOpen
  \bibfield  {author} {\bibinfo {author} {\bibfnamefont {T.}~\bibnamefont
  {Kr{\"u}ger}}, \bibinfo {author} {\bibfnamefont {H.}~\bibnamefont
  {Kusumaatmaja}}, \bibinfo {author} {\bibfnamefont {A.}~\bibnamefont
  {Kuzmin}}, \bibinfo {author} {\bibfnamefont {O.}~\bibnamefont {Shardt}},
  \bibinfo {author} {\bibfnamefont {G.}~\bibnamefont {Silva}}, \ and\ \bibinfo
  {author} {\bibfnamefont {E.~M.}\ \bibnamefont {Viggen}},\ }\href
  {https://www.springer.com/la/book/9783319446479} {\emph {\bibinfo {title}
  {The Lattice Boltzmann Method: Principles and Practice}}}\ (\bibinfo
  {publisher} {Springer},\ \bibinfo {year} {2016})\BibitemShut {NoStop}%
\bibitem [{\citenamefont {Succi}(2018)}]{succi2018lattice}%
  \BibitemOpen
  \bibfield  {author} {\bibinfo {author} {\bibfnamefont {S.}~\bibnamefont
  {Succi}},\ }\href
  {https://global.oup.com/academic/product/the-lattice-boltzmann-equation-9780199592357?cc=nl&lang=en&}
  {\emph {\bibinfo {title} {The Lattice Boltzmann Equation: For Complex States
  of Flowing Matter}}}\ (\bibinfo  {publisher} {Oxford University Press},\
  \bibinfo {year} {2018})\BibitemShut {NoStop}%
\bibitem [{\citenamefont {Bhatnagar}\ \emph {et~al.}(1954)\citenamefont
  {Bhatnagar}, \citenamefont {Gross},\ and\ \citenamefont
  {Krook}}]{bhatnagar1954model}%
  \BibitemOpen
  \bibfield  {author} {\bibinfo {author} {\bibfnamefont {P.}~\bibnamefont
  {Bhatnagar}}, \bibinfo {author} {\bibfnamefont {E.}~\bibnamefont {Gross}}, \
  and\ \bibinfo {author} {\bibfnamefont {M.}~\bibnamefont {Krook}},\ }\href
  {\doibase 10.1103/PhysRev.94.511} {\bibfield  {journal} {\bibinfo  {journal}
  {Phys. Rev.}\ }\textbf {\bibinfo {volume} {94}},\ \bibinfo {pages} {511}
  (\bibinfo {year} {1954})}\BibitemShut {NoStop}%
\bibitem [{\citenamefont {Shan}\ \emph {et~al.}(2006)\citenamefont {Shan},
  \citenamefont {Yuan},\ and\ \citenamefont {Chen}}]{shan2006kinetic}%
  \BibitemOpen
  \bibfield  {author} {\bibinfo {author} {\bibfnamefont {X.}~\bibnamefont
  {Shan}}, \bibinfo {author} {\bibfnamefont {X.-F.}\ \bibnamefont {Yuan}}, \
  and\ \bibinfo {author} {\bibfnamefont {H.}~\bibnamefont {Chen}},\ }\href
  {\doibase 10.1017/S0022112005008153} {\bibfield  {journal} {\bibinfo
  {journal} {J. Fluid Mech.}\ }\textbf {\bibinfo {volume} {550}},\ \bibinfo
  {pages} {413} (\bibinfo {year} {2006})}\BibitemShut {NoStop}%
\bibitem [{\citenamefont {d'Humi{\`e}res}(2002)}]{d2002multiple}%
  \BibitemOpen
  \bibfield  {author} {\bibinfo {author} {\bibfnamefont {D.}~\bibnamefont
  {d'Humi{\`e}res}},\ }\href {\doibase 10.1098/rsta.2001.0955} {\bibfield
  {journal} {\bibinfo  {journal} {Philos. T. R. Soc. A}\ }\textbf {\bibinfo
  {volume} {360}},\ \bibinfo {pages} {437} (\bibinfo {year}
  {2002})}\BibitemShut {NoStop}%
\bibitem [{\citenamefont {Latt}\ and\ \citenamefont
  {Chopard}(2006)}]{Latt2006165}%
  \BibitemOpen
  \bibfield  {author} {\bibinfo {author} {\bibfnamefont {J.}~\bibnamefont
  {Latt}}\ and\ \bibinfo {author} {\bibfnamefont {B.}~\bibnamefont {Chopard}},\
  }\href {https://www.sciencedirect.com/science/article/pii/S0378475406001583}
  {\bibfield  {journal} {\bibinfo  {journal} {Math. Comput. Simulat.}\ }\textbf
  {\bibinfo {volume} {72}},\ \bibinfo {pages} {165 } (\bibinfo {year}
  {2006})}\BibitemShut {NoStop}%
\bibitem [{\citenamefont {Mari{\'e}}\ \emph {et~al.}(2009)\citenamefont
  {Mari{\'e}}, \citenamefont {Ricot},\ and\ \citenamefont
  {Sagaut}}]{marie2009comparison}%
  \BibitemOpen
  \bibfield  {author} {\bibinfo {author} {\bibfnamefont {S.}~\bibnamefont
  {Mari{\'e}}}, \bibinfo {author} {\bibfnamefont {D.}~\bibnamefont {Ricot}}, \
  and\ \bibinfo {author} {\bibfnamefont {P.}~\bibnamefont {Sagaut}},\ }\href
  {\doibase 10.1016/j.jcp.2008.10.021} {\bibfield  {journal} {\bibinfo
  {journal} {J. Comput. Phys.}\ }\textbf {\bibinfo {volume} {228}},\ \bibinfo
  {pages} {1056} (\bibinfo {year} {2009})}\BibitemShut {NoStop}%
\bibitem [{\citenamefont {Nie}\ \emph {et~al.}(2008)\citenamefont {Nie},
  \citenamefont {Shan},\ and\ \citenamefont {Chen}}]{0295-5075-81-3-34005}%
  \BibitemOpen
  \bibfield  {author} {\bibinfo {author} {\bibfnamefont {X.~B.}\ \bibnamefont
  {Nie}}, \bibinfo {author} {\bibfnamefont {X.}~\bibnamefont {Shan}}, \ and\
  \bibinfo {author} {\bibfnamefont {H.}~\bibnamefont {Chen}},\ }\href {\doibase
  10.1209/0295-5075/81/34005} {\bibfield  {journal} {\bibinfo  {journal}
  {Europhys. Lett.}\ }\textbf {\bibinfo {volume} {81}},\ \bibinfo {pages}
  {34005} (\bibinfo {year} {2008})}\BibitemShut {NoStop}%
\bibitem [{\citenamefont {Coreixas}(2018)}]{COREIXAS_PhD_2018}%
  \BibitemOpen
  \bibfield  {author} {\bibinfo {author} {\bibfnamefont {C.}~\bibnamefont
  {Coreixas}},\ }\emph {\bibinfo {title} {High-order extension of the recursive
  regularized lattice Boltzmann method}},\ \href {\doibase
  http://oatao.univ-toulouse.fr/19861} {Ph.D. thesis},\ \bibinfo  {school} {INP
  Toulouse} (\bibinfo {year} {2018})\BibitemShut {NoStop}%
\bibitem [{\citenamefont {Geier}\ \emph {et~al.}(2006)\citenamefont {Geier},
  \citenamefont {Greiner},\ and\ \citenamefont {Korvink}}]{geier2006cascaded}%
  \BibitemOpen
  \bibfield  {author} {\bibinfo {author} {\bibfnamefont {M.}~\bibnamefont
  {Geier}}, \bibinfo {author} {\bibfnamefont {A.}~\bibnamefont {Greiner}}, \
  and\ \bibinfo {author} {\bibfnamefont {J.}~\bibnamefont {Korvink}},\ }\href
  {https://journals.aps.org/pre/abstract/10.1103/PhysRevE.73.066705} {\bibfield
   {journal} {\bibinfo  {journal} {Phys. Rev. E}\ }\textbf {\bibinfo {volume}
  {73}},\ \bibinfo {pages} {066705} (\bibinfo {year} {2006})}\BibitemShut
  {NoStop}%
\bibitem [{\citenamefont {Geier}\ \emph {et~al.}(2007)\citenamefont {Geier},
  \citenamefont {Greiner},\ and\ \citenamefont
  {Korvink}}]{geier2007properties}%
  \BibitemOpen
  \bibfield  {author} {\bibinfo {author} {\bibfnamefont {M.}~\bibnamefont
  {Geier}}, \bibinfo {author} {\bibfnamefont {A.}~\bibnamefont {Greiner}}, \
  and\ \bibinfo {author} {\bibfnamefont {J.}~\bibnamefont {Korvink}},\ }\href
  {\doibase 10.1142/S0129183107010681} {\bibfield  {journal} {\bibinfo
  {journal} {Int. J. Mod. Phys. C}\ }\textbf {\bibinfo {volume} {18}},\
  \bibinfo {pages} {455} (\bibinfo {year} {2007})}\BibitemShut {NoStop}%
\bibitem [{\citenamefont {Geier}(2008)}]{geier2008aliasing}%
  \BibitemOpen
  \bibfield  {author} {\bibinfo {author} {\bibfnamefont {M.}~\bibnamefont
  {Geier}},\ }\href {\doibase 10.1002/fld.1634} {\bibfield  {journal} {\bibinfo
   {journal} {Int. J. Numer. Meth. Fl.}\ }\textbf {\bibinfo {volume} {56}},\
  \bibinfo {pages} {1249} (\bibinfo {year} {2008})}\BibitemShut {NoStop}%
\bibitem [{\citenamefont {Asinari}(2008)}]{asinari2008generalized}%
  \BibitemOpen
  \bibfield  {author} {\bibinfo {author} {\bibfnamefont {P.}~\bibnamefont
  {Asinari}},\ }\href
  {https://journals.aps.org/pre/abstract/10.1103/PhysRevE.78.016701} {\bibfield
   {journal} {\bibinfo  {journal} {Phys. Rev. E}\ }\textbf {\bibinfo {volume}
  {78}},\ \bibinfo {pages} {016701} (\bibinfo {year} {2008})}\BibitemShut
  {NoStop}%
\bibitem [{\citenamefont {Geier}\ \emph {et~al.}(2009)\citenamefont {Geier},
  \citenamefont {Greiner},\ and\ \citenamefont
  {Korvink}}]{geier2009factorized}%
  \BibitemOpen
  \bibfield  {author} {\bibinfo {author} {\bibfnamefont {M.}~\bibnamefont
  {Geier}}, \bibinfo {author} {\bibfnamefont {A.}~\bibnamefont {Greiner}}, \
  and\ \bibinfo {author} {\bibfnamefont {J.}~\bibnamefont {Korvink}},\ }\href
  {\doibase 10.1140/epjst/e2009-01011-1} {\bibfield  {journal} {\bibinfo
  {journal} {Eur. Phys. J-Spec. Top.}\ }\textbf {\bibinfo {volume} {171}},\
  \bibinfo {pages} {55} (\bibinfo {year} {2009})}\BibitemShut {NoStop}%
\bibitem [{\citenamefont {Geier}\ \emph {et~al.}(2015)\citenamefont {Geier},
  \citenamefont {Sch{\"o}nherr}, \citenamefont {Pasquali},\ and\ \citenamefont
  {Krafczyk}}]{geier2015cumulant}%
  \BibitemOpen
  \bibfield  {author} {\bibinfo {author} {\bibfnamefont {M.}~\bibnamefont
  {Geier}}, \bibinfo {author} {\bibfnamefont {M.}~\bibnamefont
  {Sch{\"o}nherr}}, \bibinfo {author} {\bibfnamefont {A.}~\bibnamefont
  {Pasquali}}, \ and\ \bibinfo {author} {\bibfnamefont {M.}~\bibnamefont
  {Krafczyk}},\ }\href {\doibase 10.1016/j.camwa.2015.05.001} {\bibfield
  {journal} {\bibinfo  {journal} {Comput. Math. Appl.}\ }\textbf {\bibinfo
  {volume} {70}},\ \bibinfo {pages} {507} (\bibinfo {year} {2015})}\BibitemShut
  {NoStop}%
\bibitem [{\citenamefont {De~Rosis}\ and\ \citenamefont
  {L\'ev\^eque}(2016)}]{DeRosis_central2016}%
  \BibitemOpen
  \bibfield  {author} {\bibinfo {author} {\bibfnamefont {A.}~\bibnamefont
  {De~Rosis}}\ and\ \bibinfo {author} {\bibfnamefont {E.}~\bibnamefont
  {L\'ev\^eque}},\ }\href {\doibase 10.1016/j.camwa.2016.07.025} {\bibfield
  {journal} {\bibinfo  {journal} {Comput. Math. Appl.}\ }\textbf {\bibinfo
  {volume} {72}},\ \bibinfo {pages} {1616} (\bibinfo {year}
  {2016})}\BibitemShut {NoStop}%
\bibitem [{\citenamefont {Geier}\ \emph
  {et~al.}(2017{\natexlab{a}})\citenamefont {Geier}, \citenamefont {Pasquali},\
  and\ \citenamefont {Sch{\"o}nherr}}]{geier2017parametrization}%
  \BibitemOpen
  \bibfield  {author} {\bibinfo {author} {\bibfnamefont {M.}~\bibnamefont
  {Geier}}, \bibinfo {author} {\bibfnamefont {A.}~\bibnamefont {Pasquali}}, \
  and\ \bibinfo {author} {\bibfnamefont {M.}~\bibnamefont {Sch{\"o}nherr}},\
  }\href {\doibase 10.1016/j.jcp.2017.05.040} {\bibfield  {journal} {\bibinfo
  {journal} {J. Comput. Phys.}\ }\textbf {\bibinfo {volume} {348}},\ \bibinfo
  {pages} {862} (\bibinfo {year} {2017}{\natexlab{a}})}\BibitemShut {NoStop}%
\bibitem [{\citenamefont {Geier}\ \emph
  {et~al.}(2017{\natexlab{b}})\citenamefont {Geier}, \citenamefont {Pasquali},\
  and\ \citenamefont {Sch{\"o}nherr}}]{geier2017parametrization2}%
  \BibitemOpen
  \bibfield  {author} {\bibinfo {author} {\bibfnamefont {M.}~\bibnamefont
  {Geier}}, \bibinfo {author} {\bibfnamefont {A.}~\bibnamefont {Pasquali}}, \
  and\ \bibinfo {author} {\bibfnamefont {M.}~\bibnamefont {Sch{\"o}nherr}},\
  }\href {\doibase 10.1016/j.jcp.2017.07.004} {\bibfield  {journal} {\bibinfo
  {journal} {J. Comput. Phys.}\ }\textbf {\bibinfo {volume} {348}},\ \bibinfo
  {pages} {889} (\bibinfo {year} {2017}{\natexlab{b}})}\BibitemShut {NoStop}%
\bibitem [{\citenamefont {Geier}\ and\ \citenamefont
  {Pasquali}(2018)}]{geier2018fourth}%
  \BibitemOpen
  \bibfield  {author} {\bibinfo {author} {\bibfnamefont {M.}~\bibnamefont
  {Geier}}\ and\ \bibinfo {author} {\bibfnamefont {A.}~\bibnamefont
  {Pasquali}},\ }\href {\doibase 10.1016/j.compfluid.2018.01.015} {\bibfield
  {journal} {\bibinfo  {journal} {Comput. Fluids}\ }\textbf {\bibinfo {volume}
  {166}},\ \bibinfo {pages} {139} (\bibinfo {year} {2018})}\BibitemShut
  {NoStop}%
\bibitem [{\citenamefont {Fei}\ and\ \citenamefont
  {Luo}(2017)}]{fei2017consistent}%
  \BibitemOpen
  \bibfield  {author} {\bibinfo {author} {\bibfnamefont {L.}~\bibnamefont
  {Fei}}\ and\ \bibinfo {author} {\bibfnamefont {K.~H.}\ \bibnamefont {Luo}},\
  }\href {\doibase 10.1103/PhysRevE.96.053307} {\bibfield  {journal} {\bibinfo
  {journal} {Phys. Rev. E}\ }\textbf {\bibinfo {volume} {96}},\ \bibinfo
  {pages} {053307} (\bibinfo {year} {2017})}\BibitemShut {NoStop}%
\bibitem [{\citenamefont {Fei}\ \emph {et~al.}(2018)\citenamefont {Fei},
  \citenamefont {Luo},\ and\ \citenamefont {Li}}]{fei2018three}%
  \BibitemOpen
  \bibfield  {author} {\bibinfo {author} {\bibfnamefont {L.}~\bibnamefont
  {Fei}}, \bibinfo {author} {\bibfnamefont {K.~H.}\ \bibnamefont {Luo}}, \ and\
  \bibinfo {author} {\bibfnamefont {Q.}~\bibnamefont {Li}},\ }\href {\doibase
  10.1103/PhysRevE.97.053309} {\bibfield  {journal} {\bibinfo  {journal} {Phys.
  Rev. E}\ }\textbf {\bibinfo {volume} {97}},\ \bibinfo {pages} {053309}
  (\bibinfo {year} {2018})}\BibitemShut {NoStop}%
\bibitem [{\citenamefont {Shah}\ \emph {et~al.}(2017)\citenamefont {Shah},
  \citenamefont {Dhar}, \citenamefont {Chinige}, \citenamefont {Geier},\ and\
  \citenamefont {Pattamatta}}]{shah2017cascaded}%
  \BibitemOpen
  \bibfield  {author} {\bibinfo {author} {\bibfnamefont {N.}~\bibnamefont
  {Shah}}, \bibinfo {author} {\bibfnamefont {P.}~\bibnamefont {Dhar}}, \bibinfo
  {author} {\bibfnamefont {S.~K.}\ \bibnamefont {Chinige}}, \bibinfo {author}
  {\bibfnamefont {M.}~\bibnamefont {Geier}}, \ and\ \bibinfo {author}
  {\bibfnamefont {A.}~\bibnamefont {Pattamatta}},\ }\href {\doibase
  10.1080/10407790.2017.1377530} {\bibfield  {journal} {\bibinfo  {journal}
  {Numer. Heat Tr. B-Fund.}\ }\textbf {\bibinfo {volume} {72}},\ \bibinfo
  {pages} {211} (\bibinfo {year} {2017})}\BibitemShut {NoStop}%
\bibitem [{\citenamefont {Kumar}\ \emph {et~al.}(2017)\citenamefont {Kumar},
  \citenamefont {Mohankumar}, \citenamefont {Geier},\ and\ \citenamefont
  {Pattamatta}}]{kumar2017numerical}%
  \BibitemOpen
  \bibfield  {author} {\bibinfo {author} {\bibfnamefont {C.~S.}\ \bibnamefont
  {Kumar}}, \bibinfo {author} {\bibfnamefont {S.}~\bibnamefont {Mohankumar}},
  \bibinfo {author} {\bibfnamefont {M.}~\bibnamefont {Geier}}, \ and\ \bibinfo
  {author} {\bibfnamefont {A.}~\bibnamefont {Pattamatta}},\ }\href {\doibase
  10.1016/j.ijthermalsci.2017.08.020} {\bibfield  {journal} {\bibinfo
  {journal} {Int. J. Therm. Sci.}\ }\textbf {\bibinfo {volume} {122}},\
  \bibinfo {pages} {201} (\bibinfo {year} {2017})}\BibitemShut {NoStop}%
\bibitem [{\citenamefont {Sharma}\ \emph {et~al.}(2017)\citenamefont {Sharma},
  \citenamefont {Straka},\ and\ \citenamefont {Tavares}}]{sharma2017new}%
  \BibitemOpen
  \bibfield  {author} {\bibinfo {author} {\bibfnamefont {K.~V.}\ \bibnamefont
  {Sharma}}, \bibinfo {author} {\bibfnamefont {R.}~\bibnamefont {Straka}}, \
  and\ \bibinfo {author} {\bibfnamefont {F.~W.}\ \bibnamefont {Tavares}},\
  }\href {\doibase 10.1016/j.ijthermalsci.2017.04.020} {\bibfield  {journal}
  {\bibinfo  {journal} {Int. J. Therm. Sci.}\ }\textbf {\bibinfo {volume}
  {118}},\ \bibinfo {pages} {259} (\bibinfo {year} {2017})}\BibitemShut
  {NoStop}%
\bibitem [{\citenamefont {Fei}\ and\ \citenamefont
  {Luo}(2018{\natexlab{a}})}]{fei2018cascaded}%
  \BibitemOpen
  \bibfield  {author} {\bibinfo {author} {\bibfnamefont {L.}~\bibnamefont
  {Fei}}\ and\ \bibinfo {author} {\bibfnamefont {K.~H.}\ \bibnamefont {Luo}},\
  }\href {\doibase 10.1016/j.ijthermalsci.2018.06.017} {\bibfield  {journal}
  {\bibinfo  {journal} {Int. J. Therm. Sci.}\ }\textbf {\bibinfo {volume}
  {132}},\ \bibinfo {pages} {368} (\bibinfo {year}
  {2018}{\natexlab{a}})}\BibitemShut {NoStop}%
\bibitem [{\citenamefont {Fei}\ and\ \citenamefont
  {Luo}(2018{\natexlab{b}})}]{fei2018cascaded_2}%
  \BibitemOpen
  \bibfield  {author} {\bibinfo {author} {\bibfnamefont {L.}~\bibnamefont
  {Fei}}\ and\ \bibinfo {author} {\bibfnamefont {K.~H.}\ \bibnamefont {Luo}},\
  }\href {\doibase 10.1016/j.compfluid.2018.01.020} {\bibfield  {journal}
  {\bibinfo  {journal} {Comput. Fluids}\ }\textbf {\bibinfo {volume} {165}},\
  \bibinfo {pages} {89} (\bibinfo {year} {2018}{\natexlab{b}})}\BibitemShut
  {NoStop}%
\bibitem [{\citenamefont {Safari}\ \emph {et~al.}(2018)\citenamefont {Safari},
  \citenamefont {Krafczyk},\ and\ \citenamefont {Geier}}]{safari2018lattice}%
  \BibitemOpen
  \bibfield  {author} {\bibinfo {author} {\bibfnamefont {H.}~\bibnamefont
  {Safari}}, \bibinfo {author} {\bibfnamefont {M.}~\bibnamefont {Krafczyk}}, \
  and\ \bibinfo {author} {\bibfnamefont {M.}~\bibnamefont {Geier}},\ }\href
  {https://www.sciencedirect.com/science/article/pii/S0045793018302123}
  {\bibfield  {journal} {\bibinfo  {journal} {Comput. Fluids}\ } (\bibinfo
  {year} {2018})}\BibitemShut {NoStop}%
\bibitem [{\citenamefont {Premnath}\ and\ \citenamefont
  {Banerjee}(2009)}]{premnath2009incorporating}%
  \BibitemOpen
  \bibfield  {author} {\bibinfo {author} {\bibfnamefont {K.}~\bibnamefont
  {Premnath}}\ and\ \bibinfo {author} {\bibfnamefont {S.}~\bibnamefont
  {Banerjee}},\ }\href {\doibase 10.1103/PhysRevE.80.036702} {\bibfield
  {journal} {\bibinfo  {journal} {Phys. Rev. E}\ }\textbf {\bibinfo {volume}
  {80}},\ \bibinfo {pages} {036702} (\bibinfo {year} {2009})}\BibitemShut
  {NoStop}%
\bibitem [{\citenamefont {Premnath}\ and\ \citenamefont
  {Banerjee}(2011)}]{premnath2011three}%
  \BibitemOpen
  \bibfield  {author} {\bibinfo {author} {\bibfnamefont {K.}~\bibnamefont
  {Premnath}}\ and\ \bibinfo {author} {\bibfnamefont {S.}~\bibnamefont
  {Banerjee}},\ }\href {\doibase 10.1007/s10955-011-0208-9} {\bibfield
  {journal} {\bibinfo  {journal} {J. Stat. Phys.}\ }\textbf {\bibinfo {volume}
  {143}},\ \bibinfo {pages} {747} (\bibinfo {year} {2011})}\BibitemShut
  {NoStop}%
\bibitem [{\citenamefont {Ning}\ \emph {et~al.}(2016)\citenamefont {Ning},
  \citenamefont {Premnath},\ and\ \citenamefont {Patil}}]{FLD:FLD4208}%
  \BibitemOpen
  \bibfield  {author} {\bibinfo {author} {\bibfnamefont {Y.}~\bibnamefont
  {Ning}}, \bibinfo {author} {\bibfnamefont {K.~N.}\ \bibnamefont {Premnath}},
  \ and\ \bibinfo {author} {\bibfnamefont {D.~V.}\ \bibnamefont {Patil}},\
  }\href {\doibase 10.1002/fld.4208} {\bibfield  {journal} {\bibinfo  {journal}
  {Int. J. Numer. Meth. Fl.}\ }\textbf {\bibinfo {volume} {82}},\ \bibinfo
  {pages} {59} (\bibinfo {year} {2016})}\BibitemShut {NoStop}%
\bibitem [{\citenamefont {Hajabdollahi}\ and\ \citenamefont
  {Premnath}(2017)}]{hajabdollahi2017improving}%
  \BibitemOpen
  \bibfield  {author} {\bibinfo {author} {\bibfnamefont {F.}~\bibnamefont
  {Hajabdollahi}}\ and\ \bibinfo {author} {\bibfnamefont {K.~N.}\ \bibnamefont
  {Premnath}},\ }\href
  {https://www.sciencedirect.com/science/article/pii/S0898122117300664}
  {\bibfield  {journal} {\bibinfo  {journal} {Comput. Math. Appl.}\ } (\bibinfo
  {year} {2017})}\BibitemShut {NoStop}%
\bibitem [{\citenamefont {Ch\'{a}vez-Modena}\ \emph {et~al.}(2018)\citenamefont
  {Ch\'{a}vez-Modena}, \citenamefont {Ferrer},\ and\ \citenamefont
  {Rubio}}]{CHAVEZMODENA_CF_172_2018}%
  \BibitemOpen
  \bibfield  {author} {\bibinfo {author} {\bibfnamefont {M.}~\bibnamefont
  {Ch\'{a}vez-Modena}}, \bibinfo {author} {\bibfnamefont {E.}~\bibnamefont
  {Ferrer}}, \ and\ \bibinfo {author} {\bibfnamefont {G.}~\bibnamefont
  {Rubio}},\ }\href {\doibase 10.1016/j.compfluid.2018.03.084} {\bibfield
  {journal} {\bibinfo  {journal} {Comput. Fluids}\ }\textbf {\bibinfo {volume}
  {172}},\ \bibinfo {pages} {397 } (\bibinfo {year} {2018})}\BibitemShut
  {NoStop}%
\bibitem [{\citenamefont {De~Rosis}(2016)}]{derosis2016epl_d2q9}%
  \BibitemOpen
  \bibfield  {author} {\bibinfo {author} {\bibfnamefont {A.}~\bibnamefont
  {De~Rosis}},\ }\href {\doibase 10.1209/0295-5075/116/44003} {\bibfield
  {journal} {\bibinfo  {journal} {Europhys. Lett.}\ }\textbf {\bibinfo {volume}
  {116}},\ \bibinfo {pages} {44003} (\bibinfo {year} {2016})}\BibitemShut
  {NoStop}%
\bibitem [{\citenamefont {De~Rosis}(2017{\natexlab{a}})}]{de2017nonorthogonal}%
  \BibitemOpen
  \bibfield  {author} {\bibinfo {author} {\bibfnamefont {A.}~\bibnamefont
  {De~Rosis}},\ }\href {\doibase 10.1103/PhysRevE.95.013310} {\bibfield
  {journal} {\bibinfo  {journal} {Phys. Rev. E}\ }\textbf {\bibinfo {volume}
  {95}},\ \bibinfo {pages} {013310} (\bibinfo {year}
  {2017}{\natexlab{a}})}\BibitemShut {NoStop}%
\bibitem [{\citenamefont {De~Rosis}(2017{\natexlab{b}})}]{PhysRevE.95.023311}%
  \BibitemOpen
  \bibfield  {author} {\bibinfo {author} {\bibfnamefont {A.}~\bibnamefont
  {De~Rosis}},\ }\href {\doibase 10.1103/PhysRevE.95.023311} {\bibfield
  {journal} {\bibinfo  {journal} {Phys. Rev. E}\ }\textbf {\bibinfo {volume}
  {95}},\ \bibinfo {pages} {023311} (\bibinfo {year}
  {2017}{\natexlab{b}})}\BibitemShut {NoStop}%
\bibitem [{\citenamefont
  {De~Rosis}(2017{\natexlab{c}})}]{0295-5075-117-3-34003}%
  \BibitemOpen
  \bibfield  {author} {\bibinfo {author} {\bibfnamefont {A.}~\bibnamefont
  {De~Rosis}},\ }\href {\doibase 10.1209/0295-5075/117/34003} {\bibfield
  {journal} {\bibinfo  {journal} {Europhys. Lett.}\ }\textbf {\bibinfo {volume}
  {117}},\ \bibinfo {pages} {34003} (\bibinfo {year}
  {2017}{\natexlab{c}})}\BibitemShut {NoStop}%
\bibitem [{\citenamefont
  {De~Rosis}(2017{\natexlab{d}})}]{de2017preconditioned}%
  \BibitemOpen
  \bibfield  {author} {\bibinfo {author} {\bibfnamefont {A.}~\bibnamefont
  {De~Rosis}},\ }\href {\doibase 10.1103/PhysRevE.96.063308} {\bibfield
  {journal} {\bibinfo  {journal} {Phys. Rev. E}\ }\textbf {\bibinfo {volume}
  {96}},\ \bibinfo {pages} {063308} (\bibinfo {year}
  {2017}{\natexlab{d}})}\BibitemShut {NoStop}%
\bibitem [{\citenamefont
  {De~Rosis}(2017{\natexlab{e}})}]{de2017central_shallow}%
  \BibitemOpen
  \bibfield  {author} {\bibinfo {author} {\bibfnamefont {A.}~\bibnamefont
  {De~Rosis}},\ }\href {\doibase 10.1016/j.cma.2017.03.001} {\bibfield
  {journal} {\bibinfo  {journal} {Comput. Method. Appl. M.}\ }\textbf {\bibinfo
  {volume} {319}},\ \bibinfo {pages} {379} (\bibinfo {year}
  {2017}{\natexlab{e}})}\BibitemShut {NoStop}%
\bibitem [{\citenamefont {De~Rosis}\ \emph {et~al.}(2018)\citenamefont
  {De~Rosis}, \citenamefont {L{\'e}v{\^e}que},\ and\ \citenamefont
  {Chahine}}]{de2018advanced}%
  \BibitemOpen
  \bibfield  {author} {\bibinfo {author} {\bibfnamefont {A.}~\bibnamefont
  {De~Rosis}}, \bibinfo {author} {\bibfnamefont {E.}~\bibnamefont
  {L{\'e}v{\^e}que}}, \ and\ \bibinfo {author} {\bibfnamefont {R.}~\bibnamefont
  {Chahine}},\ }\href {\doibase 10.1080/14685248.2018.1461875} {\bibfield
  {journal} {\bibinfo  {journal} {J. Turbul.}\ }\textbf {\bibinfo {volume}
  {19}},\ \bibinfo {pages} {446} (\bibinfo {year} {2018})}\BibitemShut
  {NoStop}%
\bibitem [{\citenamefont {Saito}\ \emph {et~al.}(2018)\citenamefont {Saito},
  \citenamefont {De~Rosis}, \citenamefont {Festuccia}, \citenamefont {Kaneko},
  \citenamefont {Abe},\ and\ \citenamefont {Koyama}}]{saito2018color}%
  \BibitemOpen
  \bibfield  {author} {\bibinfo {author} {\bibfnamefont {S.}~\bibnamefont
  {Saito}}, \bibinfo {author} {\bibfnamefont {A.}~\bibnamefont {De~Rosis}},
  \bibinfo {author} {\bibfnamefont {A.}~\bibnamefont {Festuccia}}, \bibinfo
  {author} {\bibfnamefont {A.}~\bibnamefont {Kaneko}}, \bibinfo {author}
  {\bibfnamefont {Y.}~\bibnamefont {Abe}}, \ and\ \bibinfo {author}
  {\bibfnamefont {K.}~\bibnamefont {Koyama}},\ }\href {\doibase
  10.1103/PhysRevE.98.013305} {\bibfield  {journal} {\bibinfo  {journal} {Phys.
  Rev. E}\ }\textbf {\bibinfo {volume} {98}},\ \bibinfo {pages} {013305}
  (\bibinfo {year} {2018})}\BibitemShut {NoStop}%
\bibitem [{\citenamefont {Malaspinas}(2015)}]{malaspinas2015increasing}%
  \BibitemOpen
  \bibfield  {author} {\bibinfo {author} {\bibfnamefont {O.}~\bibnamefont
  {Malaspinas}},\ }\href {https://arxiv.org/abs/1505.06900} {\bibfield
  {journal} {\bibinfo  {journal} {arXiv preprint arXiv:1505.06900}\ } (\bibinfo
  {year} {2015})}\BibitemShut {NoStop}%
\bibitem [{\citenamefont {Coreixas}\ \emph {et~al.}(2017)\citenamefont
  {Coreixas}, \citenamefont {Wissocq}, \citenamefont {Puigt}, \citenamefont
  {Boussuge},\ and\ \citenamefont {Sagaut}}]{coreixas2017recursive}%
  \BibitemOpen
  \bibfield  {author} {\bibinfo {author} {\bibfnamefont {C.}~\bibnamefont
  {Coreixas}}, \bibinfo {author} {\bibfnamefont {G.}~\bibnamefont {Wissocq}},
  \bibinfo {author} {\bibfnamefont {G.}~\bibnamefont {Puigt}}, \bibinfo
  {author} {\bibfnamefont {J.-F.}\ \bibnamefont {Boussuge}}, \ and\ \bibinfo
  {author} {\bibfnamefont {P.}~\bibnamefont {Sagaut}},\ }\href {\doibase
  10.1103/PhysRevE.96.033306} {\bibfield  {journal} {\bibinfo  {journal} {Phys.
  Rev. E}\ }\textbf {\bibinfo {volume} {96}},\ \bibinfo {pages} {033306}
  (\bibinfo {year} {2017})}\BibitemShut {NoStop}%
\bibitem [{\citenamefont {De~Rosis}\ and\ \citenamefont
  {Luo}(2019)}]{derosisHermite}%
  \BibitemOpen
  \bibfield  {author} {\bibinfo {author} {\bibfnamefont {A.}~\bibnamefont
  {De~Rosis}}\ and\ \bibinfo {author} {\bibfnamefont {K.~H.}\ \bibnamefont
  {Luo}},\ }\href
  {https://journals.aps.org/pre/abstract/10.1103/PhysRevE.99.013301} {\bibfield
   {journal} {\bibinfo  {journal} {Phys. Rev. E}\ }\textbf {\bibinfo {volume}
  {99}},\ \bibinfo {pages} {013301} (\bibinfo {year} {2019})}\BibitemShut
  {NoStop}%
\bibitem [{\citenamefont {Grad}(1949)}]{grad1949kinetic}%
  \BibitemOpen
  \bibfield  {author} {\bibinfo {author} {\bibfnamefont {H.}~\bibnamefont
  {Grad}},\ }\href@noop {} {\bibfield  {journal} {\bibinfo  {journal} {Commun.
  Pure Appl. Maths}\ }\textbf {\bibinfo {volume} {2}},\ \bibinfo {pages} {331}
  (\bibinfo {year} {1949})}\BibitemShut {NoStop}%
\bibitem [{\citenamefont {Shan}\ and\ \citenamefont
  {He}(1998)}]{SHAN_PRL_80_1998}%
  \BibitemOpen
  \bibfield  {author} {\bibinfo {author} {\bibfnamefont {X.}~\bibnamefont
  {Shan}}\ and\ \bibinfo {author} {\bibfnamefont {X.}~\bibnamefont {He}},\
  }\href {\doibase 10.1103/PhysRevLett.80.65} {\bibfield  {journal} {\bibinfo
  {journal} {Phys. Rev. Lett.}\ }\textbf {\bibinfo {volume} {80}},\ \bibinfo
  {pages} {65} (\bibinfo {year} {1998})}\BibitemShut {NoStop}%
\bibitem [{\citenamefont {Guo}\ \emph {et~al.}(2002)\citenamefont {Guo},
  \citenamefont {Zheng},\ and\ \citenamefont {Shi}}]{guo2002discrete}%
  \BibitemOpen
  \bibfield  {author} {\bibinfo {author} {\bibfnamefont {Z.}~\bibnamefont
  {Guo}}, \bibinfo {author} {\bibfnamefont {C.}~\bibnamefont {Zheng}}, \ and\
  \bibinfo {author} {\bibfnamefont {B.}~\bibnamefont {Shi}},\ }\href
  {10.1103/PhysRevE.65.046308} {\bibfield  {journal} {\bibinfo  {journal}
  {Phys. Rev. E}\ }\textbf {\bibinfo {volume} {65}},\ \bibinfo {pages} {046308}
  (\bibinfo {year} {2002})}\BibitemShut {NoStop}%
\bibitem [{\citenamefont {Guo}\ and\ \citenamefont
  {Zheng}(2008)}]{guo2008analysis}%
  \BibitemOpen
  \bibfield  {author} {\bibinfo {author} {\bibfnamefont {Z.}~\bibnamefont
  {Guo}}\ and\ \bibinfo {author} {\bibfnamefont {C.}~\bibnamefont {Zheng}},\
  }\href {\doibase 10.1080/10618560802253100} {\bibfield  {journal} {\bibinfo
  {journal} {Int. J. Comput. Fluid D.}\ }\textbf {\bibinfo {volume} {22}},\
  \bibinfo {pages} {465} (\bibinfo {year} {2008})}\BibitemShut {NoStop}%
\bibitem [{\citenamefont {He}\ \emph {et~al.}(1998)\citenamefont {He},
  \citenamefont {Chen},\ and\ \citenamefont {Doolen}}]{he1998novel}%
  \BibitemOpen
  \bibfield  {author} {\bibinfo {author} {\bibfnamefont {X.}~\bibnamefont
  {He}}, \bibinfo {author} {\bibfnamefont {S.}~\bibnamefont {Chen}}, \ and\
  \bibinfo {author} {\bibfnamefont {G.~D.}\ \bibnamefont {Doolen}},\ }\href
  {\doibase 10.1006/jcph.1998.6057} {\bibfield  {journal} {\bibinfo  {journal}
  {Journal of Computational Physics}\ }\textbf {\bibinfo {volume} {146}},\
  \bibinfo {pages} {282} (\bibinfo {year} {1998})}\BibitemShut {NoStop}%
\bibitem [{\citenamefont {Huang}\ \emph {et~al.}(2018)\citenamefont {Huang},
  \citenamefont {Wu},\ and\ \citenamefont {Adams}}]{Huang2018}%
  \BibitemOpen
  \bibfield  {author} {\bibinfo {author} {\bibfnamefont {R.}~\bibnamefont
  {Huang}}, \bibinfo {author} {\bibfnamefont {H.}~\bibnamefont {Wu}}, \ and\
  \bibinfo {author} {\bibfnamefont {N.~A.}\ \bibnamefont {Adams}},\ }\href
  {\doibase 10.1103/PhysRevE.97.053308} {\bibfield  {journal} {\bibinfo
  {journal} {Phys. Rev. E}\ }\textbf {\bibinfo {volume} {97}},\ \bibinfo
  {pages} {053308} (\bibinfo {year} {2018})}\BibitemShut {NoStop}%
\bibitem [{Note1()}]{Note1}%
  \BibitemOpen
  \bibinfo {note} {See Supplemental Material at [\protect \url
  {D3Q27_CentralMomentsWithForcingScheme.m}] for performing all the symbolic
  manipulations to easily compute all formulas required for the implementation
  (the transformation matrix, the post-collision CMs, etc).}\BibitemShut
  {Stop}%
\bibitem [{\citenamefont {Latt}\ \emph {et~al.}(2008)\citenamefont {Latt},
  \citenamefont {Chopard}, \citenamefont {Malaspinas}, \citenamefont
  {Deville},\ and\ \citenamefont {Michler}}]{latt2008straight}%
  \BibitemOpen
  \bibfield  {author} {\bibinfo {author} {\bibfnamefont {J.}~\bibnamefont
  {Latt}}, \bibinfo {author} {\bibfnamefont {B.}~\bibnamefont {Chopard}},
  \bibinfo {author} {\bibfnamefont {O.}~\bibnamefont {Malaspinas}}, \bibinfo
  {author} {\bibfnamefont {M.}~\bibnamefont {Deville}}, \ and\ \bibinfo
  {author} {\bibfnamefont {A.}~\bibnamefont {Michler}},\ }\href {\doibase
  10.1103/PhysRevE.77.056703} {\bibfield  {journal} {\bibinfo  {journal} {Phys.
  Rev. E}\ }\textbf {\bibinfo {volume} {77}},\ \bibinfo {pages} {056703}
  (\bibinfo {year} {2008})}\BibitemShut {NoStop}%
\bibitem [{\citenamefont {Kupershtokh}\ and\ \citenamefont
  {Medvedev}(2006)}]{KUPERSHTOKH2006581}%
  \BibitemOpen
  \bibfield  {author} {\bibinfo {author} {\bibfnamefont {A.}~\bibnamefont
  {Kupershtokh}}\ and\ \bibinfo {author} {\bibfnamefont {D.}~\bibnamefont
  {Medvedev}},\ }\href {\doibase https://doi.org/10.1016/j.elstat.2005.10.012}
  {\bibfield  {journal} {\bibinfo  {journal} {J. Electrostat.}\ }\textbf
  {\bibinfo {volume} {64}},\ \bibinfo {pages} {581 } (\bibinfo {year}
  {2006})}\BibitemShut {NoStop}%
\bibitem [{\citenamefont {Malaspinas}\ \emph {et~al.}(2010)\citenamefont
  {Malaspinas}, \citenamefont {Fietier},\ and\ \citenamefont
  {Deville}}]{malaspinas2010lattice}%
  \BibitemOpen
  \bibfield  {author} {\bibinfo {author} {\bibfnamefont {O.}~\bibnamefont
  {Malaspinas}}, \bibinfo {author} {\bibfnamefont {N.}~\bibnamefont {Fietier}},
  \ and\ \bibinfo {author} {\bibfnamefont {M.}~\bibnamefont {Deville}},\ }\href
  {\doibase 10.1016/j.jnnfm.2010.09.001} {\bibfield  {journal} {\bibinfo
  {journal} {J. Non-Newton. Fluid}\ }\textbf {\bibinfo {volume} {165}},\
  \bibinfo {pages} {1637} (\bibinfo {year} {2010})}\BibitemShut {NoStop}%
\bibitem [{\citenamefont {Dellar}(2013)}]{dellar2013moment}%
  \BibitemOpen
  \bibfield  {author} {\bibinfo {author} {\bibfnamefont {P.~J.}\ \bibnamefont
  {Dellar}},\ }in\ \href {\doibase 10.1007/978-3-642-33134-3_9} {\emph
  {\bibinfo {booktitle} {Numerical Mathematics and Advanced Applications
  2011}}}\ (\bibinfo  {publisher} {Springer},\ \bibinfo {year} {2013})\ pp.\
  \bibinfo {pages} {83--90}\BibitemShut {NoStop}%
\bibitem [{\citenamefont {Dellar}(2002)}]{dellar2002lattice}%
  \BibitemOpen
  \bibfield  {author} {\bibinfo {author} {\bibfnamefont {P.~J.}\ \bibnamefont
  {Dellar}},\ }\href {\doibase 10.1006/jcph.2002.7044} {\bibfield  {journal}
  {\bibinfo  {journal} {J. Comput. Phys.}\ }\textbf {\bibinfo {volume} {179}},\
  \bibinfo {pages} {95} (\bibinfo {year} {2002})}\BibitemShut {NoStop}%
\bibitem [{\citenamefont {Pattison}\ \emph {et~al.}(2008)\citenamefont
  {Pattison}, \citenamefont {Premnath}, \citenamefont {Morley},\ and\
  \citenamefont {Abdou}}]{pattison2008progress}%
  \BibitemOpen
  \bibfield  {author} {\bibinfo {author} {\bibfnamefont {M.}~\bibnamefont
  {Pattison}}, \bibinfo {author} {\bibfnamefont {K.}~\bibnamefont {Premnath}},
  \bibinfo {author} {\bibfnamefont {N.}~\bibnamefont {Morley}}, \ and\ \bibinfo
  {author} {\bibfnamefont {M.}~\bibnamefont {Abdou}},\ }\href {\doibase
  10.1016/j.fusengdes.2007.10.005} {\bibfield  {journal} {\bibinfo  {journal}
  {Fusion Eng. Des.}\ }\textbf {\bibinfo {volume} {83}},\ \bibinfo {pages}
  {557} (\bibinfo {year} {2008})}\BibitemShut {NoStop}%
\bibitem [{\citenamefont {Orszag}\ and\ \citenamefont
  {Tang}(1979)}]{orszag1979small}%
  \BibitemOpen
  \bibfield  {author} {\bibinfo {author} {\bibfnamefont {S.~A.}\ \bibnamefont
  {Orszag}}\ and\ \bibinfo {author} {\bibfnamefont {C.-M.}\ \bibnamefont
  {Tang}},\ }\href {\doibase 10.1017/S002211207900210X} {\bibfield  {journal}
  {\bibinfo  {journal} {J. Fluid Mech.}\ }\textbf {\bibinfo {volume} {90}},\
  \bibinfo {pages} {129} (\bibinfo {year} {1979})}\BibitemShut {NoStop}%
\bibitem [{\citenamefont {Mininni}\ \emph {et~al.}(2006)\citenamefont
  {Mininni}, \citenamefont {Pouquet},\ and\ \citenamefont
  {Montgomery}}]{mininni2006small}%
  \BibitemOpen
  \bibfield  {author} {\bibinfo {author} {\bibfnamefont {P.}~\bibnamefont
  {Mininni}}, \bibinfo {author} {\bibfnamefont {A.}~\bibnamefont {Pouquet}}, \
  and\ \bibinfo {author} {\bibfnamefont {D.}~\bibnamefont {Montgomery}},\
  }\href {\doibase 10.1103/PhysRevLett.97.244503} {\bibfield  {journal}
  {\bibinfo  {journal} {Phys. Rev. Lett.}\ }\textbf {\bibinfo {volume} {97}},\
  \bibinfo {pages} {244503} (\bibinfo {year} {2006})}\BibitemShut {NoStop}%
\bibitem [{\citenamefont {Greene}(1988)}]{greene1988geometrical}%
  \BibitemOpen
  \bibfield  {author} {\bibinfo {author} {\bibfnamefont {J.~M.}\ \bibnamefont
  {Greene}},\ }\href {\doibase 10.1029/JA093iA08p08583} {\bibfield  {journal}
  {\bibinfo  {journal} {J. Geophys. Res.-Space}\ }\textbf {\bibinfo {volume}
  {93}},\ \bibinfo {pages} {8583} (\bibinfo {year} {1988})}\BibitemShut
  {NoStop}%
\bibitem [{\citenamefont {Shan}\ and\ \citenamefont
  {Chen}(1993)}]{shan1993lattice}%
  \BibitemOpen
  \bibfield  {author} {\bibinfo {author} {\bibfnamefont {X.}~\bibnamefont
  {Shan}}\ and\ \bibinfo {author} {\bibfnamefont {H.}~\bibnamefont {Chen}},\
  }\href {https://journals.aps.org/pre/abstract/10.1103/PhysRevE.47.1815}
  {\bibfield  {journal} {\bibinfo  {journal} {Phys. Rev. E}\ }\textbf {\bibinfo
  {volume} {47}},\ \bibinfo {pages} {1815} (\bibinfo {year}
  {1993})}\BibitemShut {NoStop}%
\bibitem [{\citenamefont {Latva-Kokko}\ and\ \citenamefont
  {Rothman}(2005)}]{PhysRevE.71.056702}%
  \BibitemOpen
  \bibfield  {author} {\bibinfo {author} {\bibfnamefont {M.}~\bibnamefont
  {Latva-Kokko}}\ and\ \bibinfo {author} {\bibfnamefont {D.~H.}\ \bibnamefont
  {Rothman}},\ }\href {\doibase 10.1103/PhysRevE.71.056702} {\bibfield
  {journal} {\bibinfo  {journal} {Phys. Rev. E}\ }\textbf {\bibinfo {volume}
  {71}},\ \bibinfo {pages} {056702} (\bibinfo {year} {2005})}\BibitemShut
  {NoStop}%
\bibitem [{\citenamefont {Reis}\ and\ \citenamefont
  {Phillips}(2007)}]{Reis_2007}%
  \BibitemOpen
  \bibfield  {author} {\bibinfo {author} {\bibfnamefont {T.}~\bibnamefont
  {Reis}}\ and\ \bibinfo {author} {\bibfnamefont {T.~N.}\ \bibnamefont
  {Phillips}},\ }\href {\doibase 10.1088/1751-8113/40/14/018} {\bibfield
  {journal} {\bibinfo  {journal} {Journal of Physics A: Mathematical and
  Theoretical}\ }\textbf {\bibinfo {volume} {40}},\ \bibinfo {pages} {4033}
  (\bibinfo {year} {2007})}\BibitemShut {NoStop}%
\bibitem [{\citenamefont {Leclaire}\ \emph {et~al.}(2012)\citenamefont
  {Leclaire}, \citenamefont {Reggio},\ and\ \citenamefont
  {Trepanier}}]{LECLAIRE20122237}%
  \BibitemOpen
  \bibfield  {author} {\bibinfo {author} {\bibfnamefont {S.}~\bibnamefont
  {Leclaire}}, \bibinfo {author} {\bibfnamefont {M.}~\bibnamefont {Reggio}}, \
  and\ \bibinfo {author} {\bibfnamefont {J.-Y.}\ \bibnamefont {Trepanier}},\
  }\href {\doibase https://doi.org/10.1016/j.apm.2011.08.027} {\bibfield
  {journal} {\bibinfo  {journal} {Applied Mathematical Modelling}\ }\textbf
  {\bibinfo {volume} {36}},\ \bibinfo {pages} {2237 } (\bibinfo {year}
  {2012})}\BibitemShut {NoStop}%
\bibitem [{\citenamefont {Saito}\ \emph {et~al.}(2017)\citenamefont {Saito},
  \citenamefont {Abe},\ and\ \citenamefont {Koyama}}]{saito2017lattice}%
  \BibitemOpen
  \bibfield  {author} {\bibinfo {author} {\bibfnamefont {S.}~\bibnamefont
  {Saito}}, \bibinfo {author} {\bibfnamefont {Y.}~\bibnamefont {Abe}}, \ and\
  \bibinfo {author} {\bibfnamefont {K.}~\bibnamefont {Koyama}},\ }\href
  {\doibase 10.1103/PhysRevE.96.013317} {\bibfield  {journal} {\bibinfo
  {journal} {Phys. Rev. E}\ }\textbf {\bibinfo {volume} {96}},\ \bibinfo
  {pages} {013317} (\bibinfo {year} {2017})}\BibitemShut {NoStop}%
\bibitem [{\citenamefont {He}\ \emph {et~al.}(1999)\citenamefont {He},
  \citenamefont {Zhang}, \citenamefont {Chen},\ and\ \citenamefont
  {Doolen}}]{he1999three}%
  \BibitemOpen
  \bibfield  {author} {\bibinfo {author} {\bibfnamefont {X.}~\bibnamefont
  {He}}, \bibinfo {author} {\bibfnamefont {R.}~\bibnamefont {Zhang}}, \bibinfo
  {author} {\bibfnamefont {S.}~\bibnamefont {Chen}}, \ and\ \bibinfo {author}
  {\bibfnamefont {G.~D.}\ \bibnamefont {Doolen}},\ }\href {\doibase
  10.1063/1.869984} {\bibfield  {journal} {\bibinfo  {journal} {Phys. Fluids}\
  }\textbf {\bibinfo {volume} {11}},\ \bibinfo {pages} {1143} (\bibinfo {year}
  {1999})}\BibitemShut {NoStop}%
\bibitem [{\citenamefont {Wang}\ \emph {et~al.}(2016)\citenamefont {Wang},
  \citenamefont {Liu},\ and\ \citenamefont {Zhang}}]{WANG2016340}%
  \BibitemOpen
  \bibfield  {author} {\bibinfo {author} {\bibfnamefont {N.}~\bibnamefont
  {Wang}}, \bibinfo {author} {\bibfnamefont {H.}~\bibnamefont {Liu}}, \ and\
  \bibinfo {author} {\bibfnamefont {C.}~\bibnamefont {Zhang}},\ }\href
  {\doibase https://doi.org/10.1016/j.jocs.2016.04.012} {\bibfield  {journal}
  {\bibinfo  {journal} {J. Comput. Sci.}\ }\textbf {\bibinfo {volume} {17}},\
  \bibinfo {pages} {340 } (\bibinfo {year} {2016})},\ \bibinfo {note} {discrete
  Simulation of Fluid Dynamics 2015}\BibitemShut {NoStop}%
\bibitem [{\citenamefont {Lee}\ and\ \citenamefont {Kim}(2013)}]{LEE20131466}%
  \BibitemOpen
  \bibfield  {author} {\bibinfo {author} {\bibfnamefont {H.~G.}\ \bibnamefont
  {Lee}}\ and\ \bibinfo {author} {\bibfnamefont {J.}~\bibnamefont {Kim}},\
  }\href {\doibase https://doi.org/10.1016/j.camwa.2013.08.021} {\bibfield
  {journal} {\bibinfo  {journal} {Comput. Math. Appl.}\ }\textbf {\bibinfo
  {volume} {66}},\ \bibinfo {pages} {1466 } (\bibinfo {year}
  {2013})}\BibitemShut {NoStop}%
\bibitem [{\citenamefont {Coreixas}\ \emph {et~al.}(2019)\citenamefont
  {Coreixas}, \citenamefont {Chopard},\ and\ \citenamefont
  {Latt}}]{COREIXAS_ARXIV_1904_2019}%
  \BibitemOpen
  \bibfield  {author} {\bibinfo {author} {\bibfnamefont {C.}~\bibnamefont
  {Coreixas}}, \bibinfo {author} {\bibfnamefont {B.}~\bibnamefont {Chopard}}, \
  and\ \bibinfo {author} {\bibfnamefont {J.}~\bibnamefont {Latt}},\ }\href
  {https://arxiv.org/abs/1904.12948} {\bibfield  {journal} {\bibinfo  {journal}
  {arXiv preprint arXiv:1904.12948}\ } (\bibinfo {year} {2019})}\BibitemShut
  {NoStop}%
\bibitem [{\citenamefont {Leclaire}\ \emph {et~al.}(2013)\citenamefont
  {Leclaire}, \citenamefont {Pellerin}, \citenamefont {Reggio},\ and\
  \citenamefont {Trepanier}}]{LECLAIRE2013159}%
  \BibitemOpen
  \bibfield  {author} {\bibinfo {author} {\bibfnamefont {S.}~\bibnamefont
  {Leclaire}}, \bibinfo {author} {\bibfnamefont {N.}~\bibnamefont {Pellerin}},
  \bibinfo {author} {\bibfnamefont {M.}~\bibnamefont {Reggio}}, \ and\ \bibinfo
  {author} {\bibfnamefont {J.-Y.}\ \bibnamefont {Trepanier}},\ }\href {\doibase
  https://doi.org/10.1016/j.ijmultiphaseflow.2013.07.001} {\bibfield  {journal}
  {\bibinfo  {journal} {Int. J. Multiphas. Flow}\ }\textbf {\bibinfo {volume}
  {57}},\ \bibinfo {pages} {159 } (\bibinfo {year} {2013})}\BibitemShut
  {NoStop}%
\bibitem [{\citenamefont {Liu}\ \emph {et~al.}(2012)\citenamefont {Liu},
  \citenamefont {Valocchi},\ and\ \citenamefont {Kang}}]{PhysRevE.85.046309}%
  \BibitemOpen
  \bibfield  {author} {\bibinfo {author} {\bibfnamefont {H.}~\bibnamefont
  {Liu}}, \bibinfo {author} {\bibfnamefont {A.~J.}\ \bibnamefont {Valocchi}}, \
  and\ \bibinfo {author} {\bibfnamefont {Q.}~\bibnamefont {Kang}},\ }\href
  {\doibase 10.1103/PhysRevE.85.046309} {\bibfield  {journal} {\bibinfo
  {journal} {Phys. Rev. E}\ }\textbf {\bibinfo {volume} {85}},\ \bibinfo
  {pages} {046309} (\bibinfo {year} {2012})}\BibitemShut {NoStop}%
\bibitem [{\citenamefont {Grunau}\ \emph {et~al.}(1993)\citenamefont {Grunau},
  \citenamefont {Chen},\ and\ \citenamefont {Eggert}}]{grunau1993lattice}%
  \BibitemOpen
  \bibfield  {author} {\bibinfo {author} {\bibfnamefont {D.}~\bibnamefont
  {Grunau}}, \bibinfo {author} {\bibfnamefont {S.}~\bibnamefont {Chen}}, \ and\
  \bibinfo {author} {\bibfnamefont {K.}~\bibnamefont {Eggert}},\ }\href
  {\doibase 10.1063/1.858769} {\bibfield  {journal} {\bibinfo  {journal} {Phys
  Fluids A-Fluid}\ }\textbf {\bibinfo {volume} {5}},\ \bibinfo {pages} {2557}
  (\bibinfo {year} {1993})}\BibitemShut {NoStop}%
\bibitem [{\citenamefont {Leclaire}\ \emph {et~al.}(2017)\citenamefont
  {Leclaire}, \citenamefont {Parmigiani}, \citenamefont {Malaspinas},
  \citenamefont {Chopard},\ and\ \citenamefont {Latt}}]{PhysRevE.95.033306}%
  \BibitemOpen
  \bibfield  {author} {\bibinfo {author} {\bibfnamefont {S.}~\bibnamefont
  {Leclaire}}, \bibinfo {author} {\bibfnamefont {A.}~\bibnamefont
  {Parmigiani}}, \bibinfo {author} {\bibfnamefont {O.}~\bibnamefont
  {Malaspinas}}, \bibinfo {author} {\bibfnamefont {B.}~\bibnamefont {Chopard}},
  \ and\ \bibinfo {author} {\bibfnamefont {J.}~\bibnamefont {Latt}},\ }\href
  {\doibase 10.1103/PhysRevE.95.033306} {\bibfield  {journal} {\bibinfo
  {journal} {Phys. Rev. E}\ }\textbf {\bibinfo {volume} {95}},\ \bibinfo
  {pages} {033306} (\bibinfo {year} {2017})}\BibitemShut {NoStop}%
\bibitem [{\citenamefont {Brackbill}\ \emph {et~al.}(1992)\citenamefont
  {Brackbill}, \citenamefont {Kothe},\ and\ \citenamefont
  {Zemach}}]{BRACKBILL1992335}%
  \BibitemOpen
  \bibfield  {author} {\bibinfo {author} {\bibfnamefont {J.}~\bibnamefont
  {Brackbill}}, \bibinfo {author} {\bibfnamefont {D.}~\bibnamefont {Kothe}}, \
  and\ \bibinfo {author} {\bibfnamefont {C.}~\bibnamefont {Zemach}},\ }\href
  {\doibase https://doi.org/10.1016/0021-9991(92)90240-Y} {\bibfield  {journal}
  {\bibinfo  {journal} {J. Comput. Phys.}\ }\textbf {\bibinfo {volume} {100}},\
  \bibinfo {pages} {335 } (\bibinfo {year} {1992})}\BibitemShut {NoStop}%
\end{thebibliography}%

\end{document}